\documentclass[pop,numberedappendix,iop]{aeb_emulateapj_2015}
\usepackage{amsmath,cases}
\usepackage{color}
\usepackage{longtable}
\usepackage{ulem}
\usepackage{grffile}
\usepackage{multirow}

\usepackage[mathscr]{euscript}

\definecolor{epcol}{rgb}{0.398, 0.0, 0.797}

\begin{document}

\title{
SHARP:
A Spatially Higher-order, Relativistic Particle-in-Cell Code}

\author
{
Mohamad Shalaby\altaffilmark{1,2,3}, 
Avery E. Broderick\altaffilmark{1,2}, 
Philip Chang\altaffilmark{4},
\\
Christoph Pfrommer\altaffilmark{5,6}, 
Astrid Lamberts\altaffilmark{7}
and
Ewald Puchwein\altaffilmark{8}}

\altaffiltext{1}{Department of Physics and Astronomy, University of Waterloo, 200 University Avenue West, Waterloo, ON, N2L 3G1, Canada}
\altaffiltext{2}{Perimeter Institute for Theoretical Physics, 31 Caroline Street North, Waterloo, ON, N2L 2Y5, Canada}
\altaffiltext{3}{Department of Physics, Faculty of Science, Cairo University, Giza 12613, Egypt}
\altaffiltext{4}{Department of Physics, University of Wisconsin-Milwaukee, 1900 E. Kenwood Boulevard, Milwaukee, WI 53211, USA}
\altaffiltext{5}{Leibniz-Institut f{\"u}r Astrophysik Potsdam (AIP), An der Sternwarte 16, 14482 Potsdam, Germany}
\altaffiltext{6}{Heidelberg Institute for Theoretical Studies, Schloss-Wolfsbrunnenweg 35, 69118 Heidelberg, Germany}
\altaffiltext{7}{Theoretical Astrophysics, California Institute of Technology, Pasadena, CA 91125, USA}
\altaffiltext{8}{Institute of Astronomy and Kavli Institute for Cosmology, University of Cambridge, Madingley Road, Cambridge, CB3 0HA, UK}

\email{mshalaby@live.ca}

\shorttitle{SHARP: A Spatially Higher-order, Relativistic Particle-in-Cell Code}
\shortauthors{Shalaby et al.}

\begin{abstract}

Numerical heating in particle-in-cell (PIC) codes currently precludes the accurate simulation of cold, relativistic plasma over long periods, severely limiting their applications in astrophysical environments.
We present a spatially higher-order accurate relativistic PIC
algorithm in one spatial dimension, which conserves charge and momentum exactly.
We utilize the smoothness implied by the usage of higher-order interpolation functions to achieve a spatially higher-order accurate algorithm (up to fifth order).
We validate our algorithm against several test problems --
thermal stability of stationary plasma, stability of linear plasma waves, and
two-stream instability in the relativistic and non-relativistic regimes.
Comparing our simulations to exact solutions of the dispersion
relations, we demonstrate that SHARP can \textit{quantitatively} reproduce
important kinetic features of the linear regime.
Our simulations have a superior ability to control energy non-conservation and avoid numerical heating in comparison to common second-order schemes.
We provide a natural definition for convergence of a general PIC algorithm: the complement of physical modes captured by the simulation, i.e., those that lie above the Poisson noise, must grow commensurately with the resolution.  This implies that it is necessary to simultaneously increase the number of particles per cell and decrease the cell size.
We demonstrate that traditional ways for testing for convergence fail, leading to plateauing of the energy error. This new PIC code enables us to faithfully study the long-term evolution of plasma problems that require absolute control of the energy and momentum conservation.
\end{abstract}

\section{Introduction}
\label{sec:intro}

\begin{figure}
\center
\includegraphics[width=8.4cm]{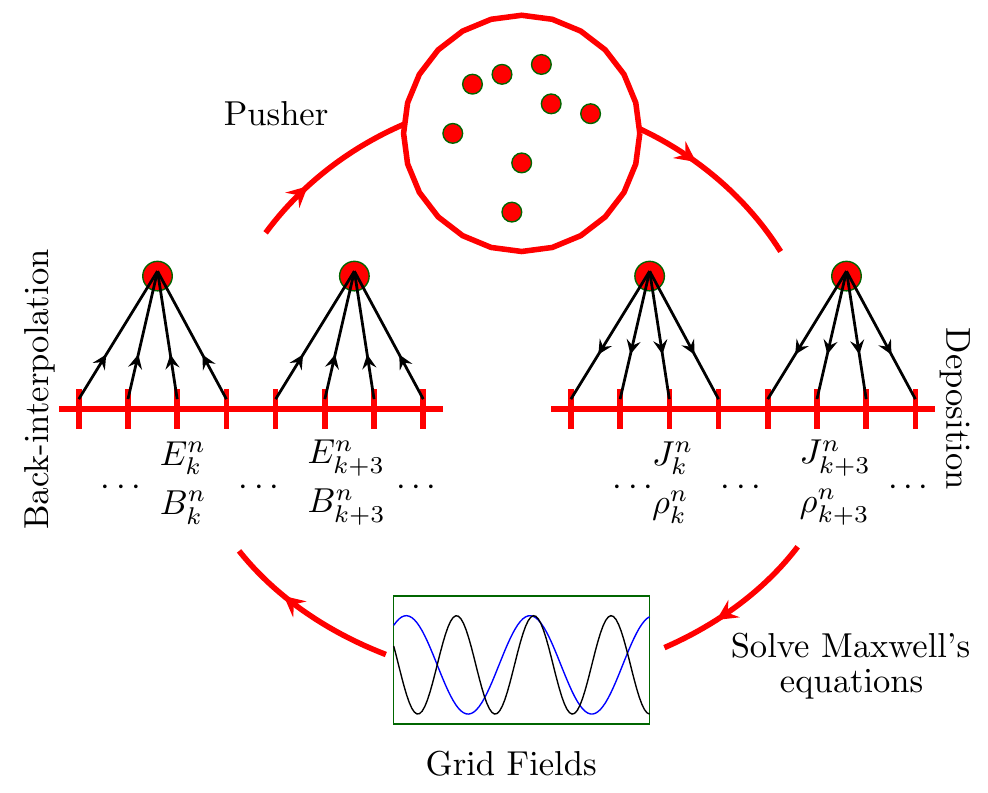}
\caption{
\label{fig:pic-loop}
Schematic representation for the general loop in the PIC method:
starting from the three o'clock position and moving clockwise, the macro-particles' position and velocity data ($x,v$) are deposited onto a physical grid to construct charge and current densities ($\rho_k$ and $J_k$) at control points ($k$) of the grid (Section~\ref{subsec:deposition}). 
These are used to solve Maxwell's equations, which yield the self-consistently computed (electric and magnetic) fields at these control points of the physical grid (Section~\ref{subsec:MEs}).
The updated fields are, then, back-interpolated on the macro-particles to construct the Lorentz force on macro-particles (Section~\ref{subsec:BI}), which is then used to evolve them via a particle pusher (Section~\ref{subsec:pusher}).
}
\end{figure}

The PIC method is a very powerful numerical tool to study the evolution of plasmas, it is used to model plasmas ranging from laboratory experiments to astrophysical environments.
First proposed in one spatial dimension (1D) by \citet{Buneman+1959} and \citet{Dawson+1962}, the general idea of this algorithm is straightforward: it follows the trajectory of particles with $N$-body methods, while solving Maxwell's equation on a Eulerian grid. The communication between grid points and particles is achieved through interpolation. The general loop (described in Figure~\ref{fig:pic-loop}) consists of first interpolating particle positions and velocities to a spatial grid to solve for the resulting charge and current density. Maxwell's equations are then solved on the grid with these source terms to find the self-consistent electromagnetic (electrostatic in 1D) fields. Fields are then interpolated back to the particle positions to calculate the Lorentz force, and hence, acceleration, to forward evolve the particles in time using a so-called pusher.
This reduces the number of computational operations from $\sim O(N^2)$ (such as in the case of $N$-body methods) to $\sim O(N)$, where $N$ is the number of particles in the simulation. This also results in eliminating all wave modes in the electromagnetic fields on scales smaller than the cell size on the grid upon which they are computed.

A major test of the accuracy and fidelity of different PIC schemes is their ability to preserve conserved quantities such as energy, momentum, and charge.  
Often, this is required to accurately study subdominant, relativistic populations that typically arise in astrophysical contexts. Examples include nonthermal particle populations accelerated at shocks and reconnection events \citep{Tristan-Relativistic_Shocks,Tristan-Reconnection}, propagation of cosmic rays \citep{Tristan-Cosmic-ray}, interaction of accretion disks and coronae \citep{Stone+disk-coronae}, and TeV blazar driven beam instabilities \citep{blazarI}.
In the latter, this problem is especially severe, with the beams being both numerically and energetically subdominant while being highly relativistic; even a small degree of heating in the background can impact or overwhelm the evolution of beam plasma instabilities.

Direct interpolation of particle data to construct charge and current densities on the grid, in general, leads to a violation of charge conservation. However, modern algorithms typically use charge conserving methods to perform such step while maintaining the charge conservation \citep[e.g.,][]{Eastwood+1991,Villasenor+Buneman+1992,Esirkepov+2001,Umeda+2003}.
Energy and momentum conservation on the other hand appear to be mutually exclusive \citep[see, e.g.,][]{Brackbill+2016}.
Due to their importance, several schemes have been developed to ameliorate their non-conservation using different underlying methodologies.

Recently introduced implicit methods \citep{Ghena+Chacona+Barnesb+2011,Lapenta+Markidis+2011,Markidis+Lapenta+2011} can in theory preserve total energy exactly (though in practice may not), while violating momentum conservation.
For computationally simpler explicit, momentum conserving schemes, energy conservation is improved by filtering the deposited grid moments (charge and current densities), as is done in the TRISTAN-MP code \citep{Tristan+1993,Tristan-mp+2005}.
However, filtering, when used with a momentum conserving scheme, leads to a violation of momentum conservation and non-vanishing self-forces (see Appendix~\ref{app:filtering}).
In lieu of filtering, energy conservation is also improved by using higher-order interpolation functions, which is the approach that we will adopt below for both forward- (from particles to fields) and back-interpolation (from fields to particles) steps. This has the added advantage that momentum conservation is maintained. 

In this paper, we describe an implementation of the PIC algorithm in 1D that uses high-order spline functions (up to the fifth order) for the forward- and back-interpolation steps of the algorithm. We couple this with an exact Poisson solver and a second-order symplectic integrator, i.e., leap frog, to produce a second-order accurate code called SHARP-1D.
SHARP-1D displays superior energy-conservation properties while conserving the momentum exactly.
The smoothness coming from the usage of high-order interpolation functions is utilized to construct an up to fifth-order spatially accurate algorithm.\footnote{Due to the usage of lower order interpolation functions, current algorithms perform the back-interpolation step with second-order spatial accuracy \citep[e.g.,][]{Lapenta+Markidis+2011, photon-plasm+2015, Brackbill+2016,  Lapenta+2016}.}
Despite the high spatial order accuracy, our code remains second-order accurate because it is limited by the accuracy of the particle pusher. This will be addressed in future work.

The paper is organized as follows. In Section~\ref{sec:pic}, we describe the basic equations of the PIC method, our choice of discretization of the equations, and discuss sources for the numerical error. After discussing the order of accuracy of the solution, we describe our choice of code units and the implementation of SHARP-1D.
In Section~\ref{sec:conservations}, we discuss the conservation properties of different PIC algorithms.
In Section~\ref{validation-sec}, we demonstrate the different capabilities of SHARP-1D by validating it against several test problems: thermal stability of plasma, plasma oscillation frequency and linear-Landau damping of standing plasma waves, and two-stream instabilities in both relativistic and non-relativistic regimes. We compare several of the results of SHARP-1D to the results of TRISTAN-MP simulations in Section~\ref{sec:comparison}.
Finally, we study the convergence properties for our algorithm in Section~\ref{sec:convergence} and discuss the performance of  SHARP-1D in Section~\ref{sec:implementation}. We conclude in Section~\ref{sec:conclusion}.

\section{The PIC method}

\label{sec:pic}

The evolution of the particles that comprise a plasma is described by the Boltzmann and Maxwell's equations.  In the absence of collisions, the particles are described by the Vlasov equation, which in one spatial dimension is
\begin{equation}
  \partial_t f_s(x,u,t) + \frac{u}{\gamma} \partial_x f_s(x,u,t) + \frac{q_{s} E(x,t)}{m_{s}} \partial_u f_s(x,u,t) = 0,
  \label{dis}
\end{equation}
where $s$ denotes a particle species, characterized by its charge $q_s$ and mass
$m_s$, $u = \gamma v$ is the spatial component of the four-velocity, $\gamma =
\sqrt{1+(u/c)^2}$ is the Lorentz factor, $E(x,t)$ is the electric field, and
$f_s(x,u,t)$ is the phase-space distribution functions of particles of species $s$.

In one dimension, Maxwell's equations imply that the magnetic field is constant and along the direction of the particle motion.
Thus, it impacts the evolution of neither the particle nor the electric field, and we take it to be zero henceforth.
Therefore, Maxwell's equations reduce to
\begin{eqnarray}
\partial_x E(x,t) - \frac{\rho(x,t)}{\epsilon_0} = 0
\quad\text{and}\quad
\partial_t E(x,t) + \frac{j(x,t)}{\epsilon_0} = 0\,,
\label{ME2}
\end{eqnarray}
where $\rho$ and $j$ are the charge and current densities, respectively.
\footnote{Here, we use SI units, and $\epsilon_0$ is the permittivity of the vacuum.  To convert to CGS units, $\epsilon_0$ needs to be replaced by  $1/4\pi$.  Ultimately, we convert to a system of numerical units, obviating the distinction between these (Section \ref{subsec:norm}).}
This set of equations is closed by the following equations for $\rho$ and $j$:
\begin{equation}
  \begin{aligned}
    \rho(x,t) &= \sum_s  q_{s}  \int f_s(x,u,t) du\\
    j(x,t) &= \sum_s  q_{s}  \int  \frac{u}{\gamma} f_s(x,u,t) du\,.
  \end{aligned}
\end{equation}

\subsection{Smoothing Phase-space Distribution Function}
The distribution function for point-like particles is given by the Klimontovich distribution function:
\begin{equation}
f^K_s (x,u) = \sum_{i_s }^{N_s^p} \delta(x-x_{i_s})\delta(u-u_{i_s}),
\end{equation}
where $N_s^p$ is the number of physical (point-like) particles and $\delta(x)$ is the Dirac delta function.

Direct usage of this singular distribution function is impractical for two reasons.
First, the number of particles to be simulated is too large to be tractable, and thus, second, this distribution function would result in an overwhelming shot noise due to the finite number of particles used in practice \citep{Lipatov+2002}.
Both can be mitigated in simulations by using a smoothed approximation for the distribution function. Thus, we approximate the distribution function by
\begin{equation}
f_s(x,u) 
=
w \sum_{i_s}^{N_s} S(x-x_{i_s})\delta(u-u_{i_s})
\approx
\int dx' f^K_s(x',u) S(x',x),
\label{Macro-distribution}
\end{equation}
where $N_s$ is the number of macro-particles (defined below), and $w = N_s^p/N_s$ is the number of physical particles that a macro-particle represents.

The macro-particles (also called computational particles\footnote{Both macro-particles and particles will be interchangeably used to mean the same thing: particles used in simulations. When we refer to physical particles, it will be explicitly specified.}) have a ``shape'' $S(x,x_{i_s})$, i.e., a smoothed localized charge distribution that has $w$ physical particles centered at $x_{i_s}$, with
\begin{equation}
  \int S(x,x_{i_s}) dx =1.
\end{equation}
The charge and the mass for these computational particles are $Q_s = w q_s$ and $M_s = w m_s$, respectively. 

The macro-particles have the same plasma frequency $\omega_P$ as the physical particles they represent in simulations: they have same charge-to-mass ratio and hence $\omega_p^2 = q_s^2 N^p_s/ (V m_s \epsilon_0) = Q_s^2 N_s/ (V M_s \epsilon_0)$, where $V$ is the volume.
Both macro-particles and physical particles also have the same normalized temperature $\theta_s = k_B T_s/M_s c^2 = k_B T_s^p/m_s c^2$, where $T_s$ and $T_s^p$ are the effective temperatures of the macro-particles and physical particles, respectively. This arises from the assumption that the macro-particles are monolithic, and thus have the same {\it velocity} distribution as the underlying physical particles, and thus
\begin{equation}
  \frac{k_B T_s}{M_s c^2}
  =
  \frac{\int du (\gamma-1) f_s}{\int du f_s}
  =
  \frac{\int du (\gamma-1) f^K_s}{\int du f^K_s}
  =
  \frac{k_B T^p_s}{m_s c^2}.
\end{equation}

Inserting the phase-space distribution function in Equation~\eqref{Macro-distribution} into the Vlasov--Maxwell system (Equations \eqref{dis} and \eqref{ME2}), we obtain from the first two moments the following equations of motion for the macro-particle of species $s$:

\begin{eqnarray}
\frac{d x_{i_s}}{dt} &=& \frac{u_{i_s}}{\gamma_{i_s}}, 
\quad 
\frac{du_{i_s}}{dt} = \frac{ Q_s }{ M_s }  E_{i_s},
\label{eqm}
\\
E_{i_s} &\equiv& \int  E(x) S(x,x_{i_s}) dx,
\label{EkEi}
\end{eqnarray}
where $E(x)$ is the solution of Maxwell's equations  (\ref{ME2}) with moments given by
\begin{eqnarray}
\rho(x,t) &=& \sum_s Q_s  \sum_{i_s}^{N_s} S(x,x_{i_s} ) ,
\label{moment1}
\\
j(x,t)   &=&  \sum_s Q_s  \sum_{i_s}^{N_s} v_{i_s}  S(x,x_{i_s}),
\label{moment2}
\end{eqnarray}
where $ v_{i_s} =  u_{i_s} /\gamma_{i_s} $.
Below we explain how implicit discretization of such a system of equations is achieved in our code.

\subsection{Spatial grid}
\label{subsec:Discretiztiona} 

For a system of macro-particles in a periodic box (line) of length $L$, we divide our domain into $N_c$ cells each of size $\Delta x =
L/N_c$. Assuming $k \in \{0,1, ... , N_c-1\}$, we define the $k$th cell ($c_k$)
centered at $ x_{k+1/2} \equiv x_k+\Delta x/2$ as $x \in [x_k,x_{k+1} )$, where $x_k \equiv \Delta x \hspace{0.1cm} k $.

We adopt spline functions extending over a number of grid cells as the shape
function of the macro-particles. Therefore, the distribution of physical
particles inside these macro-particles is symmetric around their center and extends over a number of computational cells depending on the order $m$ of the spline
functions used.  For instance, $m=1$ is a top-hat distribution (shape) given by
\begin{equation}
S(x,x_i)
\rightarrow
S^1\left(\frac{ |x-x_i|}{\Delta x}\right) =
\frac{1}{\Delta x}
\begin{cases}
1, \qquad \text{If } | x-x_{i}| <  \Delta x/2,
\\ \\
0, \qquad  \text{otherwise. } 
\end{cases}
\end{equation}

We also define, $t^n = n\times\Delta t$, $E^n \equiv E(t^n)$, i.e., superscript $n$ denotes the $n$th time step for the particular quantity. We choose the time step $\Delta t$ such that $c \Delta t \leq \Delta x$, to obey the Courant--Friedrichs--Lewy (CFL) stability condition in 1D \citep{CFL+1967}.

\subsection{Charge and current deposition}
\label{subsec:deposition}

To obtain a discrete set of equations that governs the evolution of such
macro-particles, we begin by integrating the first equation in \eqref{ME2} over the $k$th
cell $c_k$:

\begin{eqnarray}
E^n_{k+1} - E^n_k
&=&
\int_{c_k} \frac{\rho^n}{\epsilon_0} dx
=
\frac{\Delta x}{\epsilon_0} \rho^n_{k+1/2}.
\label{eq:Ek-rho}
\end{eqnarray}
Here, $ \rho^n_{k+1/2}$ is the average charge density inside $c_k$ at $t=t^n$:
\begin{eqnarray}
\rho^n_{k+1/2} 
&=&
\int_{c_k} \rho^n \frac{dx}{\Delta x}
=
\sum_s  Q_s  \sum_{i_s} \int_{x_{k}}^{x_{k+1}} 
S^m\left(\frac{x - x^n_{i_s}}{\Delta x}\right)
 \frac{dx}{\Delta x}
\nonumber \\
&=&
\sum_s  \frac{Q_s}{\Delta x} \sum_{i_s}  
 W^m\left(\frac{ x_{k+1/2} - x^n_{i_s}}{\Delta x} \right) ,
 \label{rhok}
\end{eqnarray}
where
\begin{eqnarray}
 W^m\left(\frac{ x_k - x^n_{i_s}}{\Delta x} \right) &=&
 \int_{x_{k}-\Delta x/2 }^{x_{k}+\Delta x/2 } S^m\left(\frac{x - x^n_{i_s}}{\Delta x}\right) dx
\label{Eq:shape-weight}
\end{eqnarray}
defines the weighed contribution of a macro-particle at $x_{i_s}$ to the average charge density of the $k$th cell.

The explicit forms for both shape $S^m$ and weight $W^m$ functions that we use in our code are given in Appendix~\ref{app:shapes-weights}. In Figure~\ref{fig:weights}, we plot different weight functions. 

\begin{figure}
\center
\includegraphics[width=8.4cm]{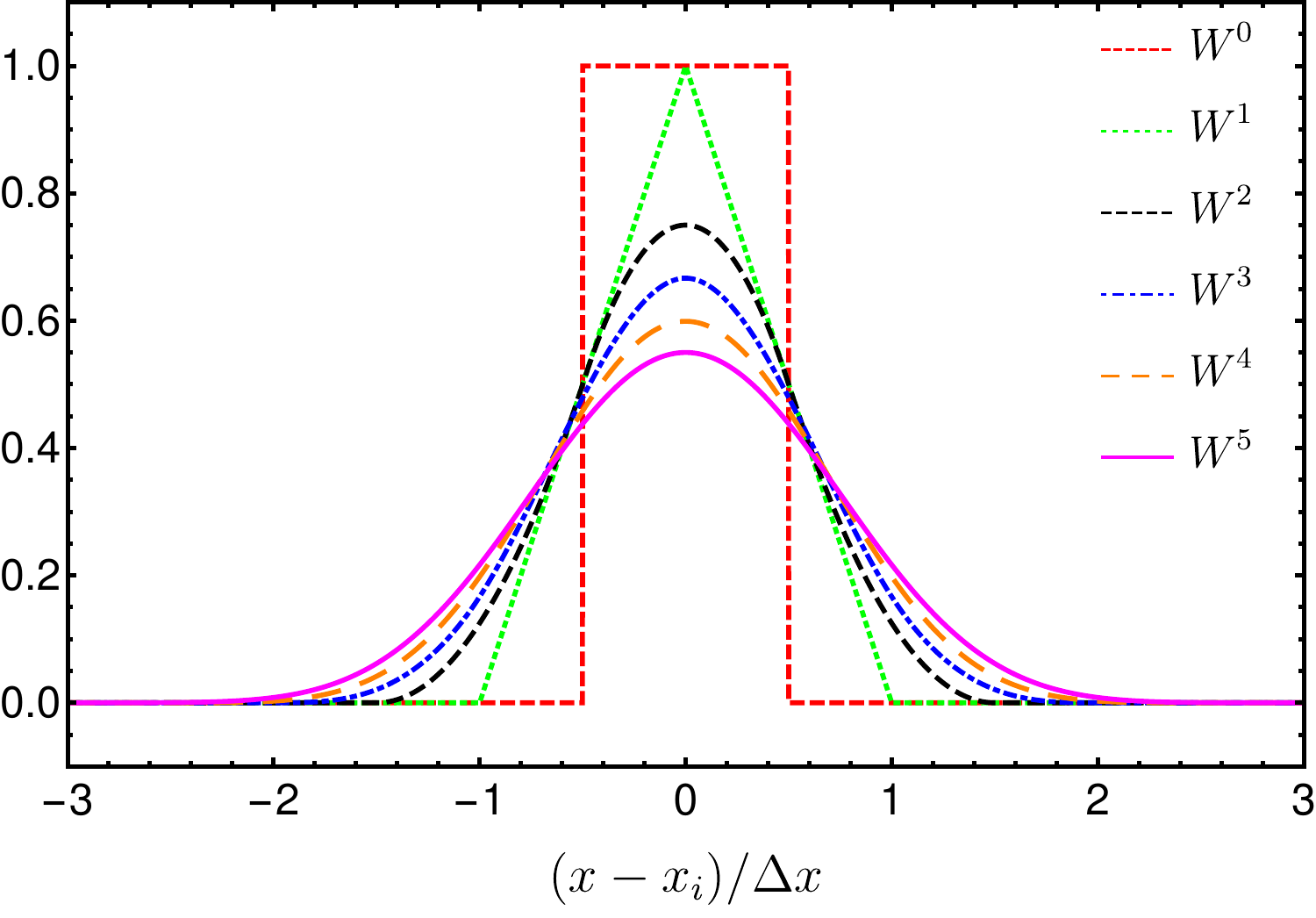}
\caption{Different weight functions implemented in the code. The relation between
  shape and weight functions is defined in
  Equation~\eqref{Eq:shape-weight}. Explicit forms for both, shape and weight
  functions are given in Appendix~\ref{app:shapes-weights}.
\label{fig:weights}
}
\end{figure}

\subsection{Solving Maxwell's equations}
\label{subsec:MEs}

Equation~\eqref{eq:Ek-rho} gives the change in the electric field at cell edges exactly. However, a complete solution also requires the boundary condition, $E_{k=0}$. Therefore, we rewrite Equation~\eqref{eq:Ek-rho} as
\begin{equation}
E^n_k = 
E^n_0 
+ 
\frac{\Delta x}{\epsilon_0}  
\sum^{k-1}_{j=0} \rho^n_{j+1/2}.
\end{equation}
We see that $E^n_k$ inherits the error of $E^n_{0}$.  To find $E^n_0$, we first
find the sum of cell-edges field $E^n_{\rm tot} \equiv \sum_k E^n_k $. The second equation in \eqref{ME2} can be re-written (for $m>0$) as
\begin{eqnarray}
\label{tt}
\partial_t E_{\rm tot}
&=& 
-
\sum_k \frac{j_k}{ \epsilon_0 } 
= 
-
\frac{1}{\epsilon_0}
\sum_s Q_s \sum_{i_s} v_{i_s}
\sum_k S^m(x_k,x_{i_s})
\nonumber \\
&=& 
- 
\sum_s \frac{Q_s}{\Delta x \epsilon_0}
\sum_{i_s} v_{i_s}
= \frac{- j_{\rm tot}}{   \epsilon_0 }.
\end{eqnarray}
The first equality uses $\sum_k \Delta x S^m(x_k,x_i) = \sum_k W^{m-1}(x_k,x_i)=1$, which is a property of spline functions implemented in our code (see Table~\ref{tab:shapes-ws}).
Note that $ j_{\rm tot}$ can be calculated exactly at each time-step using
the macro-particles' velocities.
If the plasma macro-particles have a total
current, then $\partial_t E \neq 0$, i.e., setting $\partial_t E$ to zero will artificially add a constant electric field on the grid or equivalently a counter current. 

Equation~\eqref{tt} is approximated to second, third, and fourth orders of accuracy, respectively, as
\begin{eqnarray}
E^{n}_{\rm tot}
&=&
E^{n-1}_{\rm tot}
- 
\frac{\Delta t}{\epsilon_0} j_{\rm tot}^{n-\frac{1}{2}}
+ 
O(\Delta t^3)
\label{Eq:O2-Et-jt}
\\
E^{n}_{\rm tot}
&=&
\frac{21}{23} E^{n-1}_{\rm tot}
+
\frac{3}{23} E^{n-2}_{\rm tot}
-
\frac{1}{23} E^{n-3}_{\rm tot}
\nonumber \\
&& 
-
\frac{24}{23} \frac{\Delta t}{\epsilon_0} j_{\rm tot}^{n-\frac{1}{2}}
+ 
O(\Delta t^4)
\label{Eq:O3-Et-jt}
\\
E^{n}_{\rm tot}
&=&
\frac{17}{22} E^{n-1}_{\rm tot}
+
\frac{9}{22}  E^{n-2}_{\rm tot}
-
\frac{5}{22}  E^{n-3}_{\rm tot}
\nonumber \\
&& 
+
\frac{1}{22}  E^{n-4}_{\rm tot}
-
\frac{12}{11} \frac{\Delta t}{\epsilon_0} j_{\rm tot}^{n-\frac{1}{2}}
+ 
O(\Delta t^5).
\label{Eq:O4-Et-jt}
\end{eqnarray}

Proceeding in a similar way to generate higher-order accurate, asymmetric estimates for $E^n_{\text{tot}}$ results in numerically unstable approximations.
Thus, we stop at fourth-order accurate method given in Equation~\eqref{Eq:O4-Et-jt}.

To find $E_0$ from $E_{\rm tot}$, we multiply Equation~\eqref{eq:Ek-rho} by the
index $k$, and then sum over all cells. The left-hand side is given by
\begin{equation}
\sum_{k=0}^{N_c-1} k \left( E^n_{k+1} -E^n_{k} \right)
=
\sum_{k=1}^{N_c} (k-1) E^n_k - \sum_{k=1}^{N_c-1} k E^n_k
= N_c E^n_{N_c} - E^n_{\rm tot}.
\end{equation}
Then, using the periodicity of $E_k$, i.e., $E_{N_c} = E_0$, we can write
\begin{equation}
\label{eq:E0-Etot}
E^n_0
=
\frac{E^n_{\rm tot}}{N_c}
+
\frac{\Delta x }{\epsilon_0 N_c} 
\sum_{k=0}^{N_c-1}
k 
\hspace{0.08cm}
\rho^n_{k+1/2}.
\end{equation}
Thus, for a given set of particle data $\{x^n_{i_s},v^{n-1/2}_{i_s}\}$, we are able to find the electric field at the edges of the cells, $E^n_k$. Importantly, the only source of error is the error introduced in finding $E^n_{\rm tot}$. 

The error, for fixed $\Delta t /\Delta x$ (motivated by the CFL condition), is of order $O(\Delta x^4)$ if we use Equation~\eqref{Eq:O2-Et-jt} to update $E_{\rm  tot}^n$. The error drops to $O(\Delta x^5)$ or $O(\Delta x^6)$ with Equations~\eqref{Eq:O3-Et-jt} or \eqref{Eq:O4-Et-jt}, respectively.
It is also important here to note that an $O(\Delta x^k)$ error in the force would introduce an $O(\Delta x^{k+1})$ error in the updated data of the macro-particles.

\subsection{Back-interpolation: Force on Macro-particles}
\label{subsec:BI}

Having determined the electric field using Equations \eqref{eq:Ek-rho},
\eqref{rhok}, and \eqref{Eq:O2-Et-jt} or \eqref{Eq:O3-Et-jt} or
\eqref{Eq:O4-Et-jt} to different orders of accuracy from the macro-particle positions and velocities, we now calculate the force on the individual macro-particles. The effective electric field that acts on the macro-particle (Equation~\eqref{EkEi}) can be determined from the electric field at each cell
face, $E_k^n$ as follows
\begin{eqnarray}
E^n_{i_s}
&=&
\sum_k \int_{c_k}  E^n(x) S^m(x,x^n_{i_s}) dx
\nonumber \\
&=&
\sum_k  \frac{E^n_{k+1}+ E^n_k}{2}  \int_{x_k}^{x_{k+1}}  S^m(x,x^n_{i_s}) dx + O(\Delta x^2)
\nonumber \\
&=&
\sum_k  \frac{E^n_{k+1}+ E^n_k}{2}  W^m\left[ (x_{k+1/2} - x^n_{i_s})/\Delta x  \right] + O(\Delta x^2). 
\label{EkEiapprox}
\end{eqnarray}

In Appendix~\ref{app:approximation2}, we show that the approximation in Equation~\eqref{EkEiapprox}
leads to exact momentum conservation and vanishing self-forces. Another possible approximation of $E^n_{i_s}$, that has the same order of accuracy is given by
\begin{eqnarray}
E^n_{i_s}
&=&
\sum_k  E^n_{k}  \int_{x_{k-1/2}}^{x_{k+1/2}}  S^m(x,x^n_{i_s}) dx + O(\Delta x^2)
\nonumber\\
&=&
\sum_k  E^n_k   W^m\left[ (x_k - x^n_{i_s})/\Delta x  \right] + O(\Delta x^2). 
\label{EkEiapprox2}
\end{eqnarray}

The approximation in Equation~\eqref{EkEiapprox2} is used, for instance, in \citet{photon-plasm+2015}.
It generally leads to a violation of momentum conservation and unphysical self-forces as shown in Appendix~\ref{app:approximation1}. 
The order of accuracy for the back-interpolation step is typically second order \citep[e.g.,][]{Lapenta+Markidis+2011, photon-plasm+2015, Brackbill+2016, Lapenta+2016} because of the use of lower order interpolation functions.

Using higher-order interpolation functions implies a smoother representation of the phase-space distribution function. As a consequence, we can assume a smoother representation of the electric fields\footnote{An order $m$ interpolation function means that $f$ is $m+1$ times spatially differentiable.}.
Therefore, since we implemented up to fifth-order spline interpolation function, this allows constructing up to \textit{fifth-order} accurate back-interpolation.
To derive a higher-order accurate method, we approximate the integration $I_k =  \int_{c_k}  E^n(x) S(x,x^n_{i_s}) dx$ using Simpson's rule:
\begin{eqnarray}
\label{Eq:simpson-rule}
I_k 
&=&   
\int^{x_{k+1}}_{x_k}  E^n(x) S^m(x,x^n_{i_s}) dx 
\nonumber \\
&=&   
\frac{ \Delta x}{6} \left[  
E^n_k   S^m_{k,i_s} +
4 E^n_{k+\frac{1}{2}}   S^m_{k+\frac{1}{2},i_s} +
E^n_{k+1}   S^m_{k+1,i_s} 
\right] 
\nonumber \\
&&
+ O(\Delta x^5),
\end{eqnarray}
where we define $ S^m_{k,i_s} \equiv S^m(x_{k},x^n_{i_s})$. 

To utilize Equation~\eqref{Eq:simpson-rule}, we need to approximate $E^n_{k+\frac{1}{2}}$ in terms of cell-edge fields $E^n_{k}$ as follows
\begin{eqnarray}
E_{k+\frac{1}{2}}
&=&
\frac{ -E_{k+2} +9(E_{k}+E_{k+1})-E_{k-1} }{16}
+
O(\Delta x^4) 
\\
&=&
\frac{3 (E_{k+3}+E_{k-2}) +150( E_{k+1} +E_k )- 25 ( E_{k+2}+ E_{k-1} ) }{256} 
+
\nonumber \\ &&
O(\Delta x^6).
\end{eqnarray}
Using Equation~\eqref{Eq:simpson-rule} and since $\Delta x S^m \in [0,1]$ (see Table~\ref{tab:shapes-ws}), $O(\Delta x^4)$ error in $E_{k+\frac{1}{2}}$ implies $O(\Delta x^4)$ error order in $ E_{i_s}$ and $O(\Delta x^6)$ error in $E_{k+\frac{1}{2}}$ implies $O(\Delta x^5)$ error order in $ E_{i_s}$ (because of the error order in Equation~\eqref{Eq:simpson-rule}).
Therefore, the approximate electric field $E^n_{i_s}=\sum_k I_k $ on a macro-particle at $x_{i_s}$, using the periodicity of $E_k$, can be expressed as

\begin{eqnarray}
E^n_{i_s}
&=&
\sum_k   \Delta x S^m_{k+\frac{1}{2},i_s}  
\left[  \frac{ 9(E^n_{k}+E^n_{k+1})-(E^n_{k+2}+E^n_{k-1}) }{24} \right]
\nonumber \\
& & + 
 \sum_k \frac{ E^n_k }{3} \left[    \Delta x S^m_{k,i_s}  \right]
+ 
O(\Delta x^4)
\label{EkEi-order4}
\\
&=&
\sum_k \frac{ E^n_k }{3}  \left[    \Delta x S^m_{k+\frac{1}{2},i_s}  \right]
+
\sum_k  \Delta x S^m_{k+\frac{1}{2},i_s}   
\nonumber \\ &&  \times
\left[   \frac{3 (E^n_{k+3}+E^n_{k-2}) +150( E^n_{k+1} +E^n_k )- 25 ( E^n_{k+2}+ E^n_{k-1} ) }{384}  \right]
\nonumber \\
&& 
+ 
O(\Delta x^5).
\label{EkEi-order5}
\end{eqnarray}

To summarize, here we showed how we can find the forces on the individual macro-particles for
a given discretized field on a grid. The force error can be of order $O(\Delta x^2)$, $O(\Delta x^4)$, or $O(\Delta x^5)$ if we use
Equation~\eqref{EkEiapprox}, \eqref{EkEi-order4}, or \eqref{EkEi-order5},
respectively.
Given the numerical error in finding $E_k^n$, so far we have shown how, for a
given set of particle data $\{x^n_{i_s},v^{n-1/2}_{i_s}\}$, we can find the
forces on such particles $F^n_{i_s}$. The error can be of order $O(\Delta x^2)$,
$O(\Delta x^4)$, or $O(\Delta x^5)$, by employing consecutively higher-order
equations in finding $E^n_0$.

\subsection{Pusher: particle update}
\label{subsec:pusher}

To push the individual particles, we use a leapfrog scheme to discretize the
equations of motions for the particles,
\begin{eqnarray}
u^{n+1/2}_{i_s}
&=&
u^{n-1/2}_{i_s} + \Delta t \frac{Q_s}{M_s} E^n_{i_s} + O(\Delta t^3),
\\
x^{n+1}_{i_s} 
&=& 
x^{n}_{i_s} + \Delta t v^{n+1/2}_{i_s} + O(\Delta t^3).
\end{eqnarray}

The code assumes that the initial positions of the particles are provided at $t=0$, but
the initial momenta are given at $t=-\Delta t/2$. It also assumes that at
$t=-\Delta t$ the sum of electric field at cell edges was zero, i.e., $E_{\rm tot}^{n=-1} =0$.

\subsection{Error sources in our algorithm}
\label{sec:error_orders}

Here, we summarize the different sources of error in the algorithm presented above.

\begin{enumerate}
\item The use of discretized equations to compute $E_{\rm tot}$, yielding up to $O(\Delta t^5)$ accurate schemes (Equations~\eqref{Eq:O2-Et-jt}-\eqref{Eq:O4-Et-jt}).
\item The interpolation of the field from the grid to the macro-particle to calculate the force $F_{i_s}$.  This can be done at $O(\Delta x)$ in such a fashion that total momentum is conserved exactly, or at higher accuracy, $O(\Delta x^3)$ and $O(\Delta x^4)$ at the cost of (slightly) violating momentum conservation.
\item The updating of the particle positions, which is currently performed at $O(\Delta t^3)$.
\end{enumerate}

For fixed $\Delta t/\Delta x$, as implied by the CFL condition, the above imply that our code is fundamentally second-order accurate, limited by the particle pusher.  That is, despite improving energy and momentum conservation, the order of the interpolation function does not set the order of accuracy of the overall scheme.
Implementing a higher-order symplectic integrator within our scheme would improve the convergence order; we leave this point for future work. Nevertheless, as we will show in Sections~\ref{validation-sec}-\ref{sec:convergence}, the improved spatial order produces substantial practical enhancements in the code performance.

\subsection{Normalized equations} \label{subsec:norm}

Using the  fiducial units ($c, n_0, q_0, m_0$), we define the following scales
\begin{equation}
\begin{split}
t_0 = \sqrt{ m_0 \epsilon_0/( q_0^2 n_0) }, 
\hspace{0.5cm}
\mathbb{  E}_0   =  \sqrt{n_0 m_0 c^2/\epsilon_0},
\\ 
\rho_0 =  q_0 n_0,
\hspace{0.5cm}
 j_0 =   \rho_0 c,
 \hspace{0.5cm} 
x_0 = c t_0.
\end{split}
\label{normalizations}
\end{equation}
We then define our dimensionless variables as $\bar{t} =  t/t_0$, $dt  =  \Delta t/t_0$, $\bar{x} =  x/x_0$, $h =  \Delta x/x_0$, $\bar{u} =  u/c$, $\bar{Q}_s = Q_s/q_0$, $\bar{M}_s=M_s/m_0$, $\bar{\rho} = \rho/ \rho_0$, $\bar{j}= j/j_0$ and $\bar{E}= E/\mathbb{  E}_0$. 
We also identify $t_0$ as the time scale of the plasma frequency of the entire plasma that includes contributions from all species.
This defines $n_0$ as follows
\begin{equation}
\omega_0^2 \equiv t_0^{-2} = \frac{q_0^2 n_0}{ m_0 \epsilon_0 } = \omega_p^2 =  \sum_s \frac{Q^2_s n_s}{M_s \epsilon_0}
\quad \rightarrow \quad
n_0 = \sum_s \frac{\bar{Q}_s^2 n_s}{\bar{M}_s},
\end{equation}
where $n_s$ is the number density of the macro-particles of species $s$ and
\begin{equation}
n_0  \Delta x = \sum_s \frac{\bar{Q}_s^2 n_s \Delta x}{\bar{M}_s} 
=  
\frac{1}{N_c}  \sum_s \frac{\bar{Q}_s^2 N_s}{\bar{M}_s}.
\end{equation}
In the case where all species have the same mass and charge, $n_0  \Delta x = (\sum_s N_s ) / N_c = N_{\rm t}/N_c$.
In such a case, we chose our fiducial units so that $\bar{Q}_s^2 =1$, $\bar{M}_s =1 $, therefore, $n_0 = \sum_s n_s$.

In terms of the dimensionless variables defined above, the equations that are solved by the code can be written as follows
\begin{align}
\bar{\rho}^n_{k+1/2} 
&=
\frac{N_c}{  \sum_s (\bar{Q}_s^2 N_s)/\bar{M}_s }
 \sum_s \bar{Q}_s 
\sum_{i_s}  
 W^m\left[  (\bar{x}_{k+1/2} - \bar{x}^n_{i_s})/ h \right],
 \label{density-interpolation}
\\
\bar{E}^n_{k+1} 
&=
\bar{E}^n_k + h \bar{\rho}^n_{k+1/2},
\label{Discrete-ME}
\\
\bar{E}^n_{i_s}
&=
\sum_k  \frac{\bar{E}^n_{k+1}+ \bar{E}^n_k}{2}  W^m\left[ (\bar{x}_{k+1/2} - \bar{x}^n_{i_s})/ h  \right],
\label{particle-interpolation}
\\
\begin{split}
\label{Discrete-EOM}
\bar{u}^{n+1/2}_{i_s}
&=
\bar{u}^{n-1/2}_{i_s} + dt \frac{\bar{Q}_s}{\bar{M}_s} \bar{E}^n_{i_s} ,
\\
\bar{x}^{n+1}_{i_s} 
&=
\bar{x}^{n}_{i_s} + dt \bar{v}^{n+1/2}_{i_s}  .
\end{split}
\end{align}

The general loop in the code is then
\begin{equation}
\left\{  
\begin{split}
& \bar{x}^n_{i_s}
\\
& \bar{v}_{i_s}^{n-1/2}
\end{split}
  \right\}
\xrightarrow{  \eqref{density-interpolation} }   
\bar{\rho}^n_{k+1/2}  
\xrightarrow{  \eqref{Discrete-ME} \text{ } \& \text{ }  \bar{E}_{\text{tot}}^{n-1} } 
\bar{E}_k^{n} 
\xrightarrow{ \eqref{particle-interpolation} }  
\bar{E}^n_{i_s}  
\xrightarrow{ \eqref{Discrete-EOM} } 
\left\{  
\begin{split}
& \bar{x}^{n+1}_{i_s}
\\
& \bar{v}_{i_s}^{n+1/2}
\end{split}
  \right\}.
\nonumber
\end{equation}
A schematic representation of this loop is shown in Figure~\ref{fig:pic-loop}.

\subsection{Implementation}

SHARP-1D is implemented in C\texttt{++}, and is massively parallelized using MPI.
The parallelization is done by distributing macro-particles on different processors, while reserving the first processor (rank = 0) to compute the electric field on the grid and to manage outputs.

\section{Conserved quantities in PIC}
\label{sec:conservations}

As we mentioned in the introduction, a major test of the accuracy and fidelity of PIC schemes is their ability to preserve conserved quantities such as energy, momentum, and charge.
Below we discuss such conservation laws in PIC schemes and how well they are respected when different methodologies are used. We begin by showing that our algorithm is charge conserving. Then, we compare the energy and momentum conservation properties when implicit and explicit techniques are employed\footnote{Typically discretization in PIC is done using either explicit or implicit schemes. In explicit schemes (such as the algorithm presented here), particle data are used first to calculate the fields on the grid and then particle advancement in time is carried out using these fields.
On the other hand, when implicit discretization is employed, the equations for fields on the grid and evolution equations of particles have to be solved simultaneously in order to evolve forward in time. 
}.
We also show that the usage of higher-order interpolation leads to a decrease in the aliasing, which improves energy conservation while maintaining exact momentum conservation.

\subsection{Charge conservation}

In traditional implementations of PIC, the direct interpolation from particles to grid points, in order to calculate the grid charge and current densities, leads to violation of the continuity equation on the grid.
The reason for that is calculating the current density requires the knowledge of both particles positions and velocities at the half-time step and approximating the particles positions at half-time step leads to an error in the calculated current density when particles cross cell-boundaries.

Recently, several methods were proposed where the calculation of the current density on the grid from the particles is done such that the continuity equation is satisfied on the grid at all times \citep{Eastwood+1991,Villasenor+Buneman+1992,Esirkepov+2001,Umeda+2003}.
In the presented algorithm, we locally obey the discretized continuity equation at all times (i.e., we use a charge conserving scheme). The discretized current density\footnote{In fact, our presented algorithm does not require to calculate   the current density for solving Poisson's equation. } coincides with the current density proposed in \citet{Esirkepov+2001}.

To see this, we integrate the continuity equation $\partial_t \rho(x,t)+ \partial_x J_x(x,t)=0$  over a cell of size $\Delta x$,
\begin{equation}
\partial_t \rho_{k+1/2}(t) + \frac{J_{k+1} - J_{k+1}}{\Delta x} =0.
\end{equation} 
Therefore, we can write
\begin{equation}
\frac{\rho_{k+1/2}^{n+1} - \rho_{k+1/2}^n}{\Delta t} + \frac{J^{n+1/2}_{k+1} - J^{n+1/2}_{k}}{\Delta x} + O(\Delta t^3)  =0.
\end{equation} 
Using the second equation in \eqref{ME2}, the current density at cell edges can be expressed as follows
\begin{eqnarray}
J^{n+1/2}_{k} 
& \equiv &
 \sum_s Q_s  \sum_{i_s}^{N_s} v^{n+1/2}_{i_s}  S\left( x_k,x^{n+1/2}_{i_s} \right)
\\
 &=&
 - \frac{\epsilon_0}{\Delta t }
\left[
E^{n+1}_k - E^n_k
\right]
+ O(\Delta t^3).
\label{eq:j-conser}
\end{eqnarray}
Then using Equations~\eqref{eq:j-conser} and \eqref{eq:Ek-rho}, we can write 

\begin{eqnarray}
J^{n+1/2}_{k+1} - J^{n+1/2}_{k}
&=&
- \frac{\epsilon_0}{\Delta t }
\left[
(E^{n+1}_{k+1}  -E^{n+1}_k) - (E^n_{k+1}-E^n_k )
\right]+ O(\Delta t^3).
\nonumber \\
&=&
- \frac{\Delta x}{\Delta t } 
\left[
\rho^{n+1}_{k+1/2} - \rho^{n}_{k+1/2}
\right] + O(\Delta t^3).
\end{eqnarray}

Therefore, our scheme obeys exactly the second-order accurate continuity equation.

\subsection{Energy and momentum conservation}
\label{sec:EMconservations}

In general, for PIC schemes, energy and momentum conservation appear to be
mutually exclusive \citep[see, e.g.,][]{Brackbill+2016}.
Momentum non-conservation typically comes from non-vanishing self-forces and errors in the interaction forces -- an example of such a case is shown in Appendix~\ref{app:approximation1}. These non-physical forces can produce macroscopic non-physical instabilities \citep{Langdon+1973}.
Energy non-conservation can also produce dramatic changes in the evolution of the plasmas. Since such an error has a secular (unbounded) growth, energy non-conservation imposes a serious limitation on the ability to study the nonlinear phenomena which occurs in long time scales (compared to $\omega_p^{-1}$). 

For instance, \citet{Lapenta+Markidis+2011} demonstrated that in the two-stream instability, the errors in the energy are disproportionately distributed to the fast particles. In particular, they demonstrated that even though the per-particle violation in energy conservation is small, the disproportionate distribution of energy non-conservation leads to errors in the distribution of particles, especially at the high energy end. This is important for particle acceleration in relativistic situations \citep{Lapenta+Markidis+2011}.
Also, results that probe the long-term behavior of particle distribution functions starting from linearly unstable conditions such as tenuous beam instabilities or particle acceleration are subject to these issues \citep[e.g.][]{Sironi+Giannios+2014,2015ApJ...811...57A,2015PhRvL.114h5003P}.

Traditional explicit algorithms lead to numerical increase in the total energy (numerical heating) while conserving the total momentum exactly \citep{Birdsall-Langdon,Hockney-Eastwood}.  For explicit schemes, an energy conserving algorithm was developed in \citet{Lewis+1970}. However, the total energy is conserved only in the limit of $\Delta t \rightarrow 0$. In practice, there will be numerical heating of the plasma because of the finite timestep.

On the other hand, traditional implicit algorithms  tend to decrease the total energy numerically (numerical cooling), while violating the total momentum conservation \citep{Brackbill-Forslund}. Recently, implicit algorithms that, in principle, conserve the total energy exactly, while still \textit{violating} momentum conservation, were introduced: for non-relativistic/classical plasmas by \citet{Markidis+Lapenta+2011} and \citet{Ghena+Chacona+Barnesb+2011} and relativistic plasmas by \citet{Lapenta+Markidis+2011}. However, in practice, these algorithms use the Jacobian-Free Newton Krylov (JFNK) method to solve the full implicit system, introducing an error that depends on the accuracy of the Newton or Picard iteration. This leads to violation in energy conservation that can be controlled by increasing the accuracy of such methods.

One major source of energy non-conservation is the coupling between the wave modes resolved by the grid and their aliases. Aliases are wave modes that differ by an integer number of $2 \pi / \Delta x$. The reason for such coupling is that continuous particle data (which support wave modes that include aliases of wave modes resolved by the grid) are used in the construction of phase-space moments at a discrete set control of points on the physical grid.
Therefore, for momentum conserving schemes, we improve the energy conservation by decreasing the coupling of the wave modes resolved by the grid with their aliases.

One way to decrease the effect of aliasing is done by filtering the deposited grid moments (charge and current densities), this is used, for instance, in TRISTAN \citep{Tristan+1993} and its parallel version TRISTAN-MP \citep{Tristan-mp+2005}.  Filtering, however, when used with a momentum conserving scheme results in violating the momentum conservation and non-vanishing self-forces (an example for such case is presented in Appendix~\ref{app:filtering}).

Alternatively, energy conservation is naturally improved when using higher-order interpolation functions, as employed here.
In Fourier space, using higher-order interpolation function is qualitatively equivalent to low-pass filtering.
However, for instance, filtering when used with $W^0$ produces larger energy errors in comparison to the errors produced when first-order interpolation, $W^1$, is used (cf. Section 8.7 of \citealt{Birdsall-Langdon}).
That is, higher-order shape functions considerably decrease the impact of aliasing, resulting in an improved energy conservation for the same reasons as filtering, while simultaneously maintaining higher accuracy evolution of the underlying system. This includes, potentially, exact momentum conservation.  As shown explicitly in Appendix~\ref{app:aliasing}, the discretization of the plasma into macro-particles with shape functions of order $m$ produces a spectral smearing with a width that scales as $(k\Delta x)^{-m}$, producing a corresponding an exponential decrease in aliasing for the resolved modes with shape-function order.

\section{Validation}

\label{validation-sec}

We begin the assessment of the numerical algorithm presented in Section~\ref{sec:pic} with the physical validation of simulation results against known results.
This represents the first in two critical numerical tests, the second being convergence, which is treated in Section~\ref{sec:convergence}.

Specifically, we present comparisons with the analytical or semi-analytical results on the following test problems:
\begin{enumerate}
 \item thermal stability of a uniform plasma (Section~\ref{sec:thermal_stability}),
 \item standing plasma waves -- Plasma oscillations and linear Landau damping (Section~\ref{sec:Standing_waves}), and
 \item two-stream instabilities -- non-relativistic and relativistic (Section~\ref{sec:2streams}).
\end{enumerate}
Because validation essentially consists of quantitative comparisons with known results, we are currently limited to phenomena in the linear regime.  This appears to be a wide-spread difficulty within plasma simulations.  Nevertheless, weak validation in the nonlinear regime can be found when we discuss code comparisons in Section~\ref{sec:comparison}.
Here, we assume that all plasma species have the same mass, and, up to a sign, the same charge.   Thus, we take as our fiducial units, $q^2_0 = Q^2_s$ and $m_0 = M_s$, which implies that $n_0 = \sum_s n_s$.

\subsection{Thermal stability of plasma}
\label{sec:thermal_stability}

\begin{figure*}
\center
\includegraphics[width=18.4cm]{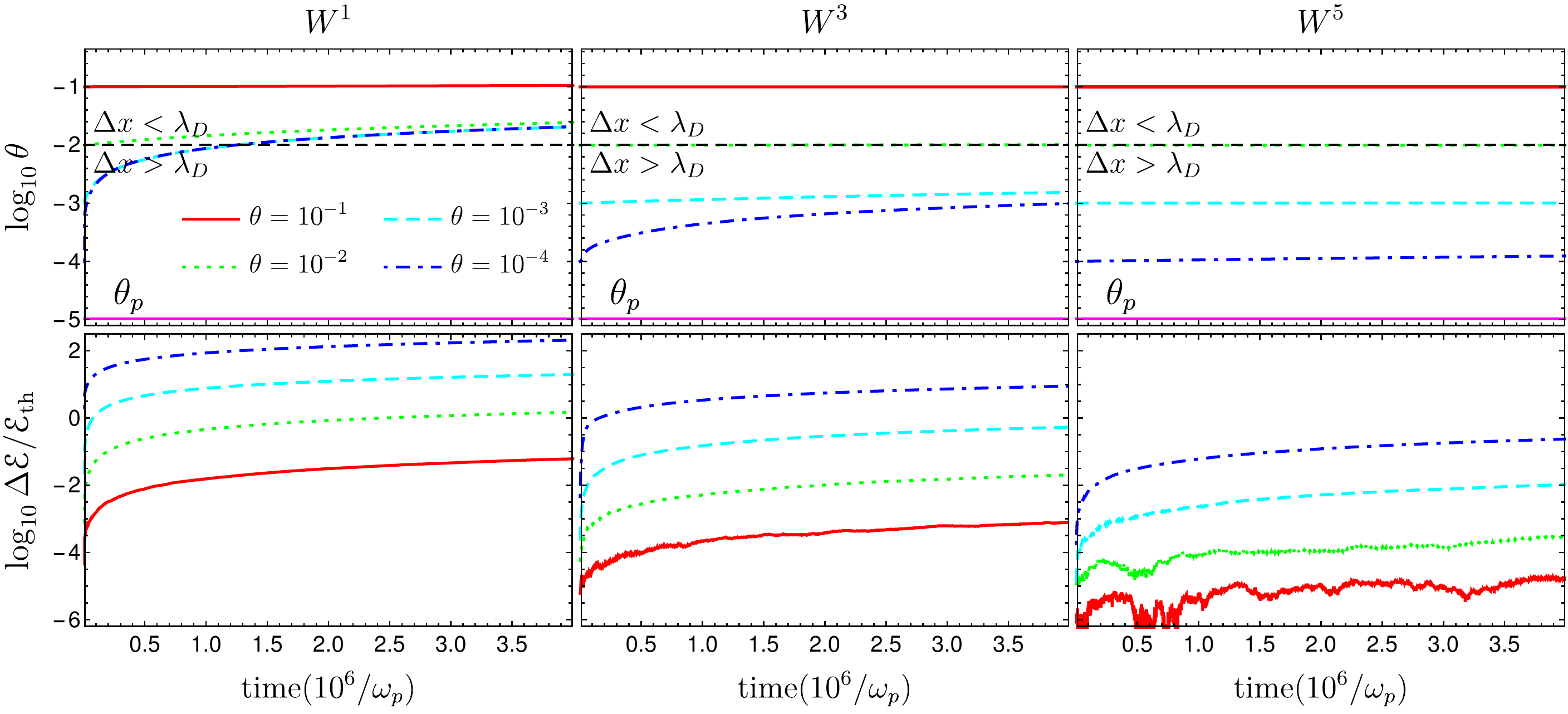}
\caption{Impact of numerical heating on the temperature (top) and energy error (bottom) evolution in simulations of a stationary thermal plasma for different interpolation functions: first-order interpolation $W^1$ (left), third-order $W^3$ (middle), and fifth-order $W^5$ (right).
Here, $\theta = k_B T / m_0 c^2$ is the normalized temperature, $\Delta \mathscr{E}$ is the energy change (error) in the total energy and $\mathscr{E}_{\rm th}$ is the initial thermal energy, i.e., excluding rest mass energy of macro-particles.
Therefore, $\Delta \mathscr{E}/\mathscr{E}_{\rm th}$ measures the fractional energy error with respect to the initial thermal energy of plasma.
For each interpolation order we perform simulations, at \textit{fixed} cell size $h=0.1$, with initial temperatures of
$\theta = 10^{-1}$ (solid-red curves), $\theta = 10^{-2}$ (dotted-green curves), $\theta = 10^{-3}$ (dashed-cyan curves), and $\theta = 10^{-4}$ (dotted-dashed blue curves).
The top panels show the long term (up to $4\times10^6 \omega_p^{-1}$) evolution of different temperatures for different interpolation orders, while the bottom panels show the evolution of the fractional energy error of the plasma.
The dashed-black line in the top panel shows the Debye temperature $\theta_D$.
The purple lines in the top panels indicate $\theta_p$ that is given by Equation~\eqref{Eq:thetap}: temperatures below $\theta_P$ are not well defined numerically.
\label{plot-Thermal_stab}
}
\end{figure*}

In the absence of microscopic radiative processes (e.g., Compton scattering, bremsstrahlung), a uniform, thermal plasma should not evolve.
In practice, even a uniform thermal plasma will numerically heat. This arises mainly as a consequence of aliasing as explained in Section~\ref{sec:EMconservations}.

There are two temperature scales that are often relevant for the numerical heating of plasma simulated with variants of the PIC algorithm.
The first is the numerical Debye temperature,
\begin{equation}
\label{eq:thetaD}
  \theta_D \equiv h^2,
\end{equation}
which is the temperature at which the Debye length is equal to square of the cell size in code units.  Note that this is a purely numerical quantity that defines those temperatures below which the discretization of Maxwell's equations no longer resolves the Debye length.
Typically, for $\theta<\theta_D$ second-order accurate codes will exhibit virulent numerical heating until $\theta\approx\theta_D$ (see, e.g., \cite{Birdsall+1980}).
Thus, $\theta_D$ presents a key numerical limitation on the classes of plasmas that have been simulated to date.
However, as shown in Section~\ref{sec:EMconservations}, implementing higher-order spatial interpolation decreases the effect of aliasing and hence considerably decrease such heating as shown in Figure~\ref{plot-Thermal_stab}.

The second is the Poisson temperature, set by the Poisson fluctuations in the reconstruction of the particle distribution.  This is set by equating the average potential and kinetic energies of randomly distributed particles (Appendix~\ref{app:thetap}), though in the cold and hot limits (i.e., non-relativistic and relativistic velocity dispersion limits, respectively) this reduces to
\begin{equation}
\label{Eq:thetap}
\theta_P = 
\theta_D \frac{N_c^2}{12 N}
\left[ 1 - \frac{6 f_m}{N_c} \right] \times
\begin{cases}
2, \qquad \theta_p \ll 1,\\ 
1, \qquad \theta_p \gg  1.
\end{cases}
\end{equation}
Temperatures below $\theta_P$ are not well defined numerically.  Note that the ordering of $\theta_P$ and $\theta_D$ is not fixed, though we will consider cases when $\theta_P<\theta_D$ exclusively. 

Here, we study the heating due to different approximations in PIC algorithms. To this end, we start all of our simulations with temperatures higher than $\theta_P$ and study the evolution of the plasma temperatures.
We perform a series of simulations for two populations of negatively and positively charged macro-particles in a periodic box of length $\bar{L}=5$ and cell size of $h = 0.1$. The total number of macro-particles $N_{\rm t}=2 \times 10^5$. Therefore, $\theta_P \approx 1.04\times10^{-5}$ and $\theta_D=10^{-2}$. 

Each simulation is started at a different initial temperature: $\theta = 10^{-1},10^{-2},10^{-3},10^{-4}$.
In all simulations presented here, we use second-order accurate back-interpolation, i.e., Equation~\eqref{EkEiapprox}. 

Since for all simulations here, $\theta \ll 1$, we start with an initial distribution function that is given by 
\begin{equation}
\label{eq:ini_gauss}
f(x,v,t=0) = \frac{N_{\rm t}}{L} \sqrt{\frac{\theta}{2 \pi c^2}}  e^{- (v/c)^2/ 2 \theta }.
\end{equation}

For the various initial temperatures, we show in Figure~\ref{plot-Thermal_stab} the evolution of the plasma temperatures (top panel) and that of the fractional energy error of the plasma (bottom panel). We show results when first-order interpolation $W^1$ (left), third-order $W^3$ (middle), and fifth-order $W^5$ (right) are used in deposition and back-interpolation steps.

As expected, for the first-order interpolation (the scheme that is most commonly employed in existing PIC codes), uncontrolled heating is observed for all $\theta\le\theta_D$. This heating subsides when $\theta\approx3\theta_D$, requiring between one and two cells per Debye length. Hence, first order PIC algorithms face severe computational requirements to resolve cold plasmas.

However, using higher-order spatial interpolation significantly reduces the temperature at which uncontrolled numerical heating occurs. By fifth order, temperatures four orders of magnitude smaller than $\theta_D$, and only an order of magnitude larger than $\theta_P$, can be resolved for millions of plasma timescales. That is, high-order spatial interpolation extends the range of temperatures and timescales that can be simulated.

The marked improvement of SHARP-1D is a direct result of the corresponding improvement in energy conservation. The bottom panels of Figure~\ref{plot-Thermal_stab} show the evolution of the growth in the energy of each simulation; in all cases, the unphysical heating can be fully attributed to the failure to conserve energy.
However, the {\it fractional} energy non-conservation is improved by nearly three orders of magnitude for each simulated temperature as the spatial interpolation order is increased from $W^1$ to $W^5$.
The net result of higher spatial order is, therefore, the ability to run simulations orders of magnitude longer with orders of magnitude lower resolutions.

\subsection{Stability of standing linear plasma waves}
\label{sec:Standing_waves}

We now turn to the stability and evolution of plasmas with linear perturbations, specifically, standing waves.  Key validation tests are the reproduction of oscillation frequencies, dispersion relations, and linear Landau damping rates.  We begin with a discussion of the anticipated values followed by quantitative comparisons of standing wave evolution.

\subsubsection{Linear Dispersion Relations and Growth Rates}

As shown in Appendix~\ref{app:Dispersions}, the linear dispersion relation for thermal plasmas are conveniently expressed in terms of a handful of dimensionless quantities:

\begin{equation}
    \hat{\omega} = \frac{\omega}{\omega_p},
\end{equation}
\begin{equation}
    \hat{k} = \frac{k v_{\rm th}}{\omega_p} = \frac{k}{k_D}, ~~~\text{and}
\end{equation}
\begin{equation}
    v_p = \frac{\hat{\omega}}{\hat{k}} = \frac{\omega}{k v_{\rm th}},
\end{equation}
where $v_{\rm th} = \theta^{1/2} c$ is the thermal velocity dispersion and $k_D$ is the wavenumber associated with the Debye length. The linear dispersion relation for a non-relativistic thermal population of uniformly distributed electrons with a fixed positively charged background, i.e., infinitely massive ions, is given by (see Appendix~\ref{app:Dispersions} for more details)

\begin{eqnarray}
\hat{k}^2+1 
&=& 
\sqrt{\frac{\pi}{2}}  \left[ \text{Erfi}  \left( v_p/\sqrt{2} \right)-i\right]  v_p e^{-\frac{v_p^2}{2}},
\label{eq:dispersion-LLD}
\end{eqnarray}
where $\text{Erfi} $ is the complex error function defined as $\text{Erfi}(v_p) = -i  \text{ Erf}(i v_p)  $.
For a given $\hat{k}$, the roots $\hat{\omega}_j = \omega_j / \omega_{p}$ can be found by solving Equation~\eqref{eq:dispersion-LLD} numerically; the real and imaginary values for $\omega_j$ yield the oscillating frequency and the growing/damping rate of the mode with wavenumber $k$, respectively, at a given thermal velocity $v_{\rm th}$.

\begin{figure}
\center
\includegraphics[width=8.4cm]{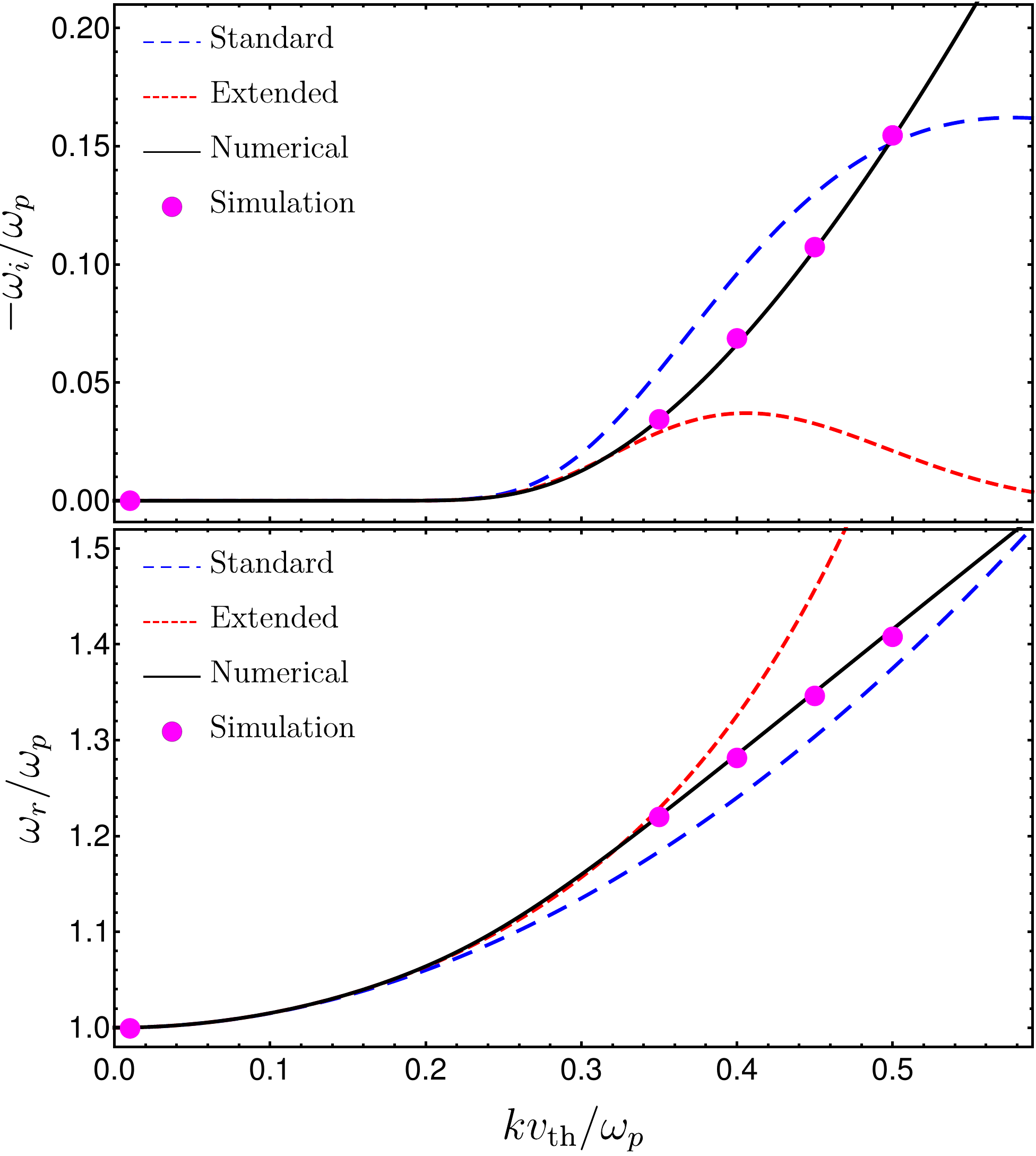}
\caption{Stability of standing plasma waves in the linear regime. The black curves show the
numerical solution to the linear dispersion relation, Equation~\eqref{eq:dispersion-LLD}.
Red and blue curves are different analytical approximate solutions for the linear dispersion
  relation~\eqref{eq:dispersion-LLD}. Purple data points show results from different simulations in Section~\ref{sec:Standing_waves}.
\label{plot-LLD}
}
\end{figure}

Approximate expressions for the roots of Equation~\eqref{eq:dispersion-LLD} are often obtained for the limit $\hat{k}\ll1$.  The most common of these, and the standard expression found in most textbooks \citep[e.g.,][and referred to here as ``Standard'']{Boyd} is
\begin{equation}
\hat{\omega}_i 
=
- \frac{1}{2}\sqrt{\frac{\pi }{2}} \frac{ 1}{  \hat{k}^3} e^{-\frac{1}{2 \hat{k}^2}-\frac{3}{2}}
\quad\text{and}\quad
\hat{\omega}_r
= 
1+ \frac{3}{2} \hat{k}^2,
\end{equation}
where $\omega_r$ and $\omega_i$ are the real and imaginary components of $\omega$, respectively.
A more accurate expression, derived by a higher-order approximation to Equation~\eqref{eq:dispersion-LLD} is given by \citet[][referred to as ``Extended'']{McKinstrie+1999}.
\begin{equation}
\label{eq:LLD-extended}
\begin{aligned}
\hat{\omega}_i
&=
-\frac{1}{2} \sqrt{\frac{\pi }{2}}  \left(\frac{1}{\hat{k}^3}-6 \hat{k}\right) e^{-\frac{1}{2 \hat{k}^2}-\frac{3}{2}-3 \hat{k}^2-12 \hat{k}^4}
\\
\hat{\omega}_r
&= 
1+ \frac{3}{2} \hat{k}^2 + \frac{15}{8} \hat{k}^4 + \frac{147}{16} \hat{k}^6.
\end{aligned}
\end{equation}
Both of these approximates are shown in comparison to the full numerical solution (``Numerical'') in
Figure~\ref{plot-LLD}.
It is immediately evident that the regime of applicability of both the Standard and Extended approximations is limited to $\hat{k}<0.25$, with the numerical consequence that neither are quantitatively accurate in the rapid damping regime, where numerical validation experiments can most easily be performed.

For convenience, we provide below a numerical fitting formula based on the formulation of \citet{McKinstrie+1999} to the full numerical solution for $\hat{k}\in[0,0.6]$ that is good to 4\% throughout and better than 0.5\% above $\hat{k}=0.3$
\begin{equation}
  \label{LLD-fits}
  \begin{aligned}
      \hat{\omega}_i
		&= 
    -\frac{1}{2} \sqrt{\frac{\pi }{2}}  \bigg(
    \frac{1}{\hat{k}^3}-6 \hat{k}
    - 40.7173 \hat{k}^3\\
    &\qquad
    - 3900.23 \hat{k}^5
    - 2462.25 \hat{k}^7
    - 274.99 \hat{k}^9
    \bigg)\\
    &\quad\times\exp\bigg(-\frac{1}{2 \hat{k}^2}-\frac{3}{2} -3 \hat{k}^2 -12 \hat{k}^4
    - 575.516 \hat{k}^6\\
    &\qquad
    + 3790.16 \hat{k}^8
    - 8827.54 \hat{k}^{10}
    + 7266.87 \hat{k}^{12}
    \bigg),
    \\
    \hat{\omega}_r
	&= 
    1+ \frac{3}{2} \hat{k}^2 + \frac{15}{8} \hat{k}^4 + \frac{147}{16} \hat{k}^6 \\
    &\qquad
    + 736.437 \hat{k}^8  
    - 14729.3 \hat{k}^{10} 
    + 105429 \hat{k}^{12}\\
    &\qquad
    - 370151 \hat{k}^{14}
    + 645538 \hat{k}^{16}
    - 448190 \hat{k}^{18}.
  \end{aligned}
\end{equation}

In the following subsections, we report a series of simulations to test the code in the
regimes $\omega_i \sim 0$ (undamped modes) and $\omega_i\neq0$ (damped
modes).
We use a fixed neutralizing background and negatively charged plasma
macro-particles whose initial distribution function is given by
\begin{equation}
\label{sim:LLD-distribution}
f(x,v,t=0) = \frac{e^{-\bar{v}^2/2\theta}}{\sqrt{2 \pi \theta }} \left[ 1+ \alpha \cos(k x ) \right] .
\end{equation}
In all simulations, we use $\theta =10^{-3}$, $\alpha =10^{-2}$, and fifth-order spatial interpolation, i.e., $W^5$. The damping rates and oscillation frequencies in different simulations of this section are also shown in Figure~\ref{plot-LLD}.

\begin{deluxetable*}{
ccccc
ccc
ccc
cc
}
\tablewidth{18.6cm}
\tablecaption{Standing plasma waves simulation parameters and results.
\label{table:standing-waves-sim}
}
\tablehead
{	
$\hat{k}$ 							& 
$N_{\rm t} $ 						&   
$\bar{\lambda}$ 		\tablenotemark{a}		& 
$\bar{\lambda} /h$ 		\tablenotemark{b}	& 
$\bar{L}/\bar{\lambda}$ 		&
\multicolumn{1}{c}{$\omega_r/\omega_p$ \tablenotemark{c}	}	& 
\multicolumn{1}{c}{$\omega^{sim}_r/\omega_p$ \tablenotemark{d} }	& 
\multicolumn{1}{c}{$R_r$	\tablenotemark{e}}	&
\multicolumn{1}{c}{$\omega_i/\omega_p$  \tablenotemark{c} }		& 
\multicolumn{1}{c}{$\omega^{sim}_i/\omega_p$ \tablenotemark{d}}	& 
\multicolumn{1}{c}{ $R_i$\tablenotemark{e}	} &
$\omega_p \hspace{.02cm} t_{\rm min} $ \tablenotemark{f} &
$\omega_p \hspace{.02cm} t_{\rm max} $ \tablenotemark{f}
}
\startdata
\rule{0pt}{8pt}
$0.01$ &  $10^9$ & $19.869$ &  $1761$ &  $2$ & $1.00015$ 
& 
$0.99925 \pm 3 \times 10^{-7} $
&  $0.9991$ & $\quad$-- & $\qquad$-- & $\quad$--  & $0$ & $10^3$
\\
\rule{0pt}{8pt}
$0.35$ & $5\times 10^8 $  & $0.568$ &  $67$ &  $12$ & $1.22095$ &  $ 1.21989 \pm 2 \times 10^{-5} $
&  $0.9991$ & $-0.03432$& $ -0.0344 \pm 4\times 10^{-5}$ 
&  $ 1.0036 $ &$1.7$ & $30.84$
\\
\rule{0pt}{8pt}
$0.40$ & $5\times 10^8 $  & $0.497$ &  $62$ &  $15$ & $1.28506$ &  $ 1.28145 \pm 4\times 10^{-5}$
&  $0.9972$ & $-0.06613$ & $ -0.0687 \pm 8 \times 10^{-5} $ 
&  $ 1.0389 $&$1.7$ & $16.80$
\\
\rule{0pt}{8pt}
$0.45$ & $5\times 10^8 $ &  $0.442$ &  $68$ & $19$ & $1.35025$ &  $  1.34617  \pm 1 \times 10^{-5}$
&  $0.9969$ & $-0.10629$ & $ -0.1066 \pm 2\times 10^{-5}$ 
& $ 1.00326 $&$1.6$ & $16.65$
\\
\rule{0pt}{8pt}
$0.50$ & $5\times 10^8 $ & $0.397$ & $68$ & $19$ & $1.41566$ & $ 1.40786 \pm 2 \times 10^{-5} $
&   
$0.9945$ &  $-0.15336$ & $ -0.1546 \pm 6 \times 10^{-5}$ 
& $1.0081$&$1.6$ & $8.77$
\enddata
\tablenotetext{a}{Wavelength of the initially excited mode in code units.}
\tablenotetext{b}{Number of cells used to resolve the initially excited wavelength.}
\tablenotetext{c}{Numerical solution of Equation~\eqref{eq:dispersion-LLD}.}
\tablenotetext{d}{Oscillation frequencies and damping rates found by fitting the changes in simulations. The error in the fitting parameters corresponds to 99\% confidence level.}
\rule{0pt}{8pt}
\tablenotetext{e}{$R_r = \omega^{\text{sim}}_r / \omega_r$ and $R_i = \omega^{\text{sim}}_i / \omega_i$, where $\omega^{\text{sim}}_r$ and $\omega^{\text{sim}}_i$ are the oscillation frequency and damping rate obtained from fitting simulation results, respectively. An example for such a fit is shown in Figure~\ref{fig:Oscillation-fitting} for the case of $\hat{k}=0.01$ and in Figure~\ref{plot-LLD-Simulations} for the case of $\hat{k}=0.45$ .}
\tablenotetext{f}{To fit our simulation results, we specify a time range [$t_{\rm min}, t_{\rm max} $] in the simulation over which we carry out the fitting.}
\end{deluxetable*}

\subsubsection{Plasma oscillations}
\label{sec:Oscillation}

At small $\hat{k}$, the linear Landau damping rate is vanishingly small, i.e., $\omega_i\approx 0$.  As a result, a linear perturbation should oscillate providing quantitative tests in the form of the oscillation frequency and evolution of the mode amplitude. We initialize the simulation with an excited mode such that
$\hat{k}=0.01$. The theoretical predictions (numerical solution of Equation~\eqref{eq:dispersion-LLD}) are $\omega_r/\omega_p = 1.00015$ and
$\omega_i=-4.7\times10^{-2167}$. The rest of our simulation parameters are given in
Table~\ref{table:standing-waves-sim}.

\begin{figure*}
\center
\includegraphics[width=18.3cm]{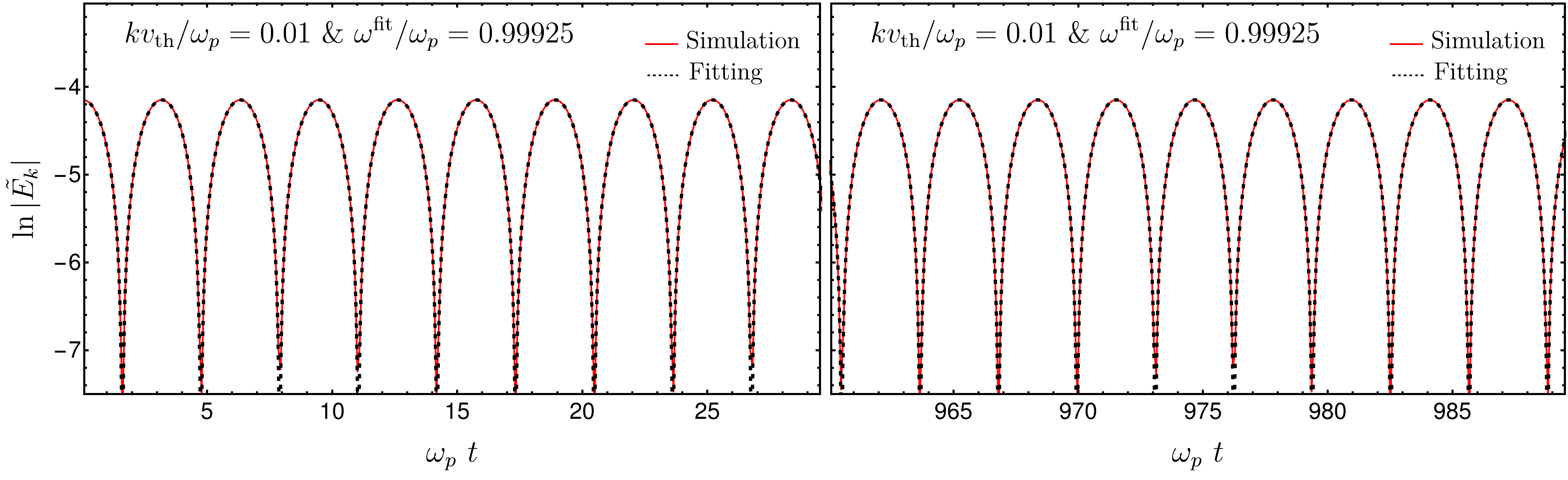}
\caption{Fitting the oscillation frequency for plasma oscillation simulations. The left panel shows that the simulation data (red-curve) is in excellent agreement with the fit. It continues to excellently fit the simulation until its end at $10^3 \omega_p^{-1}$ (right).}
\label{fig:Oscillation-fitting}
\end{figure*}

Figure~\ref{fig:Oscillation-fitting} shows the fitted values for the electric field of the initially excited oscillation modes. The oscillation frequency of the initially excited mode is found to be within 0.09\% of the theoretically predicted oscillation frequency (this is measured by fitting the oscillation of such a mode in the simulation over about 159 oscillation periods).
The oscillation frequency and amplitude are  still in excellent agreement with the theoretical prediction until the end of the simulation time $t=10^3\omega_p^{-1}.$

\begin{figure}
\center
\includegraphics[width=8.4cm]{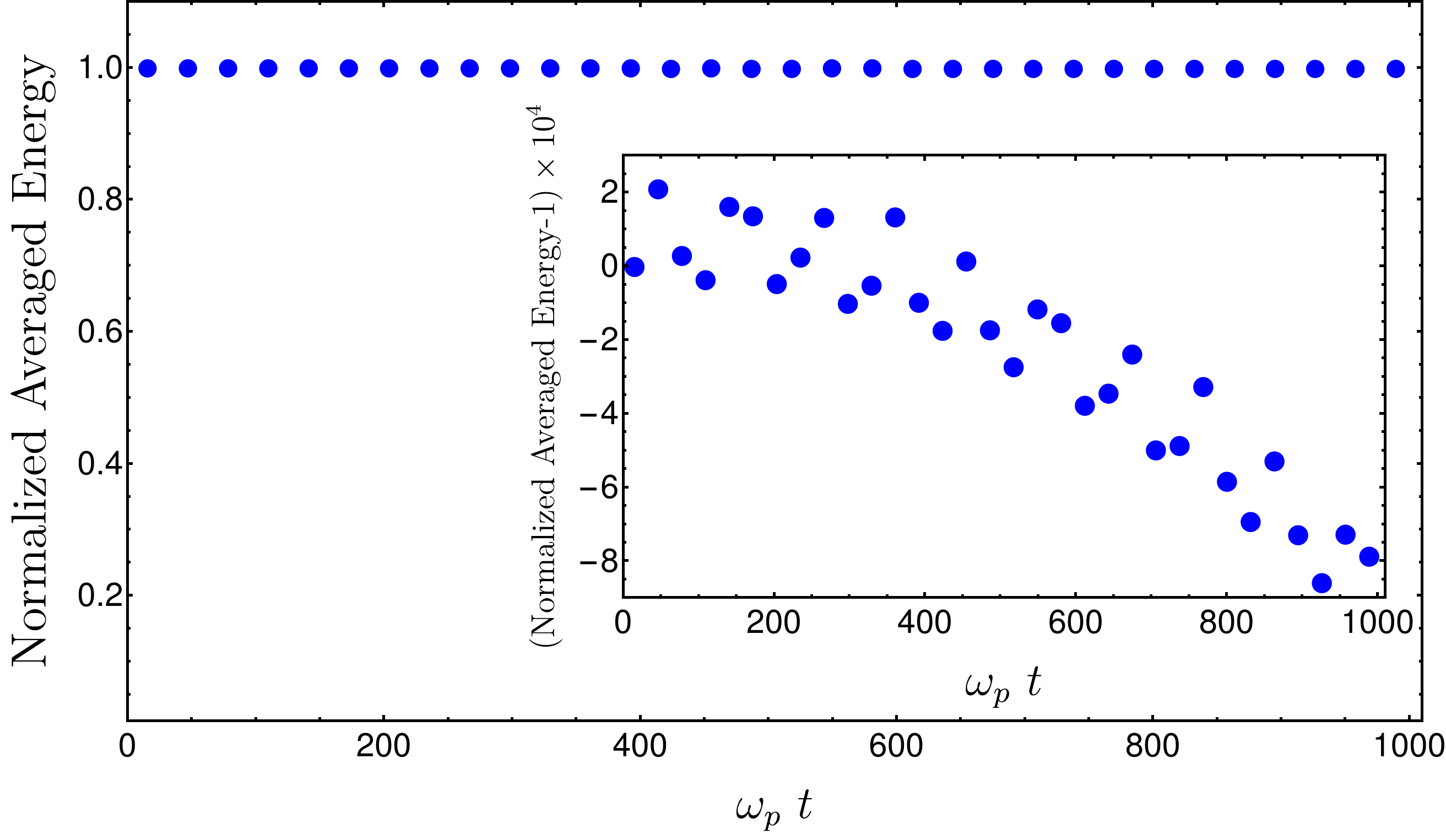}
\caption{The evolution of the averaged (over $5$ plasma periods) energy in the initially excited mode normalized to averaged energy in the first 5 plasma periods of such mode. The inset shows that, after evolving the simulation to $10^3 \omega_p^{-1}$, the level of variation on energy carried by the mode is about 0.1\% of the initial energy in that mode.
\label{fig:standing-power}
}
\end{figure}

In Figure~\ref{fig:standing-power}, we show the evolution of the averaged energy in the excited modes (over five plasma periods).
We find that most of the power stays in the excited mode for the whole simulation period. Coupled with the degree of energy conservation during the simulation, this implies that no more that 0.1\% of the initial energy in the mode leaks into other degrees of freedom of the plasma (e.g., other modes or heating of the plasma).

\subsubsection{Linear Landau damping rates}
\label{sec:LLD}

For larger $\hat{k}$, corresponding to comparatively larger wavenumbers, the damping rates become large.
By $\hat{k}=0.35$ the wave should damp by one $e$-fold in about six wave oscillation periods.  
Again, this provides a number of quantitative tests of SHARP-1D: oscillation frequencies and damping rates.  Thus, here we report on simulations at large $\hat{k}$; the values of parameters in these simulations and their results are summarized in Table~\ref{table:standing-waves-sim}. The damping rates and oscillation frequencies in different simulations are also shown in Figure~\ref{plot-LLD}.

\begin{figure}
\center
\includegraphics[width=8.4cm]{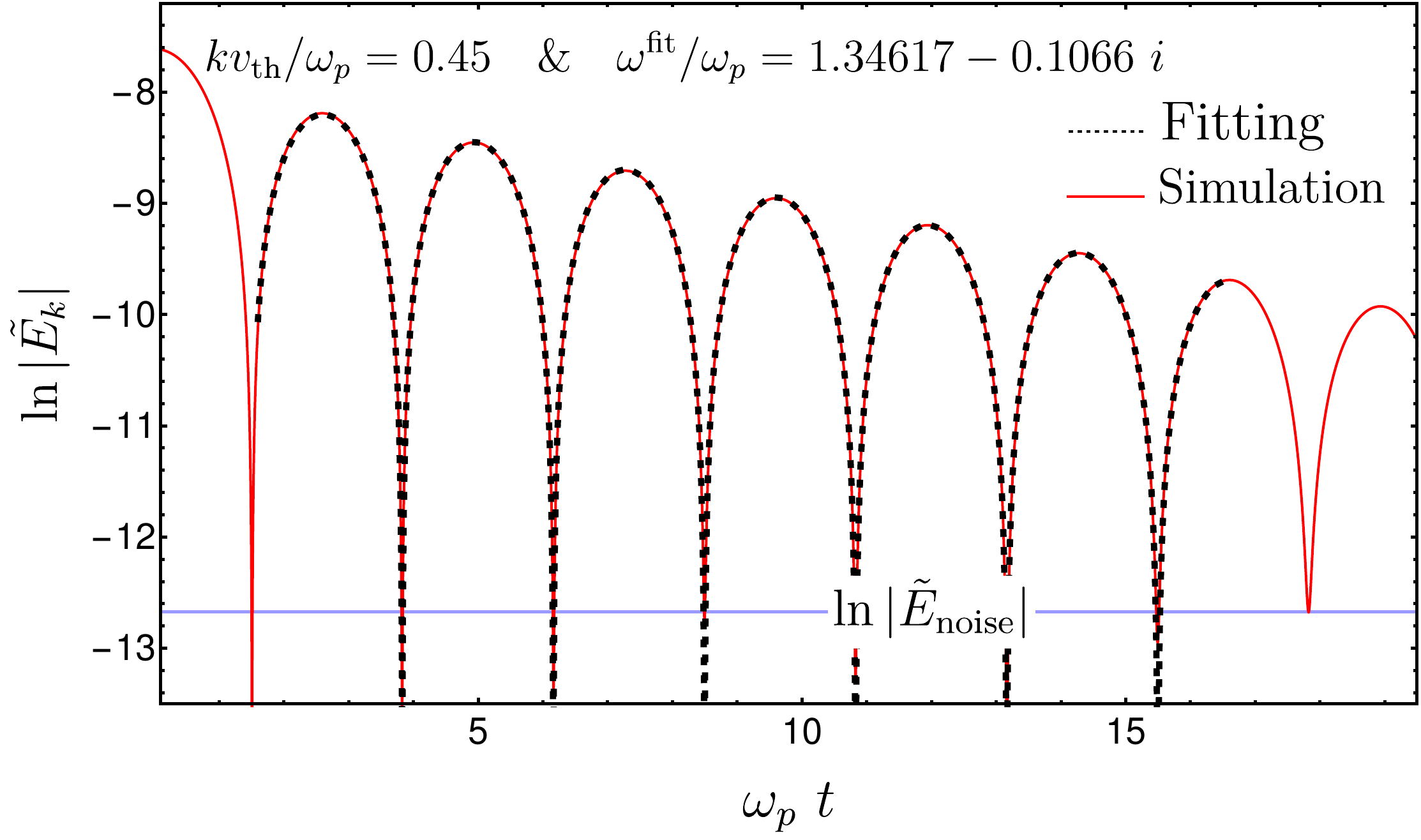}
\caption{The evolution of standing linear plasma mode in the regime of high linear landau damping rate ($\hat{k}=0.45$). The figure shows the fitting of the Fourier component of the grid electric field that corresponds to the initially excited mode. The noise level (blue line) corresponds to the estimated Fourier component of the electric field in Equation \eqref{Eq:Noise-fourier}, i.e., the noise due to the finite number of macro-particles distributed uniformly on a periodic grid.
\label{plot-LLD-Simulations}
}
\end{figure}

In all simulations, the oscillation frequencies and damping rates are found to be within 0.5\% and 0.8\% of the theoretical predictions of linear theory, respectively.
The two columns $R_r$ and $R_i$ of Table~\ref{table:standing-waves-sim} report the ratio between the oscillation frequency and damping rate in simulations to their theoretically predicted values, obtained by numerically solving Equation~\eqref{eq:dispersion-LLD}.
The damping rate and the oscillation frequency of the simulation are obtained by fitting the evolution of the Fourier component of the electric field that corresponds to the initially excited wave mode. In Figure~\ref{plot-LLD-Simulations}, we show an example (for $\hat{k}=0.45$ simulation) of the fitting carried out to find oscillation frequencies and damping rates of different simulations.

A sudden drop in the mode energy at the first period can be observed in Figure~\ref{plot-LLD-Simulations}. This is a characteristic feature of linear Landau simulations, i.e., when damping rate is comparable to the plasma frequency. Such a feature is also present when other simulation methods are used to simulate the evolution of such modes in a thermal plasma \citep[e.g.,][]{Semi-Lagrangian,DG-scheme}.

\subsection{Two-stream Instability}
\label{sec:2streams}
We now consider the quantitative accuracy with which SHARP-1D can reproduce a dynamical instability -- in 1D the primary example is the two-stream instability.  This provides an opportunity to also assess the relativistic performance of the code, through the simulation of relativistic beams. Thus, we will consider two limiting regimes: non-relativistic ($v_b\ll c$) and relativistic ($u_b\gg c$), where $v_b$ is the speed of the streams and $u_b = v_b/\sqrt{1-(v_b/c)^2}$ is the specific momentum.

As with linear Landau damping, we will begin with a general discussion of the anticipated instability properties and then move onto quantitative comparisons.

\subsubsection{Instability Growth Rates}

\begin{figure}
\center
\includegraphics[width=8.4cm]{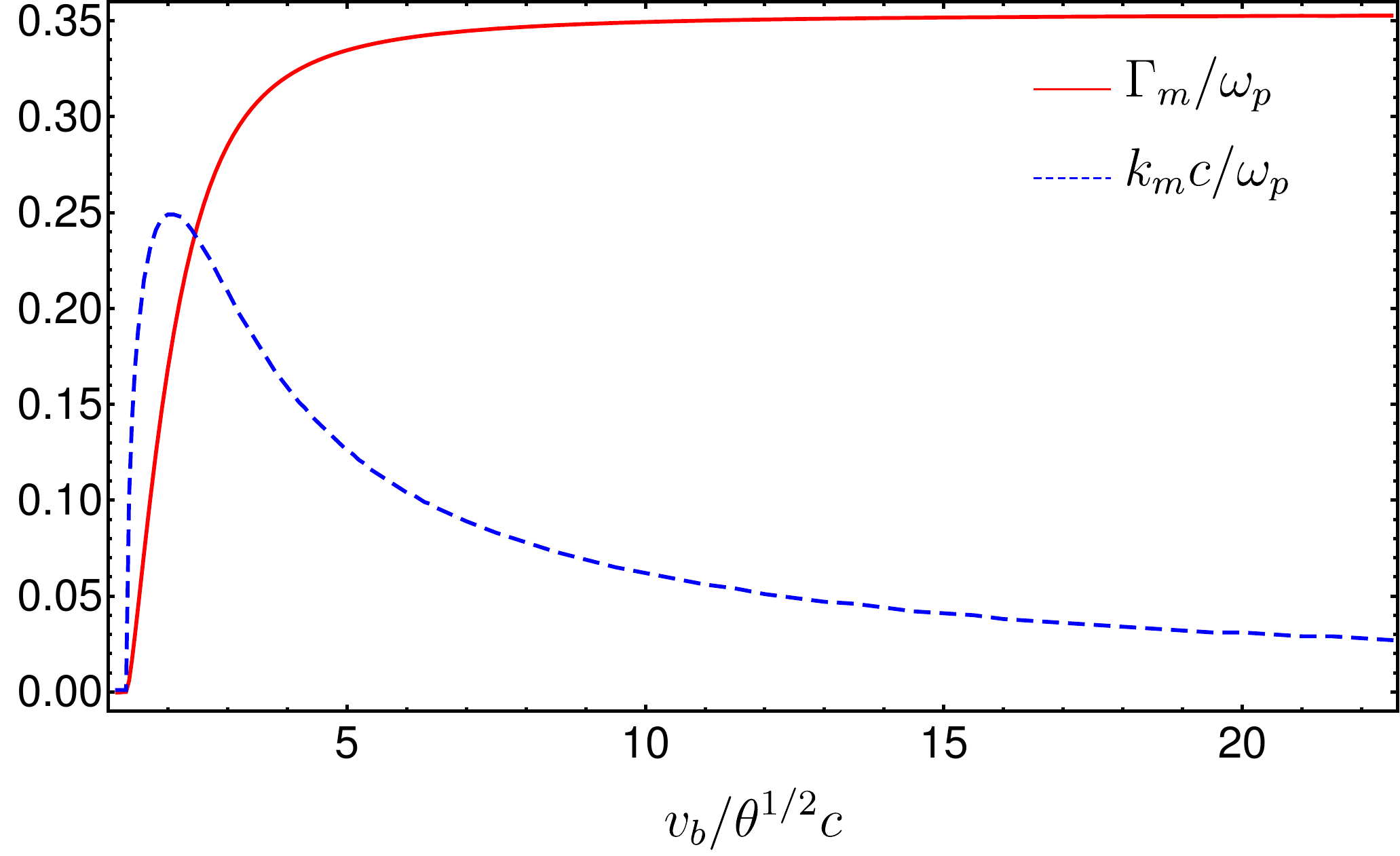}
\caption{Numerical solutions to the non-relativistic two-stream instability dispersion relation (Equation~\eqref{eq:disp-2stream-non-rel}). The blue curve shows the dependence of maximum growth rate on the stream speed (and temperature). The red curve shows the dependence of maximally growing wave mode on the stream speed (and temperature).
\label{plot-2stream-non-rel}
}
\end{figure}

In the non-relativistic regime, i.e., the streams are moving with non-relativistic speeds and have non-relativistic temperatures, the linear dispersion relation for two counter streams in such limit is given by (see Appendix~\ref{app:Dispersions} for more details)
\begin{multline}
\label{eq:disp-2stream-non-rel}
\hat{k}^2 +1
=
\sqrt{ \frac{\pi}{8}} 
\left[
( v_p + z_b )\left( \text{Erfi}  \left[ \frac{ v_p + z_b }{\sqrt{2} }\right]-i\right) e^{- v_p  z_b }
\right.\\
+
\left.
( v_p - z_b )
\left(  \text{Erfi}  \left[ \frac{ v_p - z_b }{\sqrt{2} }\right]-i  \right) 
e^{ v_p  z_b }
\right]
e^{-( v^2_p + z^2_b)/2},
\end{multline}
where $z_b = \bar{v}_b/\sqrt{\theta}$, $\bar{v}_b = v_b /c$, $\hat{k} = kc \sqrt{\theta} /\omega_p$ and $v_p = \omega/ kc\sqrt{\theta }$.
In general, this must be numerically solved to obtain the mode frequencies. 
The solution generally consist of two oscillatory modes, a growing mode, and several damping modes.
We present some of these solutions in Table~\ref{table:2stream-non-rel-sim}.
In Figure~\ref{plot-2stream-non-rel}, we show the numerical solutions for the maximum growth rates (and the mode growing with such a rate) as a function of the stream speed and its temperature.

In the relativistic limit, the beam velocity distribution exhibits a narrow peak very close to the speed of light $c$, and is thus well described by the cold-plasma limit, i.e., $\theta=0$.  Within this limit, the linear dispersion relation for two relativistic counter streaming $e^{+}$-$e^{-}$ populations (with speed $v_b$) is given by 
\begin{equation}
1
= 
\frac{\omega_p^2/2}{\gamma_b^3 (k v_b - \omega)^2} + \frac{\omega_p^2/2}{\gamma_b^3(k v_b + \omega)^2}.
\end{equation}
In this case, it is possible to obtain analytic solutions:
\begin{equation}
\frac{\omega}{\omega_p} 
=
\pm
\sqrt{
\frac{1+2 \hat{k}_b^2 \gamma _b^3 \pm \sqrt{8 \hat{k}_b^2 \gamma _b^3+1} }{2 \gamma_b^3}
},
\label{eq:disp-2-stream}
\end{equation}
where $\hat{k}_b = k v_b / \omega_p$.
When $\gamma_b^3 \hat{k}_b^2<1$, these again correspond to two oscillating modes, a growing mode and a damping mode.
The positive imaginary root is maximized at the wavenumber
\begin{equation}
\frac{ k_m c}{\omega_p}
=
\sqrt{\frac{3/8}{\bar{v}_b^2\gamma_b^3}}
\end{equation}
at which the growth rate is
\begin{equation}
\frac{ \Gamma_m }{\omega_p} 
= 
\frac{1 }{2 \sqrt{2 \gamma_b^3 } },
\label{cold-2stream}
\end{equation}
where $\Gamma_m = \Im(\omega_m)$ is the growth rate of that wavenumber.

\begin{figure}
\center
\includegraphics[width=8.6cm]{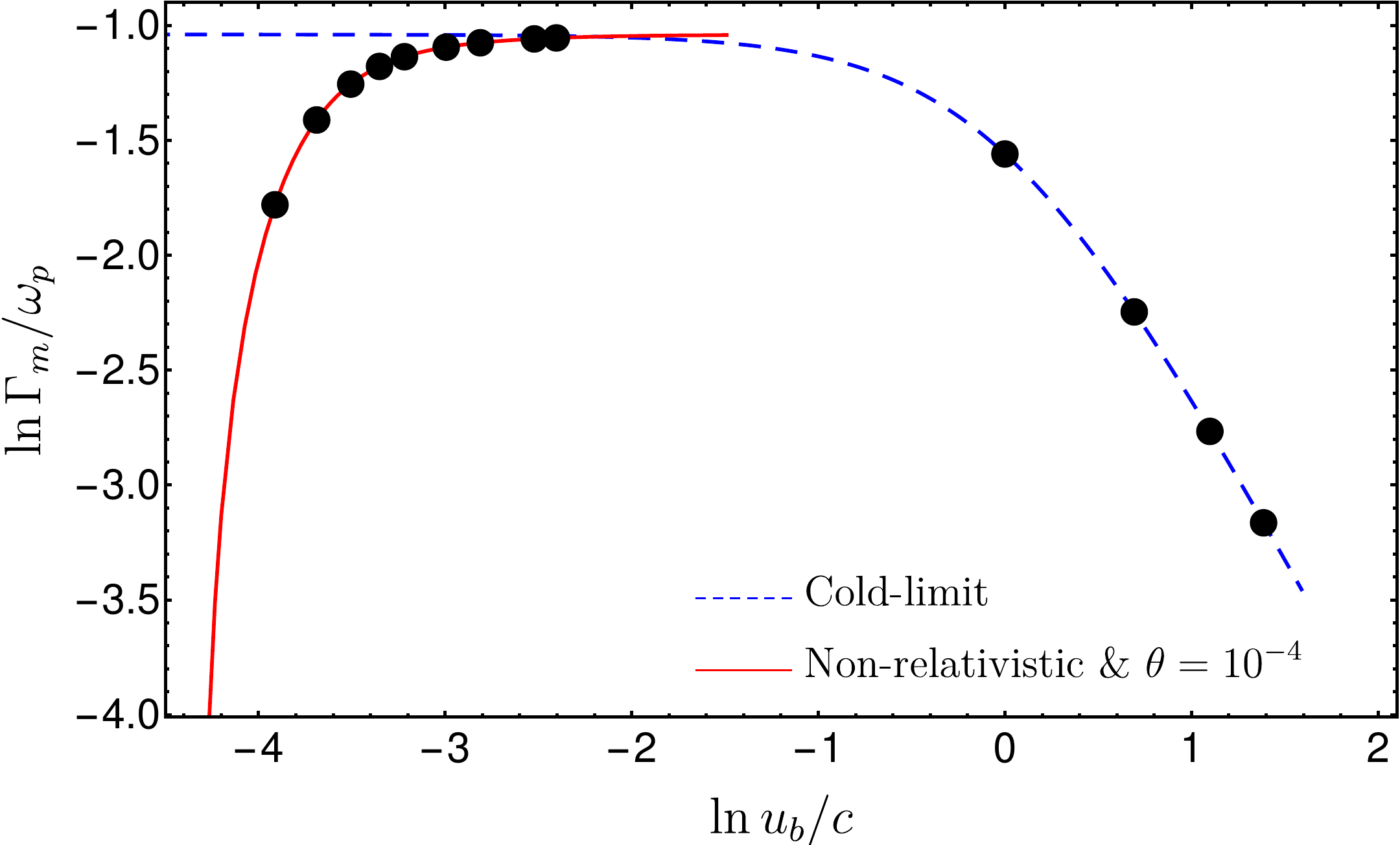}
\caption{Theoretical predictions for the two-stream instability growth rates in both relativistic and non-relativistic regimes. The black data points correspond to our simulated growth rates. Here $u_b = \gamma_b v_b$.
\label{plot:2stream-sim}
}
\end{figure}

The anticipated growth rates for both the relativistic and non-relativistic regimes are shown in Figure~\ref{plot:2stream-sim}.  At low beam velocities the finite temperatures of the beams suppress the growth rates appreciably relative to the cold-plasma limit, highlighting the importance of numerically solving the dispersion relation in the non-relativistic regime.
In both cases (non-relativistic and relativistic streams) investigated below, we use a fifth-order interpolation function $W^5$.

\subsubsection{Non-relativistic Two-stream Simulations}
\label{sec:2snr}

\begin{figure} 
\center
\includegraphics[width=8.6cm]{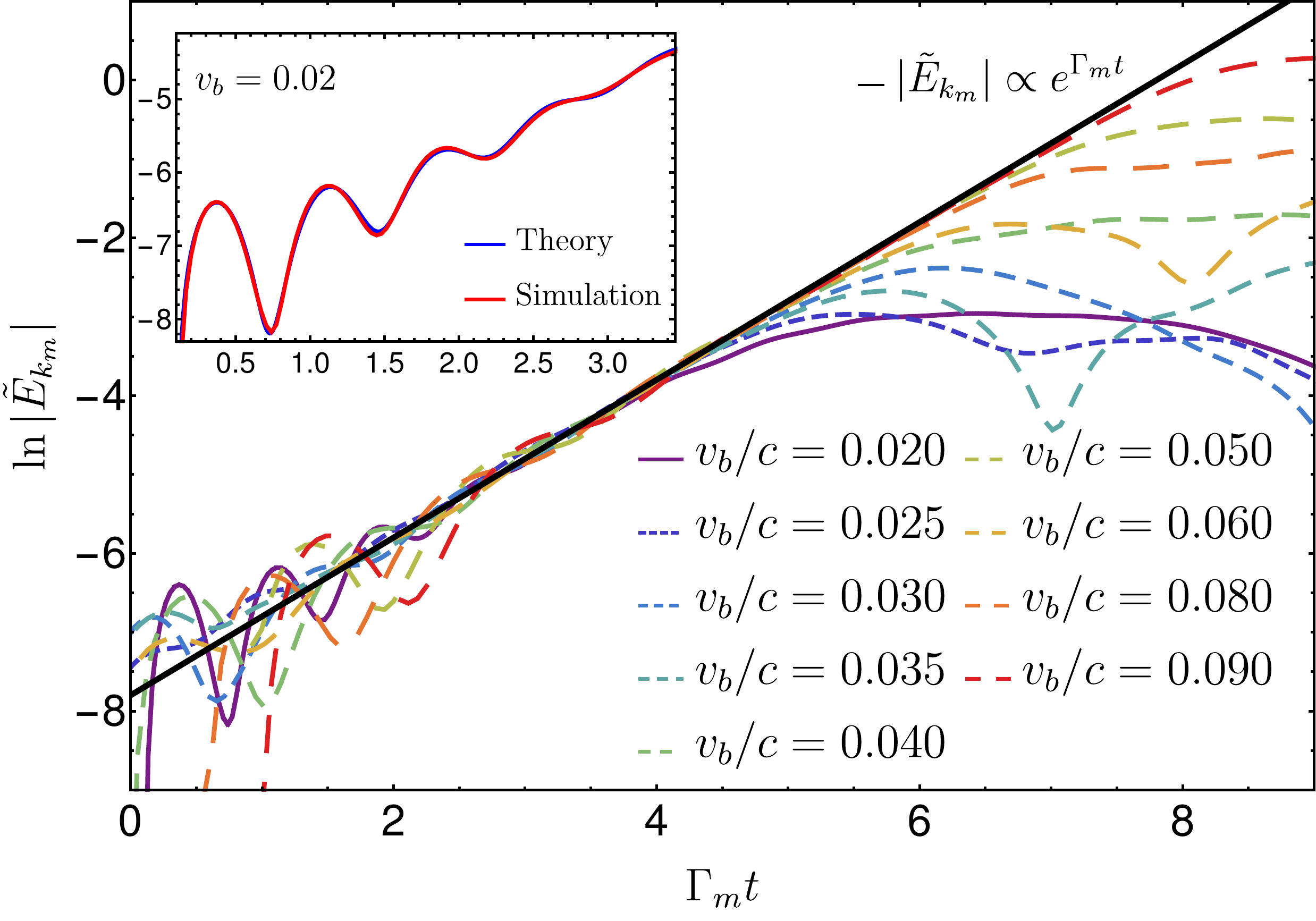}
\caption{
Two-stream instability simulation results in the non-relativistic regime.
Different curves show the growth of the maximally growing mode (predicted by the theory) with time in units of the growth rate of this mode (predicted also theoretically) for streams with different speeds. The solid-black line is a line with slope = 1 .
An excellent agreement between simulation results and theoretical predictions is therefore evident: the growth rate found in the simulation and the rates calculated by solving the linear dispersion relation for such plasma, Equation~\eqref{eq:disp-2stream-non-rel}, numerically are quantitatively similar. 
The inset shows a reconstruction of the $v_b/c=0.020$ simulation in terms of the full complement of plasma modes shown in Table~\ref{table:2stream-non-rel-sim}.
Note that $\Gamma_{m}$ is different for different beam velocities $v_b$ (see Table~\ref{table:2stream-non-rel-sim} and Figure~\ref{plot:2stream-sim}.
Since the instability grows from the noise in all simulations, we shifted the time so that linear-phase instability growth starts at the same time for all simulations.
\label{plot:2stream-non-rel-sim}
}
\end{figure}

The non-relativistic two-stream simulations are initialized with an initial
co-moving temperature of the streams at $\theta=10^{-4}$. We use $\Delta x \sim
0.001 c/\omega_p$ and $L/\lambda_m = 20$, where $\lambda_m = 2 \pi /k_m$ is the
fastest growing wavelength. We start with a uniform distribution of
macro-particles. Therefore, the initial distribution function is given by
\begin{equation}
\label{eq:2stream-non-rel-distribution}
f(x,v,t=0) = \frac{1}{  \sqrt{2 \pi \theta }}
\left[
\frac{  e^{-(\bar{v}-\bar{v}_b)^2/2\theta}  + e^{-(\bar{v}+\bar{v}_b)^2/2\theta} }{2}
\right].
\end{equation}

Other simulation parameters and the theoretical prediction are given in Table~\ref{table:2stream-non-rel-sim}.

Generally, these simulations exhibit excellent quantitative agreement with the results of the linear theory. This is clearly evident in Figure~\ref{plot:2stream-non-rel-sim}, which shows the evolution of the amplitude of the most rapidly growing mode (based on linear theory) in the non-relativistic simulations as a function of linear growth times. These should be compared to the solid black line, which shows the expected exponential growth, i.e., $e^{\Gamma_m t}$; the correspondence lasts over 4-8 $e$-folding times, i.e., 2-3.5 orders of magnitude, ending when the instability saturates non-linearly.

\begin{deluxetable*}
{
ccccc
cccc
}
\tablewidth{18.6cm}
\tabletypesize{\scriptsize}  
\tablecaption{
Non-relativistic two-stream simulation parameters. \label{table:2stream-non-rel-sim}
}
\tablehead
{	
$v_b/c$ 				& 
$N_c$ \tablenotemark{a}					& 
$N_{\rm pc}$ 	\tablenotemark{b}		& 
$\Gamma_m/\omega_p$ \tablenotemark{c}		& 
$k_m c /\omega_p$	\tablenotemark{d}		&
\multicolumn{1}{c}{$\hat{\omega}_m^1$ and $\hat{\omega}_m^2$ \tablenotemark{e}} &
\multicolumn{1}{c}{$\hat{\omega}_m^3$ \tablenotemark{e} } &
\multicolumn{1}{c}{$\hat{\omega}_m^4$ \tablenotemark{e} } &
\multicolumn{1}{c}{$\hat{\omega}_m^5$ \tablenotemark{e} }
}
\startdata
\rule{0pt}{8pt}
0.02 & 5046 & 900. & 0.168555 & 24.9 & 
$\pm 1.40842 -0.0161625 i $ &
$-0.485447 i$ & 
$-0.915823 i$ &
$-1.32416 i$
\\
\rule{0pt}{8pt}
0.025 & 5324 & 900. & 0.243771 & 23.6 & 
$\pm 1.47658 - 0.0123756 i $ &
$-0.342545i$ & 
$-0.673817 i$ &
$-0.987125 i$
\\
\rule{0pt}{8pt}
0.03 & 6013 & 899.85 & 0.284945 & 20.9 & 
$\pm 1.48041 - 0.00479742 i $ &
$-0.249662i$ & 
$-0.48542 i$ &
$-0.715823 i$
\\
\rule{0pt}{8pt}
0.035 & 6904 & 900. & 0.307734 & 18.2 & 
$\pm  1.46288 - 0.00103408i $ &
$-0.213058 i$&
$-0.356584 i$ & 
$-0.524922i$ 
\\
\rule{0pt}{8pt}
0.04 & 7903 & 900. & 0.321003 & 15.9 & 
$\pm 1.44312 - 0.00011898 i $ &
$ -0.247524 i$ &
$ -0.391793 i$ & 
$ -0.527878 i$ 
\\
\rule{0pt}{8pt}
0.05 & 9974 & 899.91 & 0.334645 & 12.6  & 
$\pm 1.41697 -3.246 \times 10^{-7} i $ &
$ -0.340762 i$ &
$ -0.388706 i$ & 
$ -0.494053 i$ 
\\
\rule{0pt}{8pt}
0.06 & 12083 & 900. & 0.34111 & 10.4 & 
$\pm 1.40108 - 1.294 \times 10^{-10} i $ &
$ -0.341415 i$ &
$ -0.467936 i$ & 
$ -0.496439 i$ 
\\
\rule{0pt}{8pt}
0.08 & 16110 & 900. & 0.346906 & 7.8 & 
$\pm 1.39159 - 3.816 \times 10^{-19} i $ &
$ -0.346906 i$ &
$ -0.396172 i$ & 
$ -0.524619 i$ 
\\
\rule{0pt}{8pt}
0.09 & 18212 & 900. & 0.348379 & 6.9 & 
$\pm 1.38628 - 1.148 \times 10^{-24} i $ &
$ -0.348379 i$ &
$ -0.548489 i$ & 
$ -0.556374 i$ 
\enddata
\tablenotetext{a}{Number of cells.}
\tablenotetext{b}{Number of computation particles per cell.}
\tablenotetext{c}{Theoretical predictions for the maximum growth rate.}
\tablenotetext{d}{Fastest growing wave mode predicted theoretically.}
\tablenotetext{e}{$\hat{\omega} = \omega / \omega_p$: other solutions of Equation~\eqref{eq:disp-2stream-non-rel} at the fastest growing wave mode $k_m$. 
}
\end{deluxetable*}

The initial oscillations correspond to other solutions of the dispersion relation, i.e., they are fully described by the linear analysis of the two-stream instability: in addition to the ultimately dominant exponentially growing mode, the linear dispersion relation admits other damping and oscillatory modes.

The inset in Figure~\ref{plot:2stream-non-rel-sim} shows a reconstruction of the $v_b/c=0.02$ simulation, which shows prominent oscillations at the beginning, in terms of the full complement of plasma modes shown in Table~\ref{table:2stream-non-rel-sim}. Thus, the simulation quantitatively reproduces all of the anticipated linear features. A similar exercise is possible with the remaining non-relativistic simulations as well.

\subsubsection{Relativistic Two-stream Simulations} \label{sec:2sr}

We also performed a series of simulations for streams moving with relativistic speeds $u_b = \gamma_b v_b$. In all simulations, the initial co-moving temperature of streams is
$\theta = 3 \times 10^{-3}$, $\Delta x \sim 0.05 c/\omega_p$ and $L/\lambda_m =
10$, where $\lambda_m = 2 \pi /k_m$ is the fastest growing wavelength. We start
with uniformly distributed macro-particles, i.e., the instability here also grows
from numerical noise. The initial distribution function is given by a combination of Maxwell--J\"uttner distributions:
\begin{equation}
\label{eq:2stream-rel-distribution}
f(x,v,t=0) 
= 
\frac{
\left[
e^{ \bar{u}\bar{u}_b /\theta}
+
e^{-\bar{u}\bar{u}_b /\theta}
\right]
 e^{-\gamma_b \gamma/\theta}
}
{ 
4 K_1\left( 1/\theta\right) 
},
\end{equation}
where $K_1$ is the Bessel function of the first kind, and $\theta$ is the temperature in the co-moving frame of each beam.
All other simulations parameters, along with  theoretical predictions, are shown in Table~\ref{table:2stream-rel-sim}.

\begin{deluxetable}{
p{.4cm}p{.5cm}p{0.6cm}p{.6cm}cc
p{1.1cm}p{.9cm}
}
\tablecaption{
Relativistic Two-stream Simulation Parameters. 
\label{table:2stream-rel-sim}
}
\tablehead{
$u_b/c$ 			& 
$v_b/c$ 				& 
$N_c$ \tablenotemark{a}					& 
$N_{\rm pc}$ 	\tablenotemark{b}			& 
$\Gamma_m/\omega_p$ \tablenotemark{c}		& 
$k_m c /\omega_p$	\tablenotemark{d}&
\multicolumn{1}{c}{$\hat{\omega}_m^1$ and $\hat{\omega}_m^2$ \tablenotemark{e}} &
\multicolumn{1}{c}{$\hat{\omega}_m^3$ \tablenotemark{e} }
}
\startdata
\rule{0pt}{8pt}
1 & 0.707 & 2441 & 491.6 & 0.2102 & 0.5147 &
$\pm 0.8142 $ & $-0.2102 i$
\\
\rule{0pt}{8pt}
2 & 0.894 & 6138 & 488.76 & 0.1057  & 0.2047&
$\pm 0.4095  $ & $-0.1057 i$  
\\
\rule{0pt}{8pt}
3 & 0.948 & 10948 & 493.24 & 0.0629 & 0.1148 &
$\pm 0.2435 $ & $-0.0629 i$
\\
\rule{0pt}{8pt}
4 & 0.970& 16668 & 497.96 & 0.0422 & 0.0754&
$\pm 0.1636 $ & $-0.0422  i$ 
\enddata
\tablenotetext{a}{Number of cells.}
\tablenotetext{b}{Number of computation particles per cell.}
\tablenotetext{c}{Theoretical predictions for the maximum growth rate.}
\tablenotetext{d}{Fastest growing wave mode predicted theoretically.}
\tablenotetext{e}{$\hat{\omega} = \omega / \omega_p$: other solutions given by Equation~\eqref{eq:disp-2-stream} at the fastest growing wave mode $k_m$.}
\end{deluxetable}

\begin{figure}
\center
\includegraphics[width=8.6cm]{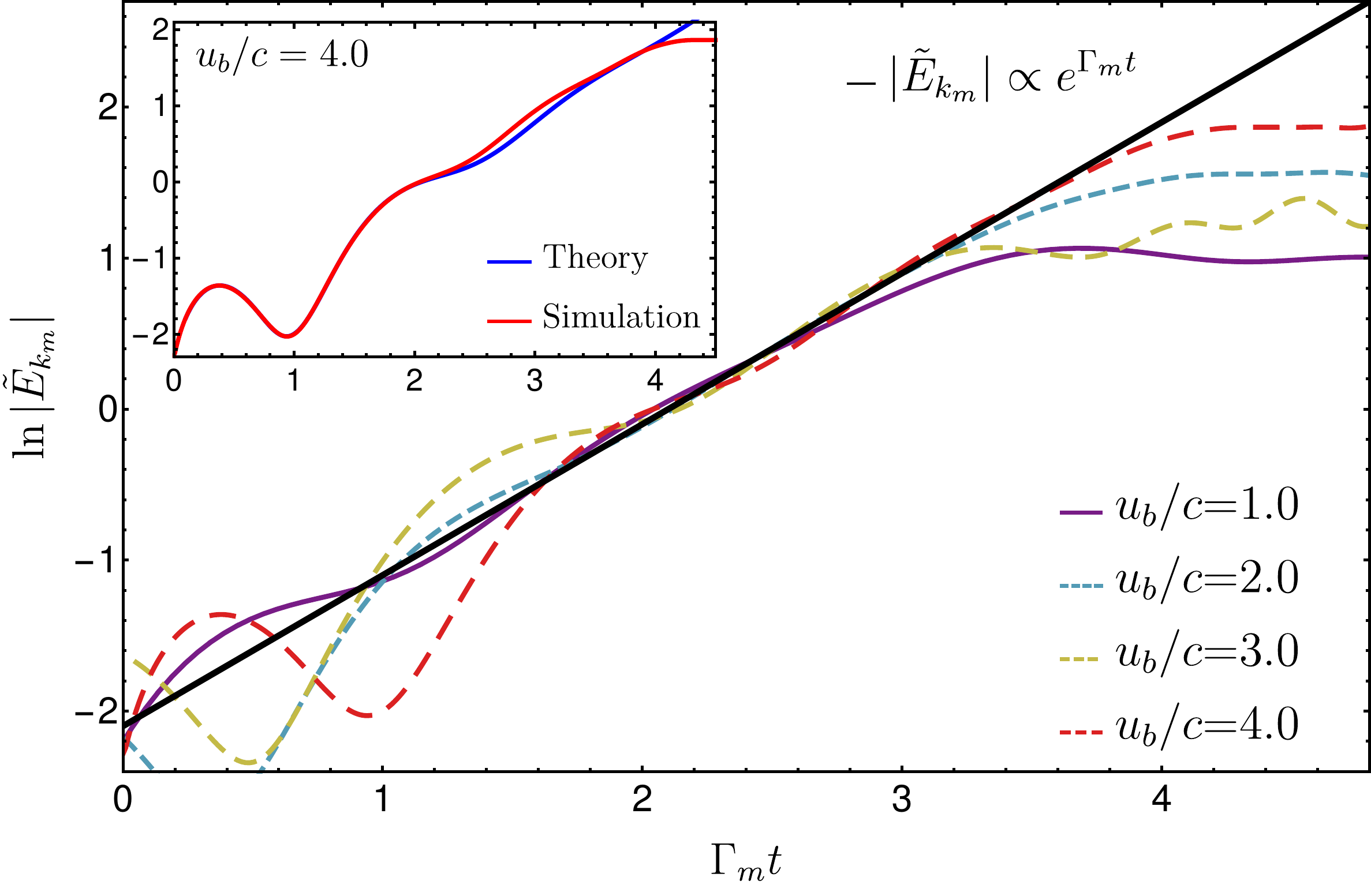}
\caption{Two-stream instability simulation results the relativistic regime. 
Different curves show the growth of the maximally growing mode (predicted by the theory) with time in units of the growth rate of this mode (predicted also theoretically) for streams with different speeds. The solid-black line is a line with slope $=1$.
An excellent agreement between simulation results and theoretical predictions is therefore evident: the growth rates found in the simulations and the rates calculated by solving the linear dispersion relation in the cold-limit Equation~\eqref{cold-2stream} are quantitatively similar. 
The inset here shows a fit for $u_b/c=4.0$ that includes the oscillatory components given in Table~\ref{table:2stream-rel-sim}.
Note that for different stream velocities, $\Gamma_{m}$ is different ( see Table~\ref{table:2stream-rel-sim} and Figure~\ref{plot:2stream-sim}).
Since the instability here grows from the noise in all simulations, we shifted the time so that linear instability growth starts at the same time for all simulations.
\label{plot:2stream-rel-sim}
}
\end{figure}

Figure~\ref{plot:2stream-rel-sim} shows again
an excellent quantitative agreement between the growth rates of the fastest growing Fourier component of the electric field in different simulations and the theoretical predictions (solid-black curve).
As in the non-relativistic case, the simulation exhibits exponential growth with the anticipated growth rate over three to four $e$-folding timescales.  As for the non-relativistic case, the initial oscillations can be identified with the oscillatory, non-growing modes (described by and below Equation~\eqref{eq:disp-2-stream}). A fit for $u_b/c=4.0$ that includes the oscillatory components is shown as an inset inside Figure~\ref{plot:2stream-rel-sim}.

\section{Comparison with TRISTAN-MP}
\label{sec:comparison}

TRISTAN-MP is a publicly available PIC code to study plasma physics relevant for astrophysical problems \citep[e.g.,][]{Tristan-Relativistic_Shocks,Tristan-Cosmic-ray,Tristan-pulsars}.
Here, we compare some of the results we obtain from SHARP-1D with those obtained using TRISTAN-MP.
In all cases, the same initial data are used.
Generally, we find a substantial improvement in the ability to conserve energy and avoid numerical heating in SHARP simulations, and a good agreement in short timescale phenomena for which energy non-conservation is not substantial.

In our test problems, the performance of SHARP compares very favorably to TRISTAN-MP, typically running roughly an order of magnitude faster.  We caution, however, that this may not be an entirely fair comparison since TRISTAN-MP, as a 3D code, may not be optimized for 1D problems.

\subsection{Numerical Particulars of TRISTAN-MP}

In addition to the initial conditions, TRISTAN-MP has a number of specific numerical parameters that impact its performance.  While we are unable to perform an exhaustive analysis of each, we did explore the result of varying a handful of these.
TRISTAN-MP is a 3D and 2D PIC code. Here, we use the 2D version with one or two cells in one of the spatial dimensions to run it in an effective 1D setup, which enables a fair comparison to SHARP-1D.

\subsubsection{Filtering}

TRISTAN-MP provides the ability to low-pass filter the deposited grid moments, e.g., current densities on the grid, damping high-frequency noise prior to using them to solve Maxwell's equations.
This reduces the coupling between the wave modes resolved on the grid with their aliases leading to improvements in the momentum and energy conservation of the algorithm. 
In TRISTAN-MP, filtering is accomplished with a three-cell stencil that generates a weighted average between the charge current density in a given cell with its neighbors. This operation may be repeated as many times as desired, smoothing the moments on progressively larger scales.

It is not a priori clear how many passes of the three-cell filtering operation are optimal in a given problem and we experimented with a number of different choices for the comparison problems presented here. We find that after a small number of filtering passes, typically three to five, the qualitative improvement is only moderate for $\theta \lesssim \theta_D$ and negligible for $\theta > \theta_D$.

This modest improvement comes with the additional computational cost, set by the addition of a substantial number of transverse grid cells required by many filtering passes. 
Thus, when comparing the numerical heating in SHARP-1D (when $W^5$ is used), we use four filtering passes with a 2D simulation box for TRISTAN-MP that is only two cells wide in the $x$-direction. 
All other comparisons employ only three filtering passes or less with a 2D simulation box for TRISTAN-MP that is only one cell wide in the $x$-direction.

\subsubsection{Electromagnetic Mode Speeds}

To suppress the numerical Cerenkov instability in TRISTAN-MP it is possible to independently set the ratio of the propagation speed of transverse electromagnetic modes to the speed of light.
This is implemented explicitly via an additional numerical coefficient in Maxwell's equations. Typically, this is set near to unity, e.g., 1.025.
However, in the 1D electrostatic case, the numerical Cerenkov instability does not exist, and we have verified that this factor does not qualitatively change any of the results from the TRISTAN-MP simulations.
  
\subsection{Thermal stability and energy conservation}

\begin{figure}
\center
\includegraphics[width=8.6cm]{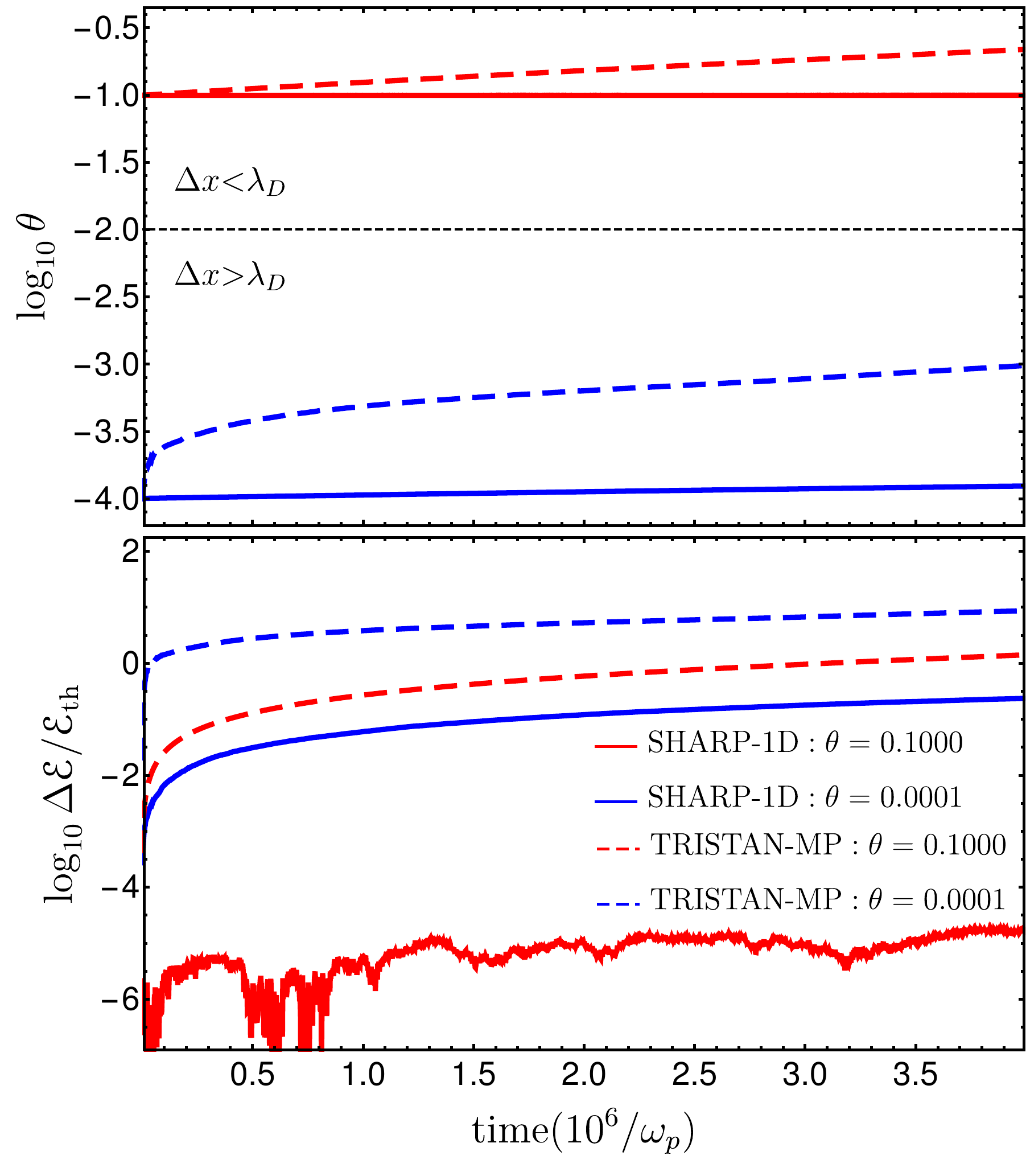}
\caption{Comparison of the numerical heating (top) and energy non-conservation (bottom) when TRISTAN-MP is used (four filters, dashed curves) and when SHARP-1D (fifth order, solid curves) is used. Here, $\theta = k_B T / m_0 c^2$ is the normalized temperature, $\Delta \mathscr{E}$ is the energy change/error in the total energy, and $\mathscr{E}_{\rm th}$ is the initial thermal energy, i.e., excluding rest mass energy of macro-particles, therefore $\Delta \mathscr{E}/\mathscr{E}_{\rm th}$ measures the fractional error with respect to the initial thermal energy of plasma. We compare the evolution of the plasmas with initial normalized temperatures of $\theta_i=10^{-1}$ (red curves) and $\theta_i=10^{-4}$ (blue curves) for the two codes.
The dashed black line in the top panel shows the Debye temperature $\theta_D$.
\label{plot-W5-R4_compare}
}
\end{figure}

In Figure~\ref{plot-W5-R4_compare} we compare the evolution of the temperature and energy error in a pair of simulations described in Section~\ref{sec:thermal_stability}.
These are chosen such that in one case the Debye length is resolved by the grid cell (red curves) and when it is not resolved (blue curves). 
In both cases, TRISTAN-MP (dashed lines) exhibits a significantly larger violation of energy conservation, differing only in the timescale over which this occurs.
When the Debye length is resolved, the numerical heating occurs more slowly, becoming untenable only after $3.5\times10^6$ plasma timescales.
On the other hand, when the Debye length is not resolved, the numerical heating dominates the initial thermal energy almost instantly.
In both cases by $4\times10^6\omega_p^{-1}$, both simulations have generated similar {\it relative} degrees of numerical heating, i.e., the ratio of the energy errors to the original thermal energy of the plasma.

In comparison, SHARP-1D (solid lines) reduces the numerical heating rate drastically. When the Debye length is resolved (high temperature) the factional errors are fixed near $10^{-5}$ throughout the simulation.  Lower temperature plasmas exhibit similar {\it absolute} heating rates, and therefore the relative heating for cold plasmas appears larger. However, even when the Debye length is unresolved by an order of magnitude, the plasma continues to be well modeled.
The origin of the improvement in the numerical heating is the improvement in the order of interpolation.

\subsection{Stability of standing linear plasma waves}

\begin{figure}
\center
\includegraphics[width=8.4cm]{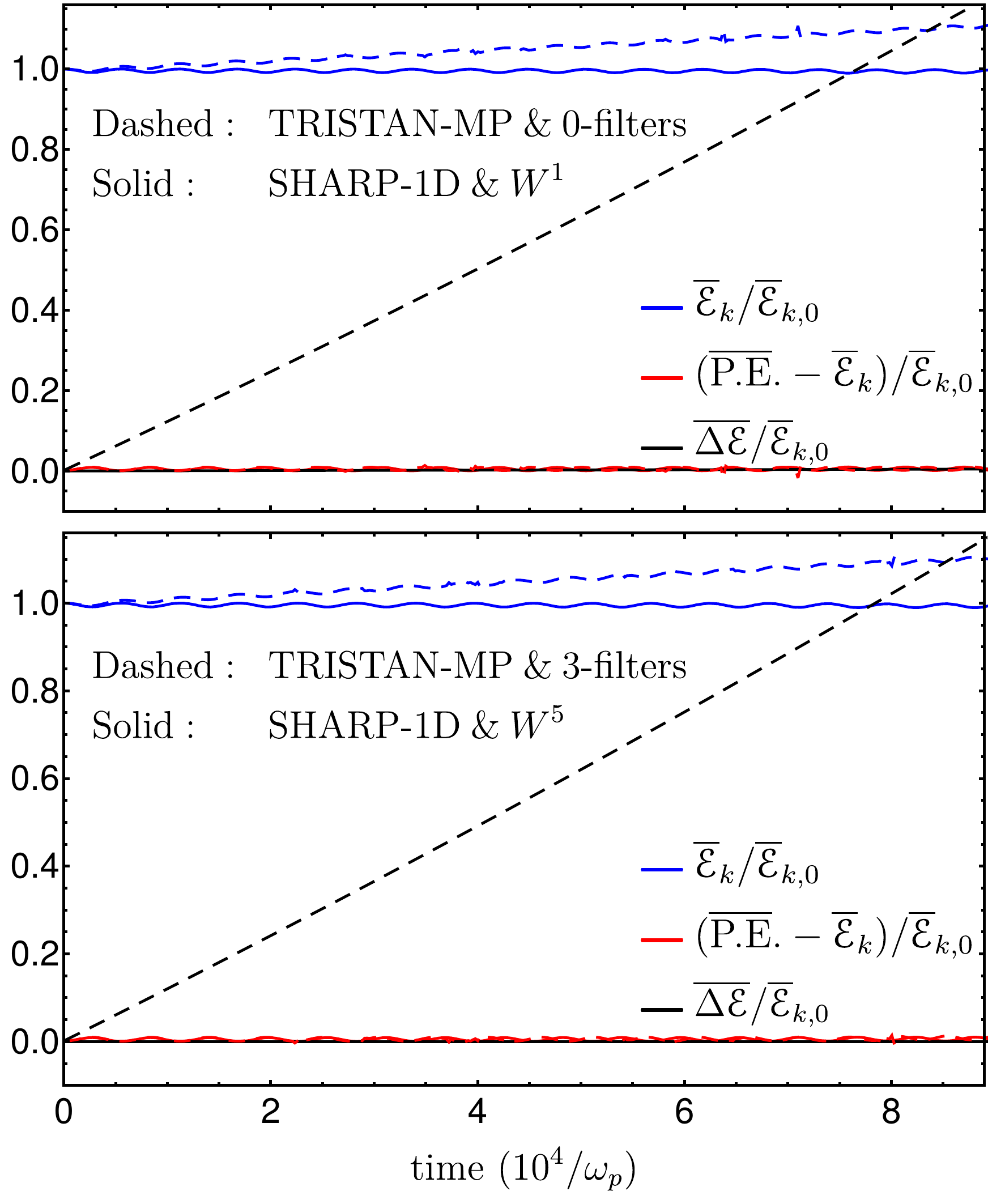}
\caption{
Evolution of the averaged energy in the initially excited mode (blue), the energy in all other modes resolved by the grid (red), and averaged energy error (black). Note that all energies, including the energy error, are normalized by the initial average energy in the excited mode, which is 16.87\% of the initial thermal energy in the plasma, i.e., excluding the rest mass energy. Results from SHARP-1D are shown as solid lines, while results from TRISTAN-MP are shown as dashed lines. The averaging is done over 37 plasma periods, while the normalization is done with respect to the initial average energy in the excited mode, i.e., in the first 37 plasma periods.  Top (bottom) panel shows a comparison of the SHARP simulations employing $W^1$ ($W^5$) to the TRISTAN-MP simulations with no (three) filtering passes.
\label{plot-oscillation_compare}
}
\end{figure}

Here, we compare the evolution of a standing plasma wave, where the linear Landau damping can be ignored, i.e., the ability of both TRISTAN-MP and SHARP-1D to maintain a small amplitude oscillating wave mode.

The simulation setup is similar to that in Section~\ref{sec:Standing_waves}: a fixed uniform background of ions with thermal electrons in a box with size $L = 40 \hspace{0.05cm} c/\omega_p$.
The electrons are initially uniformally distributed. The initially excited mode is added through a velocity perturbation: we first initialize electron velocities using Equation~\eqref{eq:ini_gauss} with $\theta=10^{-3}$, then add a position dependent velocity perturbation to individual particles' velocity by adding $\beta \cos(2 \pi x / \lambda)$ to their velocities, where the initially excited wavelength $\lambda = 20 \hspace{0.05cm} c/\omega_p$, with $\beta = 0.01 \lambda / 2 \pi$, i.e., after about $0.25$ of a plasma period this will introduce a density perturbation with an amplitude of 0.01.\footnote{Unlike the simulations in Section~\ref{sec:Standing_waves}, here we initialize a perturbation in the velocity, which is more easily done with the native initialization routines of TRISTAN-MP.  We have verified that there are no significant differences when the mode is initialized as a density perturbation, and hence do so when we later compare a Landau damped wave.}
The cell size is $\Delta x = 0.01 \hspace{0.05cm} c/\omega_p$ and since $k v_{\rm th}/\omega_p = 0.009934$, linear Landau damping can be ignored.
In all simulations described here, we fix the number of electrons per cell to $N_{pc}=1250$. We also note that in all simulations the Debye length is well resolved, i.e.,
$\lambda_D = 3.162 \hspace{.05cm} \Delta x$.

In Figure~\ref{plot-oscillation_compare} we show the long-term evolution of the energy in this isolated wave mode.
In both the SHARP-1D (solid curves) and TRISTAN-MP (dashed curves) simulations, 
the square of the amplitude ($|\tilde{E}_k|^2$) exhibits very small, long-timescale oscillations.  In the SHARP-1D simulations, these are confined to within 0.8\% of the initial value over the entire simulation.  In contrast, the TRISTAN-MP simulations also exhibit a secular growth in the mode amplitude, leading to an approximate energy increase of 10\% by $8\times10^4 \omega^{-1}_p$.

This behavior is independent of the interpolation order of SHARP-1D or number of smoothing filters employed in TRISTAN-MP.  Even when employing $W^1$ with SHARP-1D, the mode amplitude continues to execute only small oscillations about the fixed value, accurately reproducing the expectation for the linear evolution of the mode.  Because the Debye length is well resolved in this case, filtering improves the energy conservation only slightly in TRISTAN-MP, again, making little difference to the mode evolution.

The origin of the unphysical growth in the mode in the TRISTAN-MP simulations is unclear.  The heating of the background is insufficient to appreciably Landau damp or excite the mode.  We have run additional simulations with SHARP-1D employing a momentum non-conserving scheme in SHARP-1D with $W^1$, i.e., using Equation~\eqref{EkEiapprox2} (which is the back-interpolation scheme used in TRISTAN-MP), finding similar results.  That is, excluding both the differences in order and the back-interpolation of the fields as the source of the secular growth in the mode energy.  The only untested distinction remaining between SHARP-1D and TRISTAN-MP is the way in which the electric fields are updated.

\begin{figure}
\center
\includegraphics[width=8.6cm]{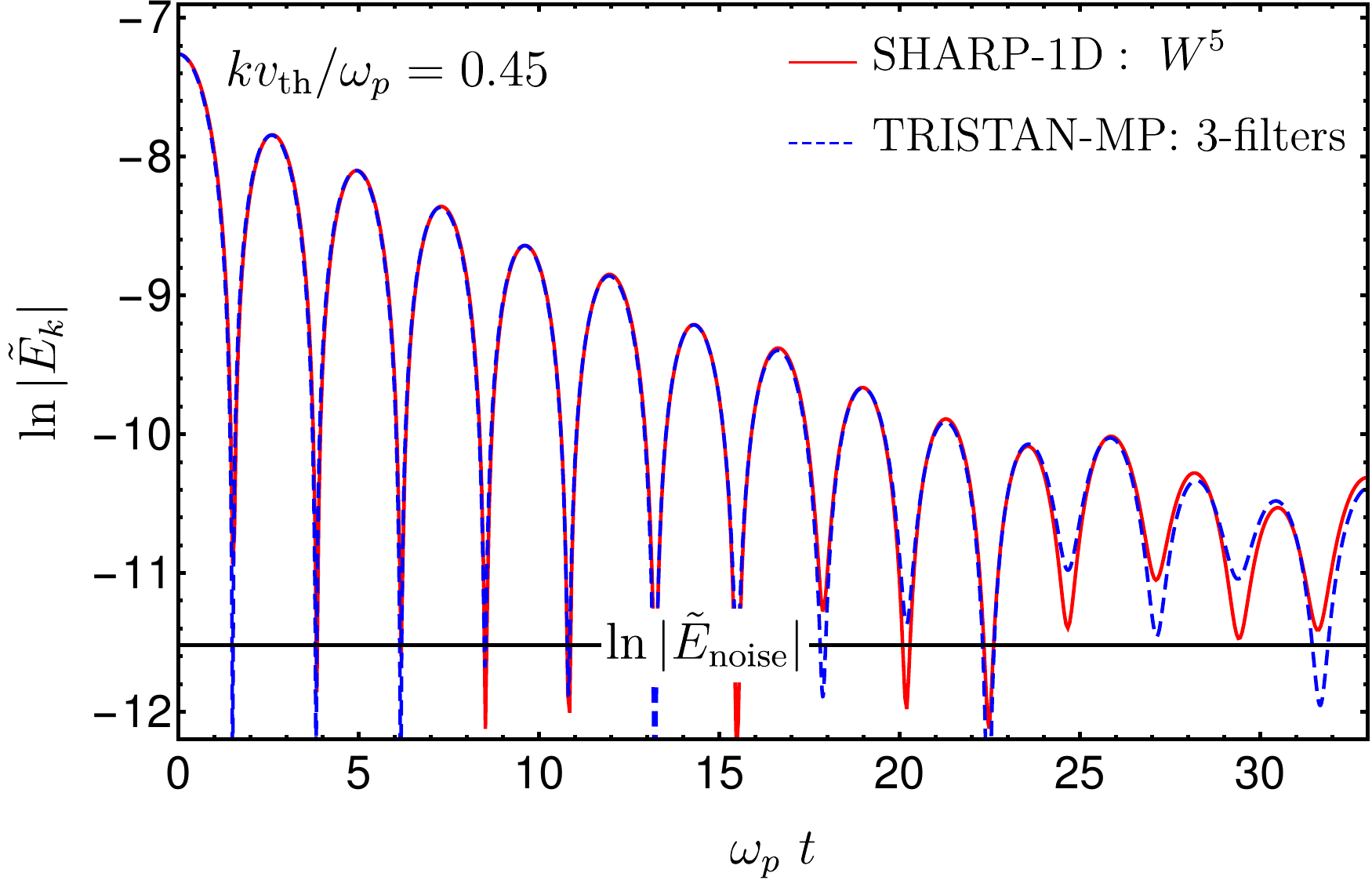}
\caption{Comparison of the evolution of a linear plasma wave in the regime where the linear Landau damping rate is high ($\hat{k}=0.45$).
Here, we compare the results of SHARP-1D with $W^5$ (solid curves) to TRISTAN-MP with three filtering passes (dashed curves).
Since, the number of particles here is lower than what was used in Section~\ref{sec:LLD}, there is a higher level of noise leading to a slightly faster damping rate than seen in Figure~\ref{plot-LLD-Simulations}. 
\label{plot-LLD_compare}
}
\end{figure}

Next, we compare the evolution of a shorter wavelength mode ($k v_{\rm th}/\omega_p = 0.45$) that has a high linear Landau damping rate. 
The simulation's setup is exactly the same as in Section~\ref{sec:LLD}, but with a lower number of macro-particles ($N_{\rm t} = 5 \times 10^7$).
The time scale for damping is much smaller than the time needed for the energy non-conservation to affect the evolution of such modes.
Thus, the result from both TRISTAN-MP (dashed blue curve) and SHARP-1D (solid red line) match exactly as shown in Figure~\ref{plot-LLD_compare}.
As noted before, the initial drop in the wave amplitude is also present when other simulation methods are used \citep[e.g.,][]{Semi-Lagrangian,DG-scheme}.

\subsection{Two-stream Instability}

\begin{figure}
\center
\includegraphics[width=8.6cm]{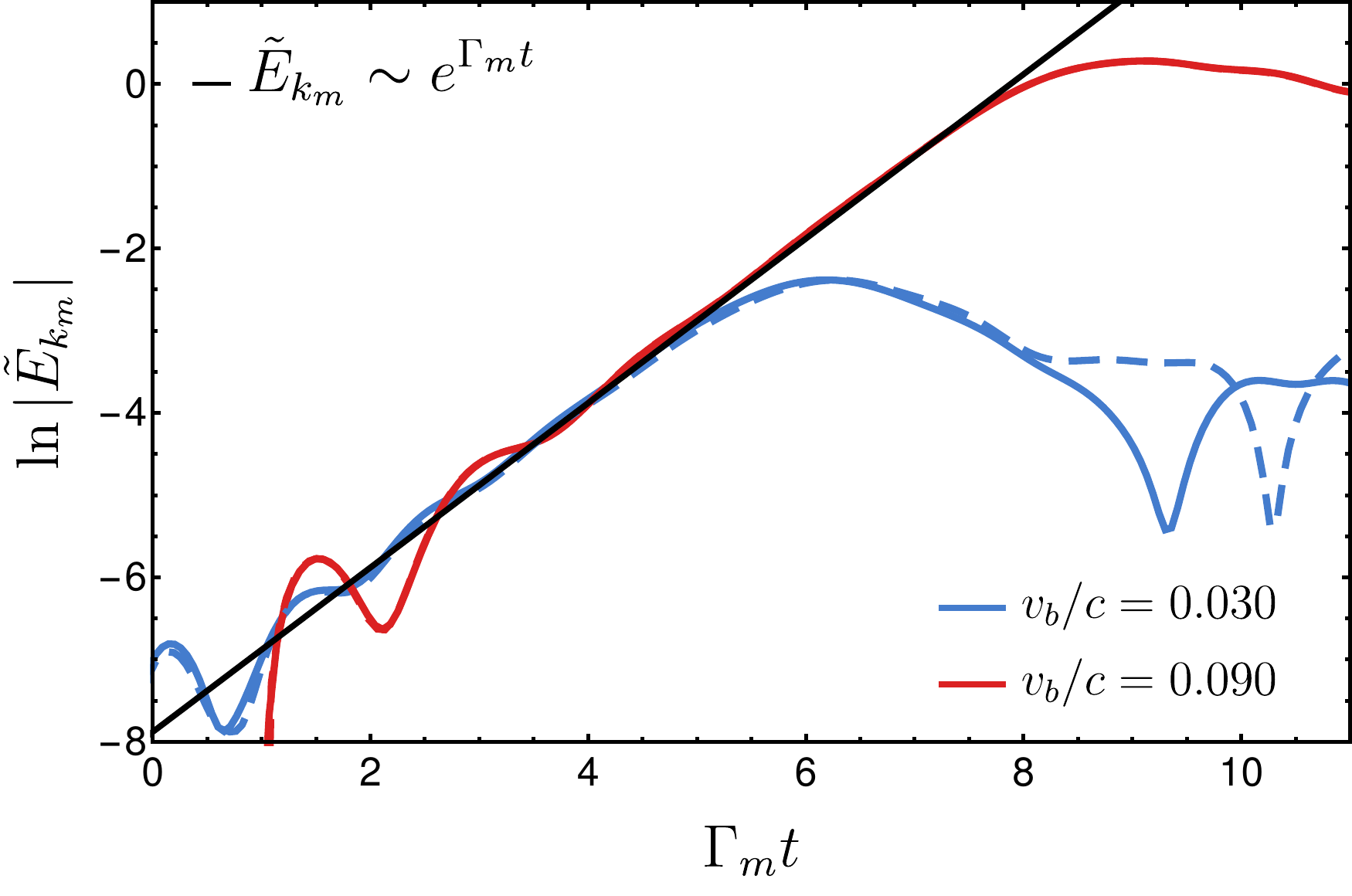}
\caption{Comparison of the non-relativistic two-streams instability. Here, we compare the results of SHARP-1D with $W^5$ (solid curves) to TRISTAN-MP with three filtering passes (dashed curves).
\label{plot-2stream_non-rel_compare}
}
\end{figure}

\begin{figure}
\center
\includegraphics[width=8.6cm]{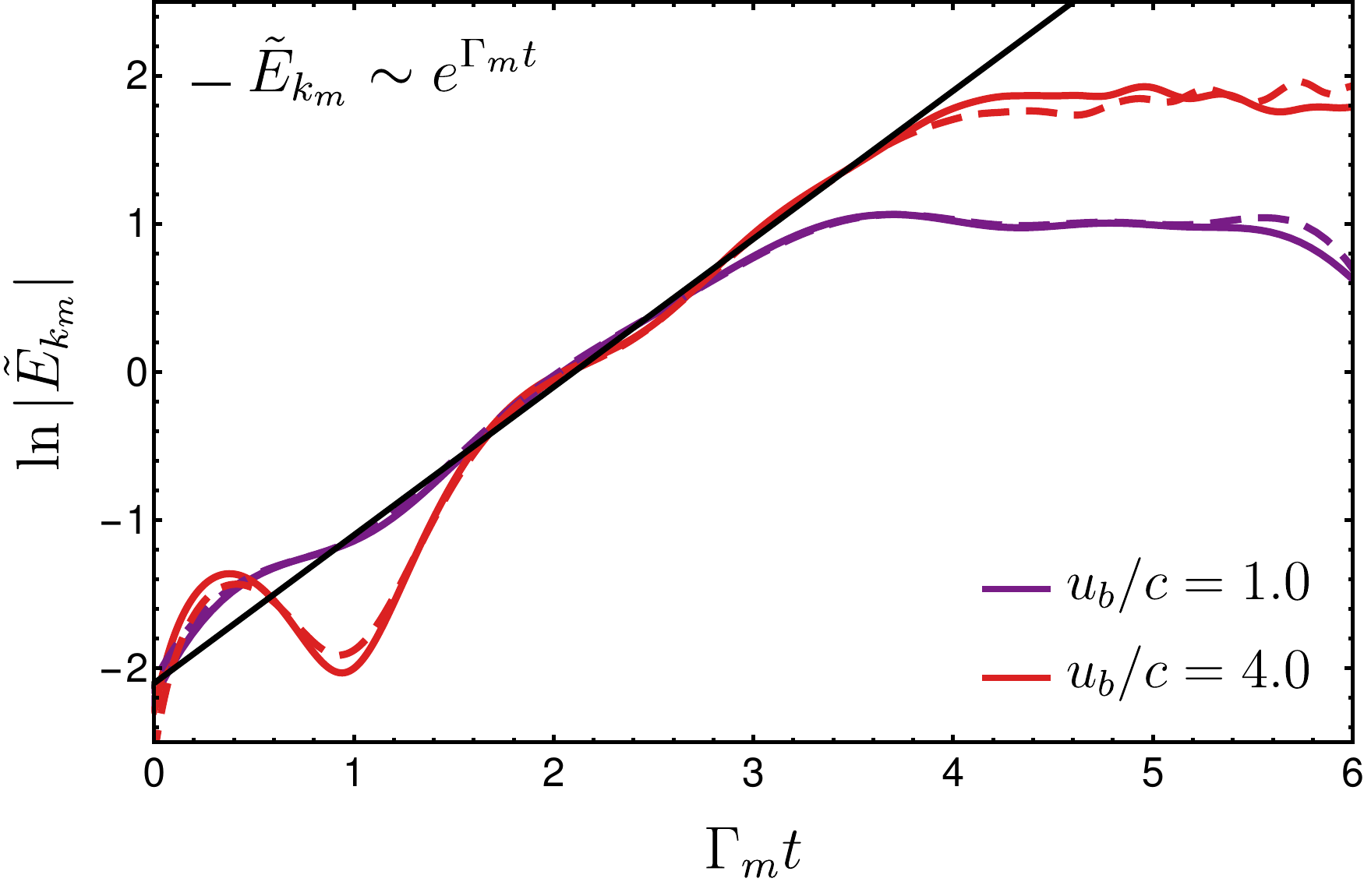}
\caption{
Comparison of the relativistic two-streams instability. Here, we compare the results of SHARP-1D with $W^5$ (solid curves) to TRISTAN-MP with three filtering passes (dashed curves).
\label{plot-2stream_rel_compare}
}
\end{figure}

We now compare the performance on dynamical instabilities, i.e., the two-stream instability as described in Section~\ref{sec:2streams}, for non-relativistic and relativistic streams. Here, again the timescale, on which such instability grows, is much shorter than the time scale needed for non-energy conservation to affect the evolution. Therefore, we find the same same linear evolution in both codes.

Figure~\ref{plot-2stream_non-rel_compare} shows a comparison for two of the non-relativistic stream velocities reported in Section~\ref{sec:2snr}.  As before, time is measured in $e$-foldings of the most unstable mode. 
In Figure~\ref{plot-2stream_rel_compare} comparisons for two relativistic stream simulations reported in Section~\ref{sec:2sr} are shown. Again the same linear evolution of the instability is found in both codes.

\section{Convergence}
\label{sec:convergence}

At the end of Section~\ref{subsec:Discretiztiona}, we discussed the accuracy and
different errors introduced in our numerical scheme. We saw that the dominant
error is of order $O(h^3)$, arising from the order of particle pusher.
Here, we assess the convergence of SHARP-1D and demonstrate that the numerical error decreases as expected. In particular, we develop a general criterion for convergence studies of PIC simulations by requiring that the ratio of the energy in the shortest wave mode to the energy in the Poisson noise of simulation to be, at least, fixed.

We then present a test case where the error in the total energy of plasmas is used as the measure of error: such a test shows that the definition for convergence motivated above leads to a decrease in the error at the expected rate and typical methods to test for convergence fail: a slower decrease in the error is observed as resolutions increase leading to a plateau in the error, where increasing the resolutions no longer lead to a decrease in the error.

\subsection{``Resolution'' in PIC Algorithms}

In general, for a PIC-type algorithm, three notions of resolution are relevant for simulations. 

\begin{enumerate}
\item Spatial resolution of the grid, i.e., $h$.  This also determines the temporal resolution.
\item Momentum resolution, set by the number of particles used to construct the charge and current density at each cell, i.e., the number of particles per cell, $N_{\rm pc}$.
\item Spectral resolution, set by the size of the ``spectral-cell,'' which, for each spatial-dimension, is given by of $2\pi / L$, where $L$ is the box-size.
\end{enumerate}
The third is rarely discussed in PIC simulations and arises when a physical phenomena (in linear or nonlinear regimes) has a narrow spectral support. In such cases, higher resolution simulations will require increasing the three types of resolutions simultaneously.
Here we will focus on the first two, leaving a complete discussion of the third for future work.

Generally, it is necessary to increase all relevant resolutions simultaneously to study algorithmic convergence.  As we will see below, this requires increasing the number of particles per cell, rather than fixing it as it is typically done. A similar requirement was found for smooth-particle hydrodynamics (SPH) simulations, where the convergence also requires increasing the number of fluid-particles within the smoothing volume of each particle to study convergence \citep{zhu+Hernquits+2015}.

\subsection{Definition of Convergence -- Equivalent Simulation}

The notions of both spatial resolution (i.e., $h$) and momentum resolution (i.e., $N_{\rm pc}$) place different constraints on the range of underlying wave modes that can effectively be simulated.  Thus, some care must be taken to ensure that as these resolutions are increased simultaneously, the underlying wave complement of the physical system resolves ever smaller scales.

The discrete nature of the macro-particles places a floor on the amplitude of a mode that can be effectively resolved.\footnote{This is what sets the thermal floor delineated by $\theta_P$ in Equation~\eqref{Eq:thetap}.}
The average potential energy of the particle distribution, or Poisson noise, is
\begin{equation}
\mathscr{E}_{\rm noise}^m 
=
 \frac{L }{12 \epsilon_0 }
\left[ 
1
-
\frac{6 f_m}{N_c} 
\right] 
q_0^2 N_p,
\end{equation}
where recall that $f_m$ is a coefficient that depends on the spatial order of the algorithm.  Any mode with an energy less than $\mathscr{E}_{\rm noise}^m$ is effectively unresolvable. 

For comparison, we compute the energy in a single plasma mode that can be revolved on the grid\footnote{The wavelength $\lambda$ is resolved by the grid, if $\lambda/L$ is an integer.}. For a single mode in plasma with charge density given by 
\begin{eqnarray}
\rho(x,t) 
&=& 
- A \left( \frac{q_0 N_p}{L}\right) \cos(\omega t) \hspace{.05cm} \cos (2 \pi  x /  \lambda),
\end{eqnarray}
the electric field is given by
\begin{eqnarray}
E(x,t) &=& E_0(t) +\int_0^x dx^{'} \frac{ \rho(x^{'} ,t) }{ \epsilon_0 }  \nonumber \\
&=& E_0(t) - \left( \frac{q_0 N_p}{L} \frac{\lambda}{2 \pi } \right) \frac{A}{\epsilon_0}  \cos(\omega t)  \sin (2 \pi  x /  \lambda),
\end{eqnarray}
where $A$ is the amplitude of the initial perturbation for a mode with wavelength $\lambda$. Hence, the total electric field energy is given by
\begin{equation}
  \begin{aligned}
    \mathscr{E} (t) 
    &= \frac{\epsilon_0 }{2}
    \int_0^L  E(x,t)^2  dx\\
    &=
    \frac{\epsilon_0  L }{2} E_0^2(t) + \frac{1}{2\epsilon_0} \left( \frac{q_0 N_p}{L}  \frac{\lambda}{2 \pi } \right)^2  A^2 \cos^2(\omega t) \frac{L}{2}.
  \end{aligned}
\end{equation}
Averaging over a full-period and assuming that $ \left\langle E^2_0  \right\rangle =0$, the averaged potential energy in the wave mode is then
\begin{equation}
\left\langle \mathscr{E}  \right\rangle
=
\frac{1}{2 \epsilon_0} \left( \frac{q_0 N_p}{L}  \frac{\lambda}{2 \pi } \right)^2  A^2   \frac{L}{4}.
\end{equation}

Whether or not a mode can be resolved is then determined by the ratio of $\left\langle \mathscr{E}  \right\rangle$ to  $\mathscr{E}_{\rm noise}^m$, 

\begin{equation}
r
=
\frac{ \left\langle \mathscr{E}  \right\rangle }{\mathscr{E}_{\rm noise}^m }
=
\frac{3}{2}
\frac{A^2  N_p}{(2 \pi L/\lambda)^2}
\left[ 
1
-
\frac{6 f_m}{N_c} 
\right]^{-1}
\approx
\frac{3 A^2}{8 \pi^2}
\frac{\lambda^2 N_p}{L^2}.
\end{equation}
When $r>1$, the mode is resolved on the grid, while when $r<1$ it is dominated by the Poisson noise in the simulation and rapidly randomized.  Because $r\propto \lambda^2$ this statement is also a function of wavelength, with the smallest wavelength modes being the most marginal.  That is, the Poisson noise limit, $\mathscr{E}_{\rm noise}^m$, sets a minimum mode wavelength, $\lambda_{\rm min}$ the simulation can resolve, independent of the spatial resolution of the grid.  Improving the fidelity of the simulation requires, therefore, concurrent increases in spatial resolution (e.g., $h$), momentum resolution (e.g., $N_{\rm pc}$), and spectral resolution (e.g., $L$).

Explicitly, requiring that modes on the smallest spatial scales are resolved, i.e., $\lambda_{\rm min}\propto h$, then translates into the requirement that $h N_{pc} / L$ is, at least, fixed.\footnote{Such requirement (a fixed ratio of the energy in the shortest wavelength mode to the Poisson energy) implies that the ratio of the energy in a specific mode to the energy in the Poisson noise of simulation increases as $\eta^2$ with improving resolution.} When spectral resolution is not important, this requires that if the spatial resolution increases by a factor $\eta$ then $N_{\rm pc}$ must grow by a similar factor, i.e.,
\begin{equation}
  h \rightarrow h/\eta
  \quad\text{and}\quad
  N_{pc} \rightarrow \eta N_{pc}\,.
\end{equation}
This is unsurprising -- convergence requires simultaneous and equal increases in the spatial and momentum resolutions.  It does mean, however, that convergence studies are numerically demanding, as they scale as $\eta^3$, even in 1D.
This is, of course, exactly the factor one would find in a purely Eulerian scheme for solving the 1D Boltzmann equation, which is similar to 2D hydrodynamics. It does make clear, however, that the
inherent randomness of the particle description does nothing to improve the convergence characteristics.

\subsection{SHARP-1D Convergence -- An Explicit Example}

\begin{figure*}
\center
\includegraphics[width=18.4cm]{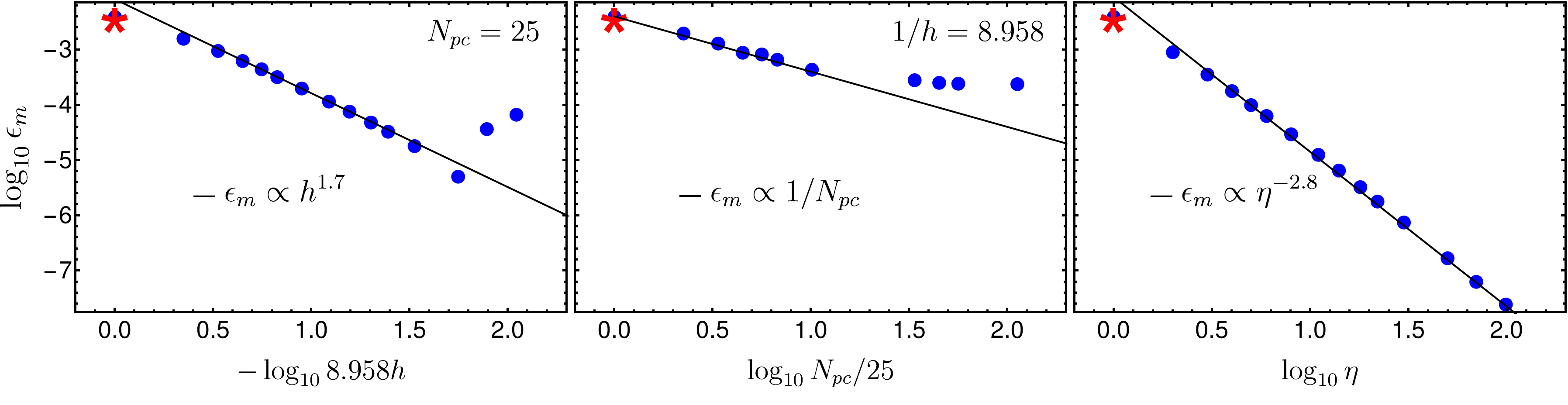}
\caption{Effect of increasing different resolutions starting with our fiducial simulation, on the maximum of the normalized energy error $\epsilon_{m} \equiv {\rm max } \left( \Delta  \mathscr{E} \right)/ \mathscr{E}_{\rm th}$, where $\Delta  \mathscr{E}$ is the energy change in the total energy, $ \mathscr{E}_{\rm th}$ is the initial thermal energy, i.e., excluding rest mass energy. The figures show the effect of increasing the spatial resolution, $h$, while fixing the momentum resolution, $N_{\rm pc}$, (right), the effect of increasing the momentum resolution while fixing the spatial resolution (middle), and the effect of simultaneously increasing both momentum and spatial resolution (left). The red-star result, which is the same simulation for all plots here, corresponds to our fiducial simulation ($N_{pc}=25$ and $1/h=8.958$). We define $\eta \equiv N_{pc}/25 = 1/(8.958~h)$.
}
\label{fig:convergence}
\end{figure*}

We now provide an explicit example of convergence testing, as described in the previous section, using SHARP-1D.  To illustrate both the convergence of SHARP-1D under this definition, and equally importantly, the lack of convergence under separate definitions often employed, we do this for an extreme range of $\eta$, extending over two orders of magnitude.

We begin with a fiducial simulation, which defines $\eta=1$.  This is comprised of a population of electrons, with total number of macro-particles of $N_p=8950$,  and a fixed neutralizing background. We use a box with normalized length $\bar{L} = 39.96175$ and the initial normalized temperature for electrons $\theta_{ini} = 10^{-3}$. We start with a single excited mode with amplitude $A=10^{-2}$ and wavelength of $\bar{\lambda} = \bar{L}/2$. Therefore, $\hat{k} = 2 \pi \sqrt{\theta}/\bar{\lambda} = 0.0099441$, i.e., the linear perturbation should oscillate without damping during the entire simulation time, $T = 100\omega_p ^{-1}$.

For our fiducial simulation, we set the cell size, $\Delta x$, such that $1/h= c / ( \Delta x \omega_p)   = 8.958$, i.e., $N_c = 358$ and $N_{pc} = N_p/N_c = 25$. 
For all simulations in this section, we use fifth-order interpolation ($W^5$).
Note that in this simulation the box is sufficiently large to spectrally resolve all relevant features of the dispersion relation, and thus we do not consider it further here.

The accuracy measure we employ is the normalized maximum error over the duration of the simulation:
\begin{equation}
\epsilon_m \equiv \frac{{\rm max} \left( \Delta \mathscr{E}  \right)}{ \mathscr{E}_{\rm th} },
\end{equation}
where $\Delta \mathscr{E}$ is the energy change in the total energy, $\mathscr{E}_{\rm th} $ is the initial thermal energy, i.e., excluding rest mass energy.
Using various definitions of the energy error, i.e., average error, result in qualitatively identical results.
Note that this is not the only accuracy measure we might use; others include the amplitude or phase of the wave, or the ability to reproduce other known solutions.  It does have the property that it is fundamentally well understood (energy is conserved), not explicitly conserved by  the code (like momentum), and universally defined.

Figure~\ref{fig:convergence} shows the impact of increasing the various relevant notions of resolution independently and together.  Increasing either the spatial or momentum resolution independently leads to a plateau in $\epsilon_m$. The location and magnitude of this plateau depends upon the value of the non-converging resolution, i.e., when converging in spatial resolution, $N_{pc}$, or when converging in momentum resolution, $h$. 
This is qualitatively distinct from the case when both resolutions are increased simultaneously, for which no plateau is evident over two orders of magnitude.

\begin{figure}
\center
\includegraphics[width=8.6cm]{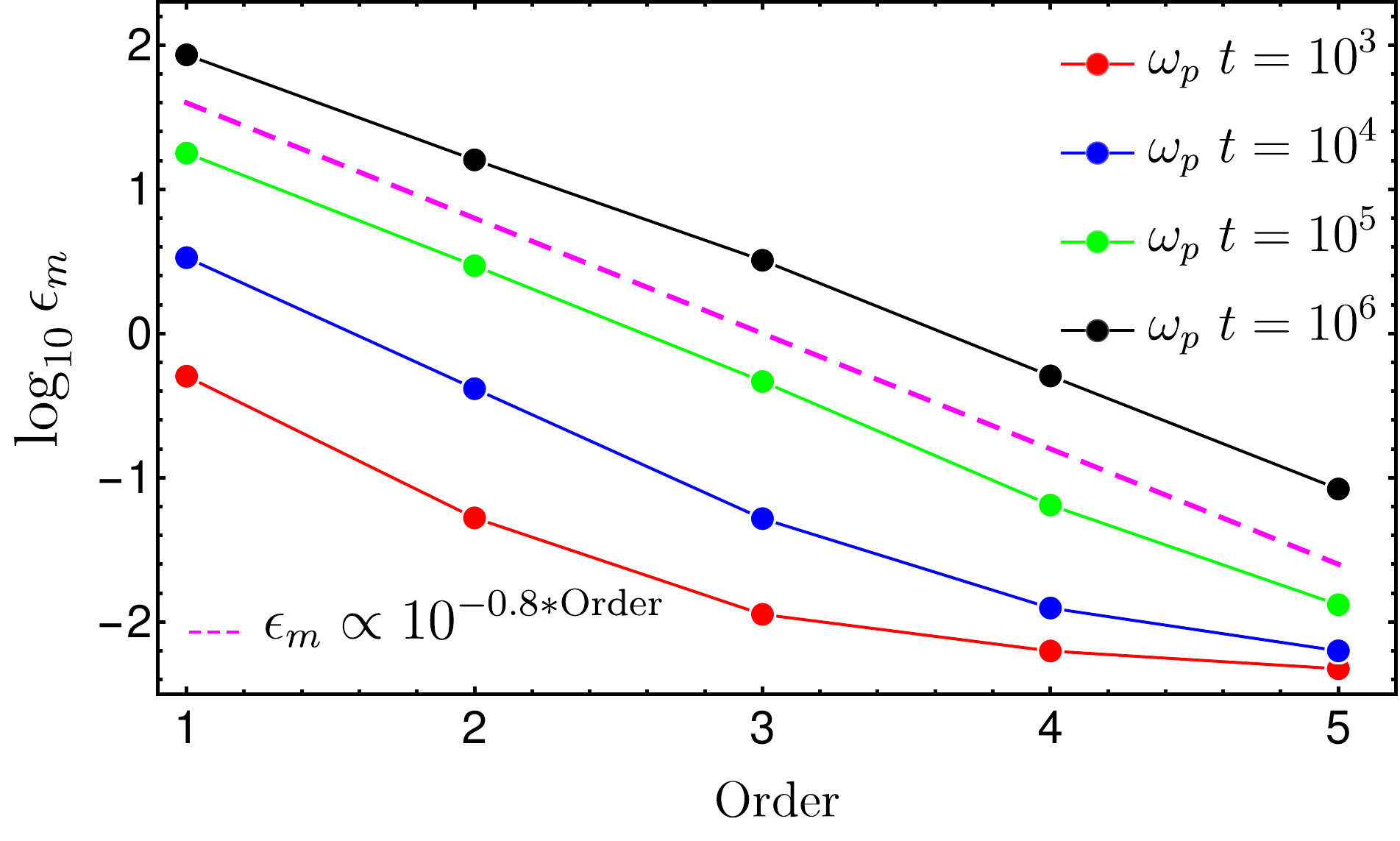}
\caption{
Effect of using higher-order interpolation functions on the normalized maximum energy error $\epsilon_m$. Red, blue, green, and black lines are, respectively, the results after running the simulation up to $10^3$, $10^4$, $10^5$, and $10^6$  $\omega_p^{-1}$.
This shows the importance of using higher-order interpolation functions in controlling the energy non-conservation, the effect is specially important for long time simulations. Here, the maximum of the normalized energy error $\epsilon_{m} \equiv {\rm max} \left( \Delta \mathscr{E} \right)/\mathscr{E}_{\rm th}$, where $\Delta \mathscr{E}$ is the energy change in the total energy, $\mathscr{E}_{\rm th}$ is the initial thermal energy, i.e., excluding rest mass energy.
}
\label{fig:eta1}
\end{figure}

Quantitatively, using our definition of convergence, SHARP-1D converges as $\eta^{-2.8} \propto h^{2.8}$. This is very similar to the anticipated $h^3$, with the implication that the algorithm performance is well understood.  Because this is ultimately set by the currently second-order symplectic integration employed in the particle pusher, implementing higher-order spatial interpolation will not improve this convergence rate.
However, the value of the higher-order spatial interpolation appears in the magnitude of the energy error (i.e., its coefficient). This is clearly evident in Figure~\ref{fig:eta1}, which shows the error in the $\eta=1$ simulation for different interpolation orders.  
Nevertheless, achieving the full benefits of the higher-order spatial implementations  will require implementing an appropriate particle pusher.

\section{Performance of SHARP-1D}
\label{sec:implementation}

\begin{figure}
\center
\includegraphics[width=8.4cm]{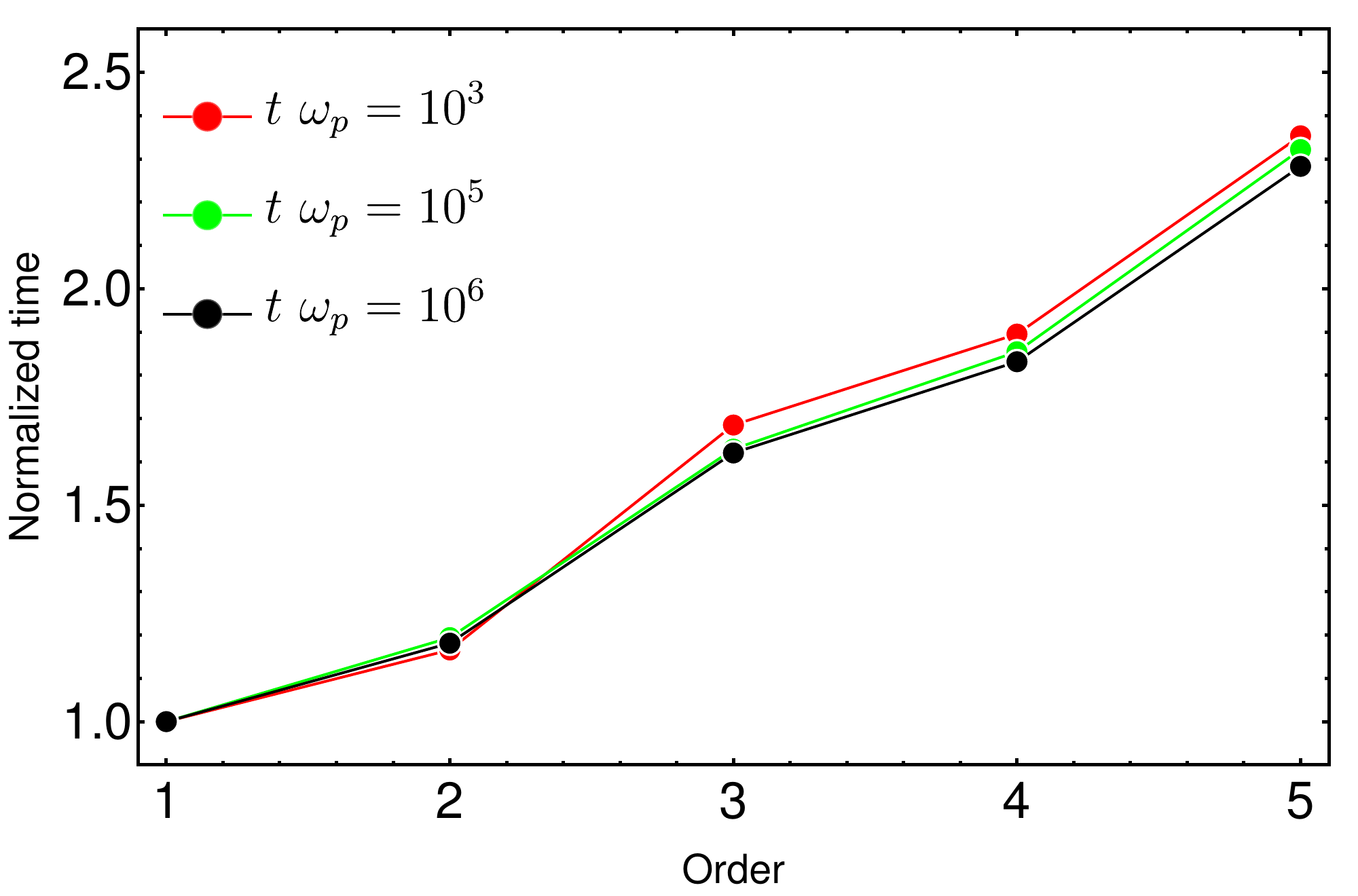}
\caption{
Increase in computational cost when higher-order interpolation functions are used in SHARP-1D, at a fixed number of macro-particles ($N_{\rm p}=8950$) and number of cells.
The normalized time is the computation time used in both deposition and back-interpolation steps when simulations are run up to $t~\omega_p =$ $10^3$ (red), $10^5$ (green) and $10^6$ (black), normalized to the corresponding time when $W^1$ is used.
Here, we used five processor in all simulations.
We have verified that the relative increase in the computational cost, we found here, is independent of the number of macro-particles and the number of processors.
\label{fig:cost}
}
\end{figure}

To quantify the increase in the computational cost of using higher-order interpolation functions, we ran a simulation using 8950 macro-particles on five processors for all implemented orders.
In Figure~\ref{fig:cost}, we show the relative increase in computational time for both deposition and back-interpolation steps after running each simulation up to $t~\omega_p =$ $10^3$ (red), $10^5$ (green), and $10^6$ (black).
The computational cost per update of using $W^5$ is $2.28$ times larger than $W^1$.
We have verified that this is independent of the number of macro-particles and the number of processors.

The advantages of using higher-order interpolation are problem dependent. 
However, if we use the error in the total energy as a measure of accuracy, we can attempt to quantify the difference by computing the relative computational cost of simulations with different interpolation orders holding the level of accuracy fixed.
Figure~\ref{fig:eta1} shows that for a simulation that runs until time $ t = 10^6 ~\omega_p^{-1}$, the energy error, $\epsilon_{m}$, is smaller by a factor of $10^3$ when $W^5$ is used instead of $W^1$.

To achieve a similar accuracy using $W^1$, i.e.,  decreasing $\epsilon_{m}$ by a factor of $10^3$, $\eta$ needs to be increased by $10^{3/2.8} \sim 11.8$ (where we employed the scaling in the right-hand panel of Figure~\ref{fig:convergence}).
Consequently, both $N_{\rm pc}$ and $N_{\rm c}$ each have to increase by $11.8$, increasing the number of steps by the same factor.
Therefore, using $W^1$, the computational cost increases by a factor of $11.8^3 \sim 1640$, $\sim730$ times that required by $W^5$. 
That is, to achieve the same level of accuracy, a simulation that uses $W^5$ is about $730$ faster than a simulation that uses $W^1$ with improved resolutions.\footnote{ 
For reference, on Intel Xeon 3.47 GHZ CPUs, the computational time to evolve 8950 macro-particles on five processors for 22396417 steps is 1732 seconds per processor when using $W^1$.}

SHARP-1D exhibits a near linear strong scaling, i.e., for fixed problem size, with the number of processors employed, $N_{\rm pr}$.  By varying $N_{\rm pr}$ between 20 and 300, we find 
\begin{equation}
t \sim N_{\rm pr}^{-0.96}.
\end{equation}

\section{Conclusion}
\label{sec:conclusion}

In this paper, we present a self-consistent discretization for the governing equations of plasma made of macro-particles in 1D (i.e., the Vlasov--Poisson equations) implemented in the SHARP-1D code.
It employs a self-consistent force on such macro-particles that is accurate up to fifth order and provides an essential step toward higher-order accurate PIC schemes.
The over all accuracy of the algorithm is, however, limited by the accuracy of the particle pusher which is still a second order accurate symplectic method (leap-frog).

SHARP-1D conserves momentum exactly, and despite its second-order accuracy, when higher-order interpolation functions are used, better energy conservation and lower numerical heating is evident. SHARP-1D simulations of a thermal plasma, whose Debye length is 10 times smaller than the cell size and which employ spatial interpolation accurate to fifth order only have an energy error, which is better than 1\% of the initial thermal energy. Moreover, it shows a negligible numerical heating over a very long time (up to millions of inverse plasma frequencies, see Figure~\ref{plot-Thermal_stab}).

We present a validation of SHARP-1D against some test problems: the thermal stability of plasmas, the stability of linear plasma modes, and the two-stream instability in the relativistic and non-relativistic regimes.

To perform such validation tests, we determine the correct modes of thermal plasmas (oscillation frequencies and damping and growing rates) by solving the corresponding linear dispersion relations numerically. This is done for thermal plasmas that are both stationary and counter streaming. For convenience, we provide a fit to the oscillation frequencies and the damping rates in the linear regime of thermal plasmas up to $\hat{k}=k v_{\rm th}/\omega_p=0.6$.

In all test problems, SHARP-1D demonstrates the ability to reproduce kinetic effects of the linear regime both qualitatively and quantitatively. This includes reproducing the correct oscillation frequencies and damping rates for different modes of the thermal plasma, and also reproducing all oscillating, growing, and damping modes in counter streaming plasmas in the relativistic and non-relativistic regimes.

Results from SHARP-1D in both relativistic and non-relativistic regimes are contrasted with results from TRISTAN-MP. A substantial improvement in the ability to conserve energy  and control numerical heating  
is shown when SHARP-1D is used.

Importantly, the improved performance due to higher spatial order does not come at the cost of increased execution time; to achieve the same level of accuracy, we have shown that, for SHARP-1D, a simulation with $W^5$ is almost three orders of magnitude faster than a simulation with $W^1$ and improved resolutions.

Finally, we develop a general criterion for convergence studies of PIC simulations by requiring that the ratio of the energy in the shortest wave mode to the energy in the Poisson noise of the simulation to be at lease fixed.

An example study for such a convergence test is presented, where the decrease in the energy error for plasma, as different relevant resolutions are increased, is used as a measure for convergence. Both the number of particles per cell and the spatial resolution of the grid are crucial resolution elements: increasing only one relevant resolution results not only in slower decrease of the error, but also in a plateau where the error  does not  decrease any longer as such resolution increases. Faster decrease of the error without any plateau is achieved when all relevant resolutions are increased simultaneously for such plasma.

This new PIC code provides a new avenue that enables the faithful study of the long-term evolution of plasma problems (in one dimension) that require absolute control of the energy and momentum conservation. Those include, e.g., the oblique instability driven by the highly anisotropic TeV pair beams that emerge from TeV gamma-rays that propagate from blazars to us or interactions of relativistic plasma components with a non-relativistic background plasma over long time scales \citep{blazarI}.

\section*{Acknowledgments}

We would like to thank Anatoly Spitkovsky for providing access to TRISTAN-MP, substantial guidance in its use, and a number of constructive comments on the manuscript.
M.S. and A.E.B. receive financial support from the Perimeter Institute for Theoretical Physics and the Natural Sciences and Engineering Research Council of Canada through a Discovery Grant. Research at Perimeter Institute is supported by the Government of Canada through Industry Canada and by the Province of Ontario through the Ministry of Research and Innovation.
P.C. gratefully acknowledges support from the NASA ATP program through NASA grant NNX13AH43G, and the NSF through grant AST-1255469.
C.P. gratefully acknowledges support by the European Research Council through ERC-CoG grant CRAGSMAN-646955 and by the Klaus Tschira Foundation.
E.P. gratefully acknowledges support by the Kavli Foundation.
Support for A.L. was provided by an Alfred P. Sloan Research Fellowship, NASA ATP Grant NNX14AH35G, and NSF Collaborative Research Grant \#1411920 and CAREER grant \#1455342.

\begin{appendix}

\section{Momentum conservation}
\label{app:0Fself}

For the purpose of comparison, we start by calculating the correct interaction force for point particle in 1D, by using  
\begin{equation}
q_0 \mathbb{E}_0 = m_0 \epsilon_0 \omega_0
\quad
\&
\quad
h N_{cp} 
= 
\frac{\omega_0 L / c}{  \sum_s (\bar{Q}_s^2 N_s)/\bar{M}_s }
=
\frac{\omega_0 q_0^2}{m_0 c \epsilon_0 \omega_0^2 }
=
\frac{ q_0^2}{\epsilon_0 q_0  \mathbb{E}_0 }
\quad
\Longrightarrow
\quad
\frac{1}{\epsilon_0 q_0  \mathbb{E}_0 }
=
\frac{h N_{cp}}{q^2_0}.
\end{equation}
Therefore, the correct interaction term in 1D for point-particles ($m=0$) (in code units) is given by
\begin{eqnarray}
\label{fm0}
\bar{F}_{\text{int}} = \frac{Q_1 Q_2}{2\epsilon_0}  \frac{1}{ q_0 \mathbb{E}_0}
\begin{cases}
+1 \qquad  x_2>x_1 \\
-1 \qquad  x_2<x_1 
\end{cases} 
=  \frac{\bar{Q}_1 \bar{Q}_2 h }{2}  N_{cp} 
\begin{cases}
+1 \qquad x_2>x_1 \\
-1 \qquad x_2<x_1 
\end{cases} .
\end{eqnarray}

On the other hand, the force on a macro-particle, with charge $q_\alpha$ and centered at $x_\alpha \epsilon [0,L]$  on a periodic box, is given by $\bar{F}_{\alpha} \equiv F_{\alpha} /(q_0 \mathbb{  E}_0 ) = \bar{Q}_{\alpha} \bar{E}_{\alpha}$, where

\begin{equation} 
\label{fexact}
\bar{E}_{\alpha}
=
\int_0^L  \bar{E}(x) S^m[(x-x_{\alpha})/\Delta x] dx 
= 
\sum_{k=0}^{N_c-1} \int_{x_{k-1/2}}^{x_{k+1/2}} \bar{E}(x) S^m  dx 
= 
\sum_{k=0}^{N_c-1} \int_{x_k}^{x_{k+1}} \bar{E}(x) S^m  dx. 
\end{equation}
By defining 
\begin{equation}
N_{cp}  \equiv \frac{N_c}{  \sum_s (\bar{Q}_s^2 N_s)/\bar{M}_s } = \frac{1}{n_0 \Delta x}
\qquad
\&
\qquad
N_{pc} \equiv \frac{1}{N_{cp}} 
\qquad
\&
\qquad
W^m_{k,i} \equiv W^m\left[  (\bar{x}_{k} - \bar{x}_{i})/ h \right].
\end{equation}
The exact equations for the fields on the grid edges are given by Equations~(\ref{density-interpolation},\ref{Discrete-ME}), these can be written as
\begin{equation}
\bar{E}_{k+1} 
=
\bar{E}_k + h \bar{\rho}_{k+\frac{1}{2}}
\qquad
\&
\qquad
\bar{\rho}_{k+\frac{1}{2}} 
=
N_{cp}
 \sum_s \bar{Q}_s 
\sum_{i_s}  
 W^m_{k+\frac{1}{2}, i_s}.
 \label{ek}
\end{equation}

The solution of Equation~\eqref{ek} can be expressed as
\begin{eqnarray}  
\label{Eksol}
\bar{E}_k &=&  \bar{E}_0 + h \sum_{j=0}^{k-1}  \bar{\rho}_{j+\frac{1}{2}} 
	 	=	 \bar{E}_0 - h \sum_{j=k}^{N_c-1} \bar{\rho}_{j+\frac{1}{2}} 
	 	= 	 \bar{E}_0 + \frac{h}{2} \sum_{j=0}^{k-1} \bar{\rho}_{j+\frac{1}{2}} -  \frac{h}{2} \sum_{j=k}^{N_c-1} \bar{\rho}_{j+\frac{1}{2}} \nonumber \\
	&=&  \bar{E}_0  -  \frac{h}{2} \bar{\rho}_{k+\frac{1}{2}} + \frac{h}{2} \sum_{j=0}^{N_c-1} A_{jk}   \bar{\rho}_{j+\frac{1}{2}},
\end{eqnarray}
where $A_{jk}$ is anti-symmetric matrix given by
\begin{align}
A_{jk} &= 
\begin{cases}
		+1 	&	j < k, \\
		0 	&	j = k, \\ 
		-1	&	j > k. \\
\end{cases}
\end{align}

Two possible, second-order accurate, approximations for Equation~\eqref{fexact} can be written as follows
\begin{numcases}
{\bar{F}_{i_s} \approx  }
 \bar{Q}_{s} \sum_{k=0}^{N_c-1} \bar{E}_k  W^m_{k,i_s} .
\label{approximation1}
\\ \nonumber \\ 
\frac{ \bar{Q}_{s} }{2} \sum_{k=0}^{N_c-1} \left[\bar{E}_k +\bar{E}_{k+1} \right] W^m_{k+\frac{1}{2},i_s} = \frac{ \bar{Q}_{\alpha} }{2} \sum_{k=0}^{N_c-1}  \bar{E}_k \left[ W^m_{k+\frac{1}{2},i_s}+ W^m_{k-\frac{1}{2},i_s} \right]. 
\label{approximation2}
\end{numcases}

\subsection{Non-momentum Conserving Second-order scheme: approximation~\eqref{approximation1}}
\label{app:approximation1}

If we use \eqref{Eksol} and net-charge neutrality, i.e., $\sum_k \bar{\rho}_k =\sum_k \bar{\rho}_{k+\frac{1}{2}} =0$, then the total force on all macro-particles is given by
\begin{eqnarray} \label{fn1}
\bar{F}_{Net} &=& \sum_s  \sum_{i_s} \bar{F}_{i_s}
 =  \sum_k \bar{E}_k \sum_s \bar{Q}_s \sum_{i_s} W^m_{k,i_s} =  N_{pc}  \sum_k \bar{E}_k \bar{\rho}_k 
=		- N_{pc}  \frac{h}{2}  \sum_k \bar{\rho}_k \bar{\rho}_{k+1/2}  +  N_{pc}  \frac{h}{2} \sum_{j,k} A_{jk} \bar{\rho}_k \bar{\rho}_{j+\frac{1}{2}} .
\end{eqnarray}
Therefore, the net-force on the system does not depend on the choice of $E_0$ (because of charge neutrality), and since using higher interpolation functions makes the variation in the interpolated density smoother, it decreases both terms in Equation~\eqref{fn1}, i.e., using higher-order interpolation \textit{improves} the momentum conservation.

\subsubsection{Origin of Momentum Non-conservation: Self-forces and wrong interaction forces}

To see the origin of momentum non-conservation, we examine the interpolated force in the case in which there are only two macro-particles. Using  $\bar{Q}_1+\bar{Q}_2=0$,
$ \sum_k W^m_{k,i_s} =1$, and
 
\[\bar{\rho}_{j+\frac{1}{2}} = N_{cp}   \left[ \bar{Q}_1  W^m_{j+\frac{1}{2},1} + \bar{Q}_2 W^m_{j+\frac{1}{2},2} \right],
\]
the force on macro-particle at $x_1$ is given by
\begin{eqnarray}
\bar{F}_1		
&=&
	\bar{Q}_1 \sum_{k=0}^{N_c-1} \left(   \bar{E}_0  -  \frac{h}{2} \bar{\rho}_{k+\frac{1}{2}} + \frac{h}{2} \sum_{j=0}^{N_c-1} A_{jk}   \bar{\rho}_{j+\frac{1}{2}} \right) W^m_{k,1}  
\nonumber \\
&=& 
	\bar{Q}_1 E_0 - \frac{ \bar{Q}_1 h}{2}  N_{cp} \sum_k   \left[ \bar{Q}_1 W^m_{k+\frac{1}{2},1} + \bar{Q}_2 W^m_{k+\frac{1}{2},2} \right]  W^m_{k,1}  
+	  
	\frac{\bar{Q}_1 h}{2}  N_{cp}  \sum_{j,k}  A_{jk}    \left[  \bar{Q}_1 W^m_{k+\frac{1}{2},1} + \bar{Q}_2 W^m_{k+\frac{1}{2},2} \right]  W^m_{k,1}  
\nonumber \\
&=& 
	\bar{Q}_1 E_0  + \frac{ \bar{Q}_1^2 h}{2}   N_{cp} \left[   \sum_{j,k}   A_{jk} W^m_{j+\frac{1}{2},1} W^m_{k,1} - \sum_k W^m_{k+\frac{1}{2},1} W^m_{k,1} \right] 
+
	\frac{ \bar{Q}_2 \bar{Q}_1 h}{2}   N_{cp} \left[   \sum_{j,k}   A_{jk} W^m_{j+\frac{1}{2},2} W^m_{k,1} - \sum_k W^m_{k+\frac{1}{2},2} W^m_{k,1} \right].
\end{eqnarray}
Therefore, the non-vanishing self-force, $\bar{F}_{\text{self}} $, and the numerical interaction force, $\bar{F}_{\text{int}}$, are given by
\begin{eqnarray}
\bar{F}_{\text{self}} 
&=& 
\frac{ \bar{Q}_1^2 h}{2}   N_{cp} \left[    \sum_{j,k}   A_{jk} W^m_{j+\frac{1}{2},1} W^m_{k,1} - \sum_k W^m_{k+\frac{1}{2},1} W^m_{k,1} \right]  \neq 0
\quad
\text{and}
\quad
\bar{F}_{\text{int}}
=
\frac{ \bar{Q}_2 \bar{Q}_1 h}{2}   N_{cp} \left[    \sum_{j,k}  A_{jk} W^m_{j+\frac{1}{2},2} W^m_{k,1} - \sum_k W^m_{k+\frac{1}{2},2} W^m_{k,1} \right].
\nonumber \\
\end{eqnarray}
Therefor, the self-force ($\sim h/N_{pc} = L/N_p$) here vanishes only in the limit of infinite number of macro-particles $N_p \rightarrow \infty$.

\subsection{Momentum Conserving Second-order Scheme: approximation~\eqref{approximation2}}

\label{app:approximation2}
If we use \eqref{Eksol} and net-charge neutrality, i.e., $\sum_k \bar{\rho}_k = \sum_k \bar{\rho}_{k+1/2} =0$, and
\begin{eqnarray}
\label{eq:Ek+Ek+1}
\bar{E}_{k} + \bar{E}_{k+1} 
&=& 
2 \bar{E}_0 + h \sum_{j=0}^{k-1} \bar{\rho}_{j+\frac{1}{2}} - h \sum_{j=k+1}^{N_c-1} \bar{\rho}_{j+\frac{1}{2}} 
= 2 \bar{E}_0 + h \sum_{j=0}^{N_c-1} A_{jk} \bar{\rho}_{j+\frac{1}{2}}.
\end{eqnarray}
The net-force is, then, given by
\begin{eqnarray} \label{fn2}
\bar{F}_{Net} 
=
\sum_s  \sum_{i_s} \bar{F}_{i_s} 
=  
\sum_k \frac{\bar{E}_{k} + \bar{E}_{k+1}}{2} \sum_s \bar{Q}_s \sum_{i_s}  W^m_{k+\frac{1}{2},i_s} 
=
N_{pc} \sum_k  \frac{\bar{E}_{k} + \bar{E}_{k+1}}{2 } \bar{\rho}_{k+\frac{1}{2}} 
=
\frac{h N_{pc} }{2 }  \sum_{j,k}   A_{jk} \bar{\rho}_{j+\frac{1}{2}} \bar{\rho}_{k+1/2} = 0.
\nonumber \\
\end{eqnarray}
Therefore, the net-force on the system, again, does not depend on the choice of $E_0$ (because of charge neutrality), and it is always exactly zero, therefore, the momentum is exactly conserved. Hence we call this a momentum conserving scheme.

\subsubsection{Vanishing of the Self-forces exactly}
\label{app:MC-exact}

To see how self-forces vanish, we examine the interpolated force in the case in which there are only two macro-particles. Using  $\bar{Q}_1+\bar{Q}_2=0$,
$  \sum_k W^m_{k,i_s} =1$ and
\[
\bar{\rho}_{j+\frac{1}{2}} = N_{cp}   \left[ \bar{Q}_1  W^m_{j+\frac{1}{2},1} + \bar{Q}_2 W^m_{j+\frac{1}{2},2} \right],
\]
the force on macro-particle at $x_1$ is given by
\begin{eqnarray}
\bar{F}_1		
&=&
\bar{Q}_1 \sum_{k=0}^{N_c-1}  \frac{\bar{E}_{k} + \bar{E}_{k+1}}{2} W^m_{k+\frac{1}{2},1}  
=   
\bar{Q}_1 \sum_{k=0}^{N_c-1}  \left[ E_0 + \frac{h  }{2}    \sum_{j=0}^{N_c-1} A_{jk} \bar{\rho}_{j+\frac{1}{2}}  \right]  W^m_{k+\frac{1}{2},1}   
\nonumber \\
&=& 
\bar{Q}_1 \bar{E}_0  +  
\frac{\bar{Q}_1 h}{2}  N_{cp}  
 \sum_{j,k}  A_{jk}  \left[ \bar{Q}_1  W^m_{j+\frac{1}{2},1} + \bar{Q}_2 W^m_{j+\frac{1}{2},2} \right] W^m_{k+\frac{1}{2},1}
=
\bar{Q}_1 \bar{E}_0  
+ 
\frac{\bar{Q}_1 \bar{Q}_2 h}{2}  N_{cp}  \sum_{j,k}   A_{jk}  W^m_{j+\frac{1}{2},2}   W^m_{k+\frac{1}{2},1}.
\end{eqnarray}
Therefore, the vanishing self-force, $\bar{F}^{\text{self}}$, and the numerical interaction force, $\bar{F}^{\text{int}} $, are given by
\begin{eqnarray}
\label{appr2-f12}
\bar{F}^{\text{self}} 
=
\frac{\bar{Q}_1^2 h}{2}  N_{cp}  \sum_{j,k}  A_{jk}  W^m_{j+\frac{1}{2},1}   W^m_{k+\frac{1}{2},1}
= 0
\qquad
\&
\qquad
\bar{F}^{\text{int}}
=
\frac{\bar{Q}_1 \bar{Q}_2 h}{2}  N_{cp}  \sum_{j,k}   A_{jk}  W^m_{j+\frac{1}{2},2}   W^m_{k+\frac{1}{2},1}.
\end{eqnarray}
We can see from comparing Equations~\eqref{fm0} and \eqref{appr2-f12} that the numerically calculated interaction force is shorter in range, and that the usage of higher-order interpolation (larger macro-particles) results in smoothing the interaction forces.

\subsection{Filtering for Momentum Conserving Scheme: Non-vanishing Self-forces}

\label{app:filtering}

Here, we show that filtering of the grid charge densities, e.g. 1D equivalence of filtering implemented in TRISTAN-MP, lead to a non-vanishing self-forces and a violation of momentum conservation.

After one-filter, the filtered charge densities $\bar{\rho}^{\rm f}_k$ are given by
\begin{equation}
\label{eq:1-filter}
\bar{\rho}^{\rm f}_{k+\frac{1}{2}} 
= 
\frac{
\bar{\rho}_{k-\frac{1}{2}}+
2\bar{\rho}_{k+\frac{1}{2}}+
\bar{\rho}_{k+\frac{3}{2}}
}{4}.
\end{equation}
Therefore, for a momentum conserving scheme, e.g., Equation~\eqref{approximation2}, if we used \eqref{eq:Ek+Ek+1} and replaced $\rho_{k+1/2} $ by $\rho_{k+1/2}^{\rm f}$, the net force is given by
\begin{equation}
\bar{F}_{Net} 
=
\sum_k \frac{\bar{E}_{k} + \bar{E}_{k+1}}{2} \sum_s \bar{Q}_s \sum_{i_s}  W^m_{k+\frac{1}{2},i_s} 
=
\frac{N_{pc} h}{2}  
\sum_{k,j} 
A_{jk}
\bar{\rho}_{k+\frac{1}{2}}^{\rm f}
\bar{\rho}_{k+\frac{1}{2}}
=
\frac{N_{pc} h}{8}  
\sum_{k,j} 
A_{jk}
\left(
\bar{\rho}_{k-\frac{1}{2}}+
\bar{\rho}_{k+\frac{3}{2}}
\right)
\bar{\rho}_{k+\frac{1}{2}} 
\neq
0.
\end{equation}
Which shows that filtering the grid deposited moment (grid charge density) leads to violation of momentum conservation in otherwise a momentum conserving scheme.

To see the origin of such violation, we look, as we did before, at the interpolated force in case of having only two macro-particles, using  $\bar{Q}_1+\bar{Q}_2=0$,
$ \sum_k W^m_{k,i_s} =1$ and
\[
\bar{\rho}_{j+\frac{1}{2}} 
= 
N_{cp}   
\left[ 
\bar{Q}_1  W^m_{j+\frac{1}{2},1} + \bar{Q}_2 W^m_{j+\frac{1}{2},2} 
\right]
\quad
\Rightarrow
\quad
\bar{\rho}^{\rm f}_{j+\frac{1}{2}} 
=
\frac{N_{cp}}{4}
\left[
\bar{Q}_1  
\left( 
W^m_{j-\frac{1}{2},1} + 
2 W^m_{j+\frac{1}{2},1} +
W^m_{j+\frac{3}{2},1}
\right)
+
\bar{Q}_2
\left( 
W^m_{j-\frac{1}{2},2} + 
2 W^m_{j+\frac{1}{2},2} +
W^m_{j+\frac{3}{2},2}
\right)
\right],
\]
the force on macro-particle at $x_1$ is given by
\begin{eqnarray}
\bar{F}_1		
&=&
\bar{Q}_1 \sum_{k=0}^{N_c-1}  \frac{\bar{E}_{k} + \bar{E}_{k+1}}{2} W^m_{k+\frac{1}{2},1}  
=   
\bar{Q}_1 \sum_{k=0}^{N_c-1}  \left[ E_0 + \frac{h  }{2}    \sum_{j=0}^{N_c-1} A_{jk} \bar{\rho}^{\rm f}_{j+\frac{1}{2}}  \right]  W^m_{k+\frac{1}{2},1} 
\nonumber \\
&=& 
\bar{Q}_1 \bar{E}_0  
+  
\frac{\bar{Q}_1 h}{4}  N_{cp} 
\sum_{j,k}   A_{jk}  
\left[ \bar{Q}_1  W^m_{j+\frac{1}{2},1} + \bar{Q}_2 W^m_{j+\frac{1}{2},2} \right] W^m_{k+\frac{1}{2},1}
\nonumber \\
&&
+
\frac{ \bar{Q}_1 h }{8} N_{cp}
\sum_{j,k}   A_{jk}  
\left[ \bar{Q}_1 
\left(
W^m_{j+\frac{3}{2},1} + W^m_{j-\frac{1}{2},1}
\right) W^m_{k+\frac{1}{2},1} 
+
\bar{Q}_2 
\left(
W^m_{j+\frac{3}{2},2} + W^m_{j-\frac{1}{2},2}
\right) W^m_{k+\frac{1}{2},1} 
\right]
\nonumber \\
&=&
\bar{Q}_1 \bar{E}_0  
+ 
\frac{\bar{Q}_1 \bar{Q}_2 h}{4}  N_{cp} 
\left[ 
\sum_{j,k}  A_{jk}  W^m_{j+\frac{1}{2},2}   W^m_{k+\frac{1}{2},1}
+
\frac{1}{2}
\sum_{j,k}  A_{jk} 
\left(
W^m_{j+\frac{3}{2},2} + W^m_{j-\frac{1}{2},2}
\right) W^m_{k+\frac{1}{2},1} 
\right]
+
\frac{ \bar{Q}^2_1 h }{8} N_{cp}
\sum_{j,k}  A_{jk} 
\left(
W^m_{j+\frac{3}{2},1} + W^m_{j-\frac{1}{2},1}
\right) W^m_{k+\frac{1}{2},1}.
\nonumber \\
\end{eqnarray}

Therefore, filtering leads to wrong interaction forces and a non-vanishing self-force given by
\[
\bar{F_1}^{\rm self}
=
\frac{ \bar{Q}^2_1 h }{8} N_{cp}
 \sum_{j,k}
 A_{jk}
\left(
W^m_{j+\frac{3}{2},1} + W^m_{j-\frac{1}{2},1}
\right) W^m_{k+\frac{1}{2},1} 
\neq 0.
\]

\section{Shape and weight functions explicit form	}
\label{app:shapes-weights}

If we define $ y = (x - x_i) / \Delta x = (\bar{x} - \bar{x}_i)/h $, the shape functions, $S^m(y)$, assumed for the macro-particles and the corresponding weight functions, $W^m(y)$, used in the interpolation steps in the code are given in Table~\ref{tab:shapes-ws}.
\begin{deluxetable*}{p{0.5cm}p{8cm}p{8cm}} 
\tabletypesize{\footnotesize}
\tablecolumns{3} 
\tablewidth{0pt}
\tablecaption{Shape and weight functions implemented in the code.
\label{tab:shapes-ws}}
\tablehead
{	
$m$			 	& 
$S^m(y)$ 		&
$W^m(y)$			
}	
\startdata
\\
0
&
$\delta (y)$ 
& 
$
\begin{cases}
1 &0\leq |y|\leq \dfrac{1}{2} \\
 0 & \text{Otherwise} 
\end{cases}
$ 
\\
\\
1 
&
$ \dfrac{1}{\Delta x}
\begin{cases}
1 &0\leq |y|\leq \dfrac{1}{2} \\\\
 0 & \text{Otherwise} 
\end{cases}
$
&
$
\begin{cases}
 1-|y| & 0\leq |y|<1 \\\\
 0 & \text{Otherwise}  
\end{cases}
$
\\
\\
2 
&
$ \dfrac{1}{\Delta x}
\begin{cases}
 1-|y| & 0\leq |y|<1 \\\\
 0 & \text{Otherwise}  
\end{cases}
$
&
$
\begin{cases}
 \dfrac{3}{4}-y^2 & 0<|y|<\dfrac{1}{2} \\\\
 \dfrac{1}{8} (3-2 |y|)^2 & \dfrac{1}{2}\leq |y|<\dfrac{3}{2} \\\\
 0 & \text{Otherwise}  
\end{cases}
$
\\
\\
3 
&
$ \dfrac{1}{\Delta x}
\begin{cases}
 \dfrac{3}{4}-y^2 & 0<|y|<\dfrac{1}{2} \\\\
 \dfrac{1}{8} (3-2 |y|)^2 & \dfrac{1}{2}\leq |y|<\dfrac{3}{2} \\\\
 0 & \text{Otherwise}  
\end{cases}
$
&
$
\begin{cases}
\dfrac{2}{3} - y^2 + |y|^3/2  &   0<|y|<1 \\\\
 \dfrac{1}{6} (2-|y|)^3 & 1\leq |y| <2 \\\\
 0 & \text{Otherwise}  
\end{cases}
$
\\
\\
4 
&
$ \dfrac{1}{\Delta x}
\begin{cases}
\dfrac{2}{3} - y^2 + |y|^3/2  &   0<|y|<1 \\\\
 \dfrac{1}{6} (2-|y|)^3 & 1\leq |y| <2 \\\\
 0 & \text{Otherwise}  
\end{cases}
$
&
$\begin{cases}
\dfrac{115}{192} -\dfrac{5 y^2}{8} + \dfrac{y^4}{4} &  0<|y|<\dfrac{1}{2} \\\\
 \dfrac{1}{96} \left[55 + 20 |y| -120 y^2 +80 |y|^3 - 16 y^4 \right] & \dfrac{1}{2}\leq |y|<\dfrac{3}{2} \\\\
 \dfrac{1}{384} (5-2 |y|)^4 & \dfrac{3}{2}\leq |y|<\dfrac{5}{2} \\\\
 0 & \text{Otherwise}  
\end{cases}
$
\\
\\ 
5 
&
$ \displaystyle
 \dfrac{1}{\Delta x}
\begin{cases}
\dfrac{115}{192} -\dfrac{5 y^2}{8} + \dfrac{y^4}{4} &  0<|y|<\dfrac{1}{2} \\\\
 \dfrac{1}{96} \left[55 + 20 |y| -120 y^2 +80 |y|^3 - 16 y^4 \right] & \dfrac{1}{2}\leq |y| <\dfrac{3}{2} \\\\
 \dfrac{1}{384} (5-2 |y|)^4 & \dfrac{3}{2}\leq |y|<\dfrac{5}{2} \\\\
 0 & \text{Otherwise} 
\end{cases}
$
&
$ \displaystyle
\begin{cases}
\dfrac{11}{20} -\dfrac{y^2}{2} +\dfrac{y^4}{4}-\dfrac{|y|^5}{12} & 0\leq |y| \leq 1 \\\\
\dfrac{17}{40} +\dfrac{5 |y|}{8} -\dfrac{7 y^2}{4} +\dfrac{5 |y|^3}{4} -\dfrac{3 y^4}{8}  +\dfrac{|y|^5}{24}& 1<|y|<2 \\\\
\dfrac{1}{120} (3-|y|)^5 & 2\leq |y|<3 \\\\
 0 & \text{Otherwise}
\end{cases}
$
\enddata
\end{deluxetable*}

\section{Aliasing}
\label{app:aliasing}

The Fourier components of grid quantities, in 1D, $g_k$ are such that $g_k=g_{k+pk_g}$, where $p$ is some integer and $k_g = 2 \pi /\Delta x$ is the wave mode associated with the  cell size $\Delta x$ on that physical grid. Therefore, for a continuous particles number density $n(x)$, the Fourier component of the grid charge density is given by \citep{Birdsall-Langdon}
\begin{equation}
\label{eq:rho_k-n_k}
\tilde{\rho}_k = \sum_{p=-\infty}^{\infty} \tilde{n}(k-pk_g) \hspace{0.08cm} \tilde{S}(k-pk_g),
\end{equation} 
where $\tilde{S}(k)$ is Fourier transform of our interpolation function and $\tilde{n}(k)$ is Fourier transform of $n(x)$.
Therefore, all aliases of $k$ (wave modes that differ from $k$ by integer number of $k_g$) contribute when grid quantities are calculated. Clearly, this will feedback on the particle quantities, when the grid quantities are used to calculate the force on the particles to evolve them. The strength of the coupling between aliases (the source of this error) depends on how fast $\tilde{S}^m(k)$ falls off for large $k$, as can be seen in Equation~\eqref{eq:rho_k-n_k}.

The Fourier transform of our interpolation functions (spline functions of order $m$,
see Table~\ref{tab:shapes-ws}) is given by
\begin{equation}
\label{eq:Sk}
\tilde{S}^m(k) = 
\left[
\frac{ \sin (k \Delta x/2)}{ k \Delta x/2}
\right]^{m}.
\end{equation}
Therefore, Using higher-order interpolation functions (larger $m$) in our code leads to a decrease in the strength of the couplings between grid wave modes and their aliases), which results in improvements in energy conservation as seen in Section~\ref{sec:thermal_stability}.

\section{Poisson noise}

\label{app:noise}
Here we calculate the noise when a finite number of computational particles are used to represent a uniform distribution function. We calculate the total energy density due to such noise in Appendix~\ref{app:Poisson-noise} and then find the power spectrum for such noise in Appendix~\ref{app:average2}. In Appendix~\ref{app:thetap}, we calculate the temperature, $\theta_p$, set by the energy in such noise.

\subsection{Average Potential Energy From Uniformly Distributed Macro-particles}
\label{app:Poisson-noise}

Using Equation~\eqref{moment1} and the first equation in \eqref{ME2}, we can write the electric field associated with plasma particles on a periodic box of length $L$, i.e., $x_{i_s} \in [0,L)$ 
as follows
\begin{eqnarray}
\label{ExE0}
E(x) -E_0  
=
\sum_s \frac{Q_s}{\epsilon_0}  \sum_{i_s}^{N_s} q^m(x,x_{i_s}).
\end{eqnarray}
Where, $ q^m(x,x_{i_s}) = \int_0^x  dx^{'} S^m(x^{'},x_{i_s} ) $. The periodicity of the box implies that the plasma is neutral. Therefore, 
\begin{equation}
E_L - E_0 
=
0
=
\frac{1}{\epsilon_0}\sum_s Q_s  \sum_{i_s}^{N_s} q^m(L,x_{i_s})
=
\frac{1}{\epsilon_0} \sum_s Q_s N_s.
\end{equation}

The spatial averaging of $q^m$, for uniformly distributed macro-particles, is
\begin{equation}
\label{eq:qm-average}
\left\langle  q^m(x,x_{i_s})   \right\rangle 
=
\int_0^L \frac{du}{L} q^m(x,u)
=
\int_0^x  dx^{'}  \int_0^L\frac{du}{L}     S^m(x^{'},u )
=
\int_0^x \frac{dx^{'}  }{L} (1)
=
 \frac{x  }{L} .
\end{equation}

For such macro-particles the average of the electric field is zero:
\begin{equation}
\left\langle E(x) -E_0 \right\rangle 
=
\sum_s \left\langle \frac{Q_s}{\epsilon_0}  \sum_{i_s}^{N_s}   q^m(x,x_{i_s})  \right\rangle 
=
\sum_s   \frac{Q_s}{\epsilon_0}  N_s \int_0^L \frac{du}{L} q^m(x,u) 
=
\sum_s \frac{Q_s}{\epsilon_0}  N_s \frac{x}{L}
=
\frac{x}{\epsilon_0 L} \sum_s Q_s  N_s 
=
0.
\end{equation}

However, due to the finite number of macro-particles, the average potential energy is non-zero, to calculate such energy we need to calculate
\begin{eqnarray}
\left\langle E^2(x) - E_0^2 \right\rangle
&=&
\left\langle (E(x) - E_0)^2 \right\rangle
= 
\left\langle \left( \sum_s \frac{Q_s}{\epsilon_0}  \sum_{i_s}^{N_s} q^m(x,x_{i_s}) \right)^2 \right\rangle
\nonumber \\
&=&
\left\langle 
\sum_s \left(  \frac{Q_s}{\epsilon_0} \sum_{i_s}^{N_s} q^m(x,x_{i_s}) \right)^2 \right\rangle
+
\left\langle \sum_{s \neq s^{'}}  \left( \frac{Q_s}{\epsilon_0}  \sum_{i_s}^{N_s} q^m(x,x_{i_s}) \right)  \left( \frac{Q_{s^{'}}}{\epsilon_0}  \sum_{i_{s^{'}}}^{N_{s^{'}}} q^m(x,x_{i_{s^{'}}}) \right)\right\rangle
\nonumber \\
&=&
\frac{1}{\epsilon_0^2}
\sum_s \left\langle \left(  Q_s  \sum_{i_s}^{N_s} q^m(x,x_{i_s}) \right)^2 \right\rangle
+
\frac{1}{\epsilon_0^2}
\sum_{s \neq s^{'}} \left\langle  \left( Q_s  \sum_{i_s}^{N_s} q^m(x,x_{i_s}) \right) \right\rangle \left\langle  \left( Q_{s^{'}}  \sum_{i_{s^{'}}}^{N_{s^{'}}} q^m(x,x_{i_{s^{'}}}) \right)\right\rangle
\nonumber \\
&=&
\frac{1}{\epsilon_0^2}
\sum_s Q^2_s
\left\langle   
\sum_{i_s }^{N_s} [q^m(x,x_{i_s})]^2  
+
\sum_{i_s\neq j_s}^{N_s}  q^m(x,x_{i_s}) q^m(x,x_{j_s}) 
\right\rangle
+
\frac{1}{\epsilon_0^2}
\frac{x^2}{L^2} \sum_{s \neq s^{'}} Q_s N_s Q_{s^{'}} N_{s^{'}} 
\nonumber \\
&=&
\sum_s \frac{ Q^2_s  }{\epsilon_0^2}
\left[ 
N_s \int_0^L \frac{du}{L} [q^m(x,u)]^2  
+
 N_s(N_s-1)  \left[ \int_0^L \frac{du}{L}  q^m(x,u) \right]^2 
\right]
+
\frac{1}{\epsilon_0^2}
\frac{x^2}{L^2} \sum_{s \neq s^{'}} Q_s N_s Q_{s^{'}} N_{s^{'}} 
\nonumber \\
&=&
\frac{1}{\epsilon_0^2}
\frac{x^2}{L^2} 
\left( 
\sum_{s \neq s^{'}} Q_s N_s Q_{s^{'}} N_{s^{'}} 
+
\sum_s  N_s^2 Q^2_s 
\right)
+
\sum_s  \frac{ Q^2_s  N_s  }{\epsilon_0^2}
\left[ 
\int_0^L \frac{du}{L} [q^m(x,u)]^2  
-
\frac{x^2}{L^2}  
\right]
\nonumber \\
&=&
\frac{x^2}{\epsilon_0^2 L^2} 
\left( \sum_{s } Q_s N_s \right)^2
+
\sum_s   \frac{ Q^2_s  N_s }{\epsilon_0^2}
\left[ 
\int_0^L \frac{du}{L} [q^m(x,u)]^2  
-
\frac{x^2}{L^2}  
\right]
=
\sum_s    \frac{ Q^2_s  N_s  }{\epsilon_0^2}
\left[ 
\int_0^L  [q^m(x,u)]^2  \frac{du}{L}
-
\frac{x^2}{L^2}  
\right].
\nonumber \\
\label{ex2}
\end{eqnarray}

For the shape functions implemented in SHARP-1D (their explicit forms are given in Appendix~\ref{app:shapes-weights}), the integral in \eqref{ex2} is given by
\begin{eqnarray}
\int_0^L \frac{du}{L} [q^m(x,u)]^2  
&=&
\frac{x}{L} - \frac{ \Delta x }{L} f_m 
\qquad
and
\qquad
f_m 
=
\frac{1}{6}
\begin{cases}
0  			& m=0\\
1 			& m=1\\
1.4			& m=2\\
1.70714 		& m=3\\
1.96693	 	& m=4\\
2.19624		& m=5\\
\end{cases}.
\label{fm}
\end{eqnarray}

Therefore,
\begin{eqnarray}
\label{eq:exav}
\left\langle E^2(x) - E_0^2 \right\rangle
&=&
\sum_s \frac{N_s   Q^2_s }{\epsilon_0^2}
\left[ 
\frac{x}{L} 
-
\frac{x^2}{L^2} 
-
\frac{ \Delta x }{L} f_m  
\right].
\end{eqnarray}

The average electrostatic potential energy due to the finite number of macro-particles is, then, given by
\begin{eqnarray}
\mathscr{E}^m 
&=&
\frac{\epsilon_0}{2} \int_0^L dx \left\langle E^2(x) - E_0^2 \right\rangle
=
\sum_s \frac{N_s   Q^2_s }{2 \epsilon_0 }
\left[ 
\frac{L}{6}
-
\Delta x f_m 
\right]
=
 \frac{L }{12 \epsilon_0 }
\left[ 
1
-
\frac{6 f_m}{N_c} 
\right]
\sum_s  N_s   Q^2_s
\label{eq:Poisson}
\end{eqnarray}
Here, $N_c = L/\Delta x$ is the number of macro-cells. If we assume that all plasma species have the same mass, and absolute value of charge, we then make the choice of our fiducial units as, $q^2_0 = Q^2_s$ and $m_0 = M_s$ (that implies $n_0 = \sum_s n_s$). Therefore,

\begin{eqnarray}
\label{Poisson_Normalized}
\left\langle \mathscr{\bar{E}}^m  \right\rangle
&=&
 \frac{ \left\langle \mathscr{E}^m \right\rangle}{m_0 c^2}  
=
\frac{L^2}{12 c^2} \frac{q_0^2 n_0}{ \epsilon_0 m_0}
\left[ 
1
-
\frac{6 f_m}{N_c} 
\right]
=
\frac{L^2 \omega_0^2}{12 c^2}
\left[ 
1
-
\frac{6 f_m}{N_c} 
\right]
=
\frac{ \bar{L}^2 }{12}
\left[ 
1
-
\frac{6 f_m}{N_c} 
\right]
\end{eqnarray}

Equation~\eqref{Poisson_Normalized} shows that using higher-order shape functions decreases the noise coming from the fact that we are using a finite number of macro-particles. 
The decrease that we gain in the potential energy noise is $ f_m  \bar{L}^2 /2 N_c $.
For a given box size, this improvement is lowered, if we increase the number of cells $N_c$ because it means a decreases in cell size, which means also a decrease in the size of the macro-particles.
On the other hand, if we increase the number of cells while keeping the cell size fixed, i.e., by increasing the box-size $\bar{L}$, that improvement due to using higher-order interpolation functions increases.

\subsection{Spectrum of the Poison Noise}
\label{app:average2}

To find the spectrum of such Poisson noise, we average the Fourier transform of the grid electric fields. Using Equation~\eqref{ExE0}, the Fourier components of the electric field are given by
\begin{eqnarray}
\tilde E_n
&=&
\int_0^L \frac{dx}{L}  \hspace{0.1cm}  
\left[ E_0  + \sum_s \frac{Q_s}{\epsilon_0}  \sum_{i_s}^{N_s} q^m(x,x_{i_s}) \right] 
e^{ - 2 \pi i n x / L }
=
E_0 \delta_{n,0}
+
 \sum_s \frac{Q_s}{\epsilon_0}  \sum_{i_s}^{N_s}
\int_0^L \frac{dx}{L}  \hspace{0.1cm}  
\left[   q^m(x,x_{i_s}) \right] 
e^{ - 2 \pi i n x / L }.
\end{eqnarray}

By defining
\begin{eqnarray}
Z(x_{i_s},n)
& \equiv &
\int_0^L \frac{dx}{L}
  q^m(x,x_{i_s}) e^{ - 2 \pi i n x / L } 
\quad
\Rightarrow
\quad
Z(x_{i_s},0) = \frac{x_{i_s}}{L},
\label{eq:zm-def}
\end{eqnarray}
and using the fact that $E(x)$ is a real valued function, the absolute value for such Fourier components are given by
\begin{eqnarray}
|\tilde E_{n}|^2  
&=& 
\tilde E_{n} \tilde E_{-n}
=
E^2_0 \delta_{n,0} 
+
2 E_0 \delta_{n,0} 
\sum_s \frac{Q_s}{\epsilon_0}  \sum_{i_s}^{N_s} Z(x_{i_s},0) 
+
\left[ \sum_s \frac{Q_s}{\epsilon_0}  \sum_{i_s}^{N_s} Z(x_{i_s}, n) \right] 
\left[ \sum_s \frac{Q_s}{\epsilon_0}  \sum_{i_s}^{N_s} Z(x_{i_s},-n) \right]
\nonumber \\
&=&
E_0 
\left[ 
E_0
+
2 
\sum_s \frac{Q_s}{\epsilon_0}  \sum_{i_s}^{N_s} \frac{x_{i_s}}{L}
\right] \delta_{n,0} 
+
\sum_s \frac{Q^2_s}{\epsilon^2_0} 
\left[ 
\sum_{i_s}^{N_s} Z(x_{i_s}, n) 
\sum_{j_s}^{N_s} Z(x_{j_s},-n)
\right] 
+
\sum_{s\neq s^{'}} \frac{Q_s Q_{s^{'}}}{\epsilon^2_0} 
\left[ 
\sum_{i_s}^{N_s} Z(x_{i_s}, n) 
\sum_{i_{s^{'}}}^{N_{s^{'}}} Z(x_{i_{s^{'}}},-n)
\right] 
\nonumber \\
&=&
E_0 
\left[ 
E_0
+
2 
\sum_s \frac{Q_s}{\epsilon_0}  \sum_{i_s}^{N_s} \frac{x_{i_s}}{L}
\right] \delta_{n,0} 
+
\sum_s \frac{Q^2_s}{\epsilon^2_0} 
\left[ 
\sum_{i_s}^{N_s} Z^m(x_{i_s}, n) Z^m(x_{i_s},-n)
+
\sum_{i_s \neq j_s }^{N_s} Z^m(x_{i_s}, n) Z(x_{j_s},-n)
\right] 
\nonumber \\
&&
+
\sum_{s\neq s^{'}} \frac{Q_s Q_{s^{'}}}{\epsilon^2_0} 
\left[ 
\sum_{i_s}^{N_s} Z^m(x_{i_s}, n) 
\sum_{i_{s^{'}}}^{N_{s^{'}}} Z^m(x_{i_{s^{'}}},-n)
\right] .
\end{eqnarray}

Averaging over a periodic box and assuming the macro-particles are uniformly distributed, we can write
\begin{eqnarray}
 \langle |\tilde E_{n}|^2   \rangle
&=& 
E_0 
\left[ 
E_0
+
2 
\sum_s \frac{Q_s N_s }{2 \epsilon_0}   
\right] \delta_{n,0}
+
\sum_s \frac{Q^2_s}{\epsilon^2_0} 
\left[ 
N_s  \int_0^L \frac{du}{L} \left[  Z^m(u, n) Z^m(u,-n) \right]  
+
(N_s^2 - N_s )
 \int_0^L \frac{du}{L}   Z^m(u, n) 
 \int_0^L \frac{du}{L}   Z^m(u, -n)
\right] 
\nonumber \\
&&
+
\sum_{s\neq s^{'}} \frac{Q_s N_s Q_{s^{'}} N_{s^{'}} }{\epsilon^2_0} 
 \int_0^L \frac{du}{L}   Z^m(u, n) 
 \int_0^L \frac{du}{L}   Z^m(u, -n)
\nonumber \\
&=&
E^2 _0   \delta_{n,0}
+
\left[ 
\sum_{s\neq s^{'}} Q_s N_s Q_{s^{'}} N_{s^{'}} 
+
\sum_{s} Q^2_s N^2_s  \right]
 \int_0^L \frac{du}{L}   Z^m(u, n) 
 \int_0^L \frac{du}{L}   Z^m(u, -n) 
\nonumber \\
&&
+
\sum_s \frac{Q^2_s N_s }{\epsilon^2_0} 
\left[ 
 \int_0^L \frac{du}{L} \left[  Z^m(u, n) Z^m(u,-n) \right]  
-
 \int_0^L \frac{du}{L}   Z^m(u, n) 
 \int_0^L \frac{du}{L}   Z^m(u, -n)
\right] 
\nonumber \\
&=&
E^2 _0   \delta_{n,0} 
+
\sum_s \frac{Q^2_s N_s }{\epsilon^2_0} 
\left[ 
 \int_0^L \frac{du}{L} \left[  Z^m(u, n) Z^m(u,-n) \right]  
-
 \int_0^L \frac{du}{L}   Z^m(u, n) 
 \int_0^L \frac{du}{L}   Z^m(u, -n)
\right].
\end{eqnarray}

By using Equations~(\ref{eq:zm-def}, \ref{eq:qm-average})
\begin{eqnarray}
 &&\int_0^L \frac{du}{L}   Z^m(u, n) 
=
 \int_0^L \frac{du}{L}  \int_0^L \frac{dx}{L}  q^m(x,u) e^{ - 2 \pi i n x / L } 
=
\int_0^L \frac{dx}{L} \frac{x}{L}  e^{ - 2 \pi i n x / L } 
=
\begin{cases}
1/2,& n =  0,
\\
i/(2 \pi n),&n \neq 0,
\end{cases}
\\
\nonumber 	\\
\nonumber 	\\
&&
\int_0^L \frac{du}{L}    Z^m(u, n)  Z^m(u,- n)
=
\begin{cases}
1/3,& n =  0,
\\
\dfrac{1}{ (2 \pi n )^2} \left[ 1+ \left( \dfrac{\sin(\pi n /N_c)}{\pi n /N_c} \right)^{2m} \right]  ,&n \neq 0.
\end{cases}
\end{eqnarray}
Therefore, the averaged magnitude for the Fourier components can be written as
\begin{eqnarray}
\langle |\tilde E_{n}|^2   \rangle
=
\begin{cases}
E^2 _0  + \displaystyle \sum_s \dfrac{ Q^2_s N_s }{12 \epsilon^2_0} ,& n =  0,
\\
\displaystyle \sum_s \dfrac{ Q^2_s N_s }{(2 \pi n)^2 \epsilon^2_0}   \left[ \dfrac{\sin(\pi n /N_c)}{\pi n /N_c} \right]^{2m}
,&n \neq 0.
\end{cases}
\label{Eq:Noise-fourier}
\end{eqnarray}
If all plasma species have the same mass, and absolute value of charges, such average can be written in code units as
\begin{eqnarray}
\langle |\bar{ \tilde E}_{n}|^2   \rangle
&=&
\frac{ \langle | \tilde E_{n}|^2   \rangle }{\mathbb{  E}_0^2}   
=
\begin{cases}
\bar{E}^2 _0  + \dfrac{\bar{L}^2}{ 12 N_{\rm t}} ,& n =  0.
\\
\\
 \dfrac{ 1 }{(2 \pi n)^2 }   
\left[ \dfrac{\sin(\pi n /N_c)}{\pi n /N_c} \right]^{2m}
\dfrac{\bar{L}^2}{N_{\rm t}} 
,&n \neq 0.
\end{cases}
\end{eqnarray}

\subsection{Heating Due to Noise}

\label{app:thetap}

If the energy due to Poisson noise is converted to heat that puts a floor in the temperature PIC scheme can simulate. Here we estimate such a temperature floor $\theta_p$.
The energy due to Poisson noise is calculate in Appendix \ref{app:Poisson-noise}. It is given by
\begin{eqnarray}
\left\langle \mathscr{\bar{E}}^m  \right\rangle
&=&
\frac{ \left\langle \mathscr{ E}^m \right\rangle }{m_0 c^2}
=
\frac{ \bar{L}^2 }{12}
\left[ 
1
-
\frac{6 f_m}{N_c} 
\right],
\end{eqnarray}
where $f_m$ is defined in \eqref{fm}. This noise is due to the finite number of
macro-particles used in the simulations. If this energy is converted to thermal energy, it would lead to heating of the plasma up to a temperature $\theta_p$. If the plasmas are at thermal equilibrium, the momentum distribution is given by Maxwell--J\"uttner distribution, and hence the temperature of plasmas is related to the kinetic energy, $\mathscr{\bar{K}}$, as follows
\begin{eqnarray}
\left\langle \mathscr{\bar{K}} \right\rangle
&=& 
\frac{ \left\langle \mathscr{K} \right\rangle }{m_0 c^2}
=
\sum_s  \left\langle  (\gamma -1) \right\rangle
=
\sum_s  N_s  \left[ \theta_s +  \frac{ K_0[1/\theta_s ]}{K_1[1/\theta_s]}-1  \right]
\nonumber \\
&=&
N  \left[ \theta +  \frac{ K_0[1/\theta ]}{K_1[1/\theta]}-1  \right]
=
\begin{cases}
N \theta /2, \qquad \theta  \ll 1,
\\ 
N \theta , \quad \qquad  \theta \gg  1.
\end{cases}
\end{eqnarray}
Here, $\theta_s = k_B T_s / m_0 c^2$ is the normalized temperature of species $s$ with $N_s$ of macro-particles, $N$ is the total number of macro-particles from all species and $K_0$, $K_1$ are the Bessel functions of zeroth and first kind respectively.
Therefore, 
\begin{equation}
\label{Eq:thetap2}
\theta_p = \frac{ \bar{L}^2 }{12 N }
\left[ 
1
-
\frac{6 f_m}{N_c} 
\right] 
\begin{cases}
2, \qquad \theta_p  \ll 1,
\\ 
1 , \qquad \theta_p \gg  1.
\end{cases}
\end{equation}
Hence, if a plasma of macro-particles starts with a temperatures below $\theta_p$ the Poisson noise will non-physically heat such plasma.

\section{Dispersion relation For Non-relativistic warm plasma}
\label{app:Dispersions}
For non-relativistic ($\gamma^3 \approx 1$) warm plasma, i.e., $ 0 < \theta \ll 1 $, we can write
\[
f_0(u) du = f_0(v)dv =  \frac{n_0 d(v/c)}{\sqrt{ 2 \pi \theta }  } e^{- (v/c)^2 / 2 \theta} 
=  \frac{n_0 d\bar{v}}{\sqrt{ 2 \pi \theta }  } e^{- \bar{v}^2 / 2 \theta}
\qquad
\&
\qquad
\theta =  \frac{k_B T }{ m c^2}
.
\]
If we assume no net current in the plasmas, i.e., the momentum distribution of all species is such that $\sum_s Q_s  \int v f^s_0(v)  dv =0$, then the linear dispersion relation of uniformly distributed plasma is given by 
\begin{equation}
1 =\sum_s \chi_s(v_p),
\end{equation}
where $v_p = \hat{\omega}/\hat{k}$, $\hat{\omega} = \omega / \omega_{p}$, $ \hat{k} = kc \sqrt{\theta} / \omega_{p}$, i.e., $v_p = \omega/kc\sqrt{\theta}$,  $\omega^2_{p} = \sum_s \omega^2_{ps} $, $\omega^2_{ps} = Q_s^2 n_s / \epsilon_0 M_s $, and $\bar{v} = v/c$.

If we assume that $\Im(\hat{\omega}) >0$, i.e.,  $\Im(\omega) >0$, then
\begin{eqnarray}
\chi_s(v_p) 
&=&
\frac{Q_s^2}{ \epsilon_0 M_s k^2} \int_{-\infty }^{\infty}  \frac{f_0(v) dv }{ (v -  \omega/k)^2  } 
= 
\frac{ \omega^2_{ps}}{ k^2 c^2} \int_{ - \infty }^{ \infty } \frac{d \bar{v}}{\sqrt{ 2 \pi \theta }  } \frac{ e^{- \bar{v}^2 / 2 \theta} }{ (\bar{v} -  \omega / k c)^2  } 
= 
\frac{ \omega^2_{ps} / \omega^2_{p}}{ \hat{\omega}^2} \int_{ - \infty }^{ \infty } \frac{dz}{\sqrt{  2 \pi  }  } \frac{ e^{- z^2/2  } }{ (z -  v_p)^2  },
\end{eqnarray}
where $z \equiv \bar{v}/\sqrt{ \theta }$.
Extending the definition of $\chi_s(v_p)$ to the entire  complex plane can be done as follows \citep{Brambilla+1998}
\begin{eqnarray}
\chi_s(v_p) 
=
\frac{ \hat{\omega}_s^2}{ \hat{k}^2} 
\int_{ - \infty }^{ \infty } \frac{dz}{\sqrt{  2 \pi  }  } \frac{ e^{- z^2/2  } }{ (z -  v_p)^2  }
-
\frac{ \hat{\omega}_s^2}{ \hat{k}^2} 
 \sqrt{ \frac{  \pi }{ 2} }  v_p
\begin{cases}
0									& \text{if } \Im(v_p) > 0  \\
i   e^{-v_p^2/2}		& \text{if } \Im(v_p) = 0  \\
2 i   e^{-v_p^2/2}		& \text{if } \Im(v_p) < 0
\end{cases},
\end{eqnarray}
where $\hat{\omega}_s \equiv \omega_{ps} / \omega_{p}$. Therefore, for $\Im(v_p) \neq 0$, we can then write

\begin{eqnarray}
\chi_s(v_p) 
&=&
\frac{ \hat{\omega}_s^2}{ \hat{k}^2} 
\left[
\begin{cases}
-1+\sqrt{\dfrac{\pi}{2}}  v_p \left[ \text{Erfi}  \left( v_p/\sqrt{2} \right)-i\right] e^{-\frac{v_p^2}{2}}				& \text{if } \Im(v_p) > 0  \\
-1+\sqrt{\dfrac{\pi}{2}}  v_p \left[ \text{Erfi}  \left( v_p/\sqrt{2} \right)+i\right] e^{-\frac{v_p^2}{2}}		 		& \text{if } \Im(v_p) < 0
\end{cases} 
-
\sqrt{\dfrac{\pi}{2}}  v_p
\begin{cases}
0									& \text{if } \Im(v_p) > 0  \\
2 i  e^{-v_p^2/2}		& \text{if } \Im(v_p) < 0
\end{cases} 
\right]
\nonumber \\
&=&
\frac{ \hat{\omega}_s^2}{ \hat{k}^2} 
\left[
-1+ \sqrt{\frac{\pi}{2}}   v_p \left[ \text{Erfi}  \left( v_p/\sqrt{2} \right)-i\right] e^{-\frac{v_p^2}{2}}
\right],
\end{eqnarray}
where $\text{Erfi} $ is the complex error function, which is defined as $\text{Erfi}(v_p) = -i  \text{ Erf}(i v_p)  $.

\subsection{Standing Linear Plasma Waves}
\label{app:LLD}

In the case of thermal electrons with fixed neutralizing background ($\hat{\omega}_s=1$), the dispersion relation is then given by
\begin{eqnarray}
\hat{k}^2+1 
&=& 
\sqrt{\frac{\pi}{2}}  v_p \left[ \text{Erfi}  \left( v_p/\sqrt{2} \right)-i\right] e^{-\frac{v_p^2}{2}}
\end{eqnarray}

\subsection{Two-stream Instability}
\label{app:TwoStream}

In the case of two population of thermal electrons (both have the same number density), propagating in two opposite directions with speed $v_b$, with fixed neutralizing background, therefore $\hat{\omega}^2_s=1/2$, and the linear dispersion relation is then given by ($z_b \equiv v_b/c \sqrt{\theta}$)
\begin{eqnarray}
\hat{k}^2 +1
&=& 
\frac{1}{2} 
\left[
\sqrt{ \frac{\pi}{2}} ( v_p + z_b )\left[ \text{Erfi}  \left( \frac{ v_p + z_b }{\sqrt{2} }\right)-i\right] e^{-\frac{( v_p + z_b )^2}{2}}
+
\sqrt{ \frac{\pi}{2}} ( v_p - z_b )\left[ \text{Erfi}  \left( \frac{ v_p - z_b }{\sqrt{2} }\right)-i\right] e^{-\frac{( v_p - z_b )^2}{2}}
\right]
\nonumber \\
&=& 
\sqrt{ \frac{\pi}{8}} 
\left[
 ( v_p + z_b )\left[ \text{Erfi}  \left( \frac{ v_p + z_b }{\sqrt{2} }\right)-i\right] e^{- v_p  z_b }
+
 ( v_p - z_b )\left[ \text{Erfi}  \left( \frac{ v_p - z_b }{\sqrt{2} }\right)-i\right] e^{ v_p  z_b }
\right]
e^{-( v^2_p + z^2_b)/2}.
\end{eqnarray}

\end{appendix}

\bibliography{refs}
\bibliographystyle{apj}

\end{document}